\documentclass[aps,prb,twocolumn,superscriptaddress,showpacs,showkeys]{revtex4-1}

\usepackage{float}
\usepackage{latexsym}
\usepackage{mathtools}
\usepackage{amssymb}
\usepackage{amsmath}
\usepackage{epsfig}
\usepackage{graphicx} 
\usepackage{balance}
\usepackage{braket}
\usepackage{amsfonts}

\usepackage[font=small, justification=raggedright, singlelinecheck=false, format=plain]{caption}
\usepackage[font=small, justification=justified, singlelinecheck=false, format=plain]{subcaption}
\usepackage{float}
\usepackage{dsfont}  
\usepackage{wasysym}
\usepackage{stackengine}

\usepackage{color}
\usepackage[dvipsnames]{xcolor}

\newlength{\upit}\upit=0.1truein

\newcommand{\ltappr}{{{\lower4pthbox{$<$} } \atop \widetilde{ \ \ \ }}}
\newlength{\bxwidth}\bxwidth=1.5 truein

\newcommand{\rarrow}{\rightarrow}
\newcommand \bea {\begin{eqnarray} }
\newcommand \eea {\end{eqnarray}}

\newcommand{\bR}{{\bf{R}}}
\newcommand{\bQ}{{\bf{Q}}}

\begin{document}

\title{Classical phase diagram of the stuffed honeycomb lattice}

\author{Jyotisman Sahoo}
%\email{jsahoo@iastate.edu}
\affiliation{Department of Physics and Astronomy, Iowa State University, USA}
\author{Dmitrii Kochkov}
\affiliation{Institute for Condensed Matter Theory and Department of Physics, University of Illinois at Urbana Champaign, USA}
\author{Bryan K. Clark}
\affiliation{Institute for Condensed Matter Theory and Department of Physics, University of Illinois at Urbana Champaign, USA}
\author{Rebecca Flint}
\email{flint@iastate.edu}
\affiliation{Department of Physics and Astronomy, Iowa State University, USA}

\begin{abstract}
We investigate the classical phase diagram of the \emph{stuffed honeycomb} Heisenberg lattice, which consists of a honeycomb lattice with a superimposed triangular lattice formed by sites at the center of each hexagon. This lattice encompasses and interpolates between the honeycomb, triangular and dice lattices, preserving the hexagonal symmetry while expanding the phase space for potential spin liquids. We use a combination of iterative minimization, classical Monte Carlo and analytical techniques to determine the complete ground state phase diagram.  It is quite rich, with a variety of non-coplanar and non-collinear phases not found in the previously studied limits.  In particular, our analysis reveals the triangular lattice critical point to be a multicritical point with two new phases vanishing via second order transitions at the critical point.  We analyze these phases within linear spin wave theory and discuss consequences for the $S = 1/2$ spin liquid.
\end{abstract}
\maketitle

%%%%%%%%%%%%%%%%%%%%%%%%%%%%%%%%%%%%%%%%%%%%%%%%%%%%%%%%

\section{Introduction}

Realizing spin liquids, highly correlated and topological magnetic phases that host fractional excitations, is a key goal in correlated materials research \cite{wen91,senthil04,balents10}.  While there are now several good spin liquid candidates, particularly on the kagom\'{e} lattice \cite{shores05,helton07,yan11,han12}, we are far from realizing the full spectrum of possible spin liquids.  The search for new spin liquid materials is often frustrated by the narrow range of parameter space occupied by those spin liquid phases in realistic models. As increasing magnetic frustration stabilizes spin liquids, one possible way to find new or more stable spin liquids is to couple together two different frustrated lattices.  This paper studies the classical phase diagram of one such lattice, the \emph{stuffed honeycomb lattice}, which couples a honeycomb lattice to its dual triangular lattice.

Generically, coupled lattices have rich phase diagrams even at the classical level; for example, the related windmill lattice showcases intriguing $Z_6$ order by disorder, with a critical phase and Berezinskii-Kosterlitz-Thouless transitions at finite temperatures \cite{orth12,orth14,jeevanesan14,jeevanesan15}. Due to their non-Bravais nature, these lattices can generically host non-coplanar phases with nontrivial spin chirality; in the classical limit, this chirality can lead to Berry phases and anomalous Hall effects in metallic magnets \cite{nagaosa10}, and in the quantum limit can lead to chiral spin liquids \cite{kalmeyer87,wen89}, as found near the cuboc phase in the kagom\'{e} lattice \cite{gong15,gong14,hu14,wietek15,messio12,messio13}.

\begin{figure}[t]
\begin{center}
\includegraphics[scale = 0.18]{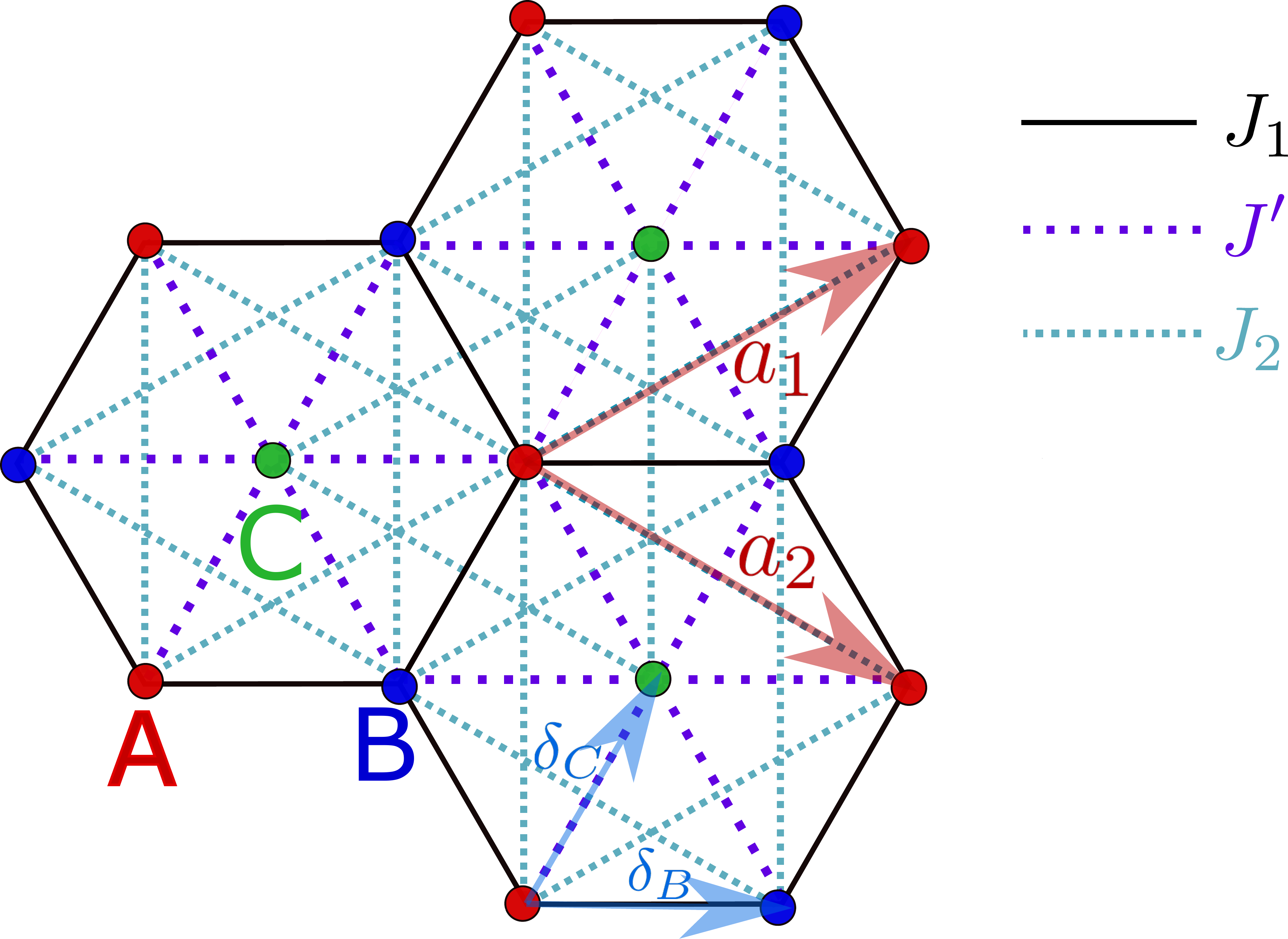}
\end{center}
\caption{\label{model}{Stuffed honeycomb lattice model. This model contains three sublattices, A (red) and B (blue) on the honeycomb sites, and C (green) on the triangular sites.  $J_1$ (solid, black) is the nearest-neighbor coupling between honeycomb sites, while $J'$ (dashed, purple) is the coupling between honeycomb and C sites. All three sublattices have next-nearest neighbor $J_2$'s that we take to be identical (dotted, blue).  The lattice vectors ${\bf a}_1$ and ${\bf a}_2$, as well as the basis vectors $\pmb{\delta}_B$ and $\pmb{\delta}_C$ are shown.}}
\end{figure}

In this paper, we discuss the classical phase diagram of the stuffed honeycomb lattice Heisenberg model, a two-dimensional model that combines the honeycomb and triangular lattices by adding a spin to the center of each hexagon of a honeycomb lattice. We consider nearest ($J_1$) and next-nearest ($J_{2H}$) neighbor couplings on the honeycomb lattice, and nearest neighbor ($J_T$) couplings on the centered triangular lattice, as well as a nearest-neighbor coupling of the two lattices, $J'$, as shown in Fig. \ref{model}. For simplicity, we define a single second neighbor coupling, $J_2 \equiv J_T = J_{2H}$; the related windmill lattice instead takes $J_{2H} = 0$ \cite{orth12}.  This model then interpolates from the honeycomb lattice at $J' = 0$ to the triangular lattice for $J' = J_1$, both of which potentially host narrow spin liquid regions, and out to the dice lattice for $J' = \infty$, all the while maintaining the hexagonal symmetry, in contrast to the usual anisotropic triangular lattices \cite{shimizu03,yamashita08,itou08,itou09,chubukov13}. As such, this model provides the perfect playground to explore the potential existing spin liquids on the honeycomb \cite{gong13} and triangular lattice \cite{zhu15,hu15} limits by enlarging their possible phase space into another dimension.  

In this paper, we focus on the classical phase diagram, which includes several non-coplanar phases, and the transitions between them. Perhaps the most interesting result is that the weakly first order transition between 120$^\circ$ and collinear orders as a function of $J_2/J_1$ on the triangular lattice is revealed to be a multi-critical point between four phases, with two new second order lines joining at that point.  Here, we see the origin of the strong fluctuations that give rise to the spin liquid in the $S = 1/2$ model.

The classical triangular lattice forms 120$^\circ$ order for $J_2/J_1 < 1/8$, and collinear order for $J_2/J_1 > 1/8$, with a weak first-order transition between the two.  For $S = 1/2$, this transition broadens into a spin-liquid region extending from $.06 < J_2/J_1 < .16$ \cite{li15,zhu15,hu15}.  While the existence of this spin liquid region is well-established, the nature of the spin liquid is not.  The spin liquid may be gapless \cite{iqbal16,saadatmund16,wietek17}, and small perturbations of many types seem to lead to different spin liquids, from gapped \cite{zhu15, hu15, wietek17,saadatmand16} to nematic \cite{hu15} to chiral \cite{hu16,wietek17,gong17}.  

The classical honeycomb lattice is bipartite for $J_2 = 0$, forming a N\'{e}el phase that gives way to a planar spiral phase for $J_2/J_1 > 1/6$ \cite{}.  Quantum fluctuations enhance the N\'{e}el phase, and it extends to $J_2/J_1 = .2$ for $S = 1/2$, while the spiral phase is destroyed in favor of a plaquette valence bond solid phase (VBS) \cite{albuquerque11,clark11,ganesh13,zhu13,gong13}.  The region near $J_2/J_1 = .2 - .25$ may form a spin liquid \cite{gong13}, or may be a deconfined critical point between the N\'{e}el and VBS phases \cite{senthilDC,albuquerque11,ganesh13}.  The potential spin liquid has been proposed to be either a gapped $\mathbb{Z}_2$ ``sublattice-pairing state'' \cite{lu11,clark11}, or a $\mathbb{Z}_2$ $d+id$ Dirac spin liquid \cite{ferrari17}.

This model was initially introduced to attempt to explain the magnetic behavior of the cluster magnet LiZn$_2$Mo$_3$O$_8$ \cite{sheckelton12,mourigal14,sheckelton14}. This material consists of a triangular lattice of Mo$_3$O$_{13}$ molecular clusters, each of which hosts a single, isotropic $S = 1/2$. Above 100K, all spins are visible in the Curie-Weiss susceptibility, while below 100K, two-thirds of the spins vanish.  This disappearance led to the proposal of a spontaneous breaking of the lattice symmetry such that a VBS or spin liquid forms on an emergent honeycomb lattice, with the leftover one-third of the spins located in the centers of the hexagons \cite{flint13}. The remaining third of these spins do not order down to the lowest temperatures. The original paper proposed octahedral cluster rotations as the mechanism for symmetry breaking, although ordering in the LiZn$_2$ layer may be a more likely mechanism \cite{privateMcQueen}.  An alternate theoretical proposal of plaquette charge ordering on a $1/6$th-filled breathing kagom\'{e} lattice extended Hubbard model exists \cite{chen16,carrasquilla17,chen18}, which also requires an enlargement of the unit cell. Neither of these proposed enlargements has been seen \cite{sheckelton15}, although the breathing kagom\'{e} lattice structure is found in the related Li$_2$In$_{1-x}$Sc$_x$Mo$_3$O$_8$ materials \cite{akbari17}. 

Another class of possible materials realizations are spin chain materials like RbFeBr$_3$, which form quasi-1D spin chains arranged in the basal plane as a stuffed honeycomb lattice \cite{adachi83}; these spins are XY-like, and are thought to form a partially disordered antiferromagnetic phase, with one-third of the spin chains disordered in the basal plane.  This model has been studied for XY \cite{plumer91,zhang93} and Heisenberg \cite{nakano17,shimada18,gonzalez18,shimada182} spins with nearest-neighbor $J_1$ and $J'$ exchange.

Engineering this lattice is another potential path, either by intercalcating extra spins into existing inorganic honeycomb lattice materials like the oxalates \cite{lu99,pilkington01,jiang03}, or more straightforwardly by forming a triangular tri-layer with ABC stacking. The C sublattice forms the center layer, with $J_2$ couplings in plane, and $J'$ couplings to nearest neighbors in the A and B layers above and below.  The nearest neighbor couplings between the outer A and B layers are $J_1$. Here, the generically somewhat artificial condition that $J_T = J_{2H} \equiv J_2$ is natural, if the three sublattices are otherwise identical. Some fine-tuning would be required to obtain $J_1 \sim J'$, as generically $J_1$ will be the smallest coupling.

%(what is LiZn2Mo3O8?) (two-thirds of the spins vanish) (original proposal of VBS order --> emergent honeycomb lattice) (remaining one-third do not order down to the lowest temperatures) (Original paper proposed octahedral rotations as the mechanism, although it is also possible that the ordering of the LiZn$_2$ layer breaks this symmetry.) (Alternate symmetry breaking proposed a Hubbard model with charge ordering on breathing kagom\'{e} lattice - does appear to be relevant for Li$_2$In$_{1-x}$Sc$_x$Mo$_3$O$_8$) (their model is a spin liquid where 2/3rds of the spins are in a fully-filled spinon band far from the spinon Fermi surface, and 1/3rd of the spins are in a half-filled band at the spinon FS - and this is a U(1) Dirac SL - that is Chen + Kim's original model, Chen and Lee have a spinon FS model that includes also orbital d.o.f.) (most of these details won't go in, of course)

The organization of the paper is as follows. The model is introduced in Sec. \ref{mod}, methods are discussed in Sec. \ref{methods}, and the full classical phase diagram is shown  in Sec. \ref{phases}. The various phases are discussed in sections \ref{hl} to \ref{dcp}.
Given the importance of the multicritical point around the triangular limit, we introduce the two off-axis non-collinear phases in a separate section, Sec. \ref{ncp}, and discuss the effect of fluctuations.  Finally, we briefly summarize in Sec. \ref{con} and suggest future directions.

% Needs to be more formal (maybe).  I want to reorganize a bit, so will end up rewriting

\begin{figure*}[t]
\begin{center}
\includegraphics[scale = 0.17]{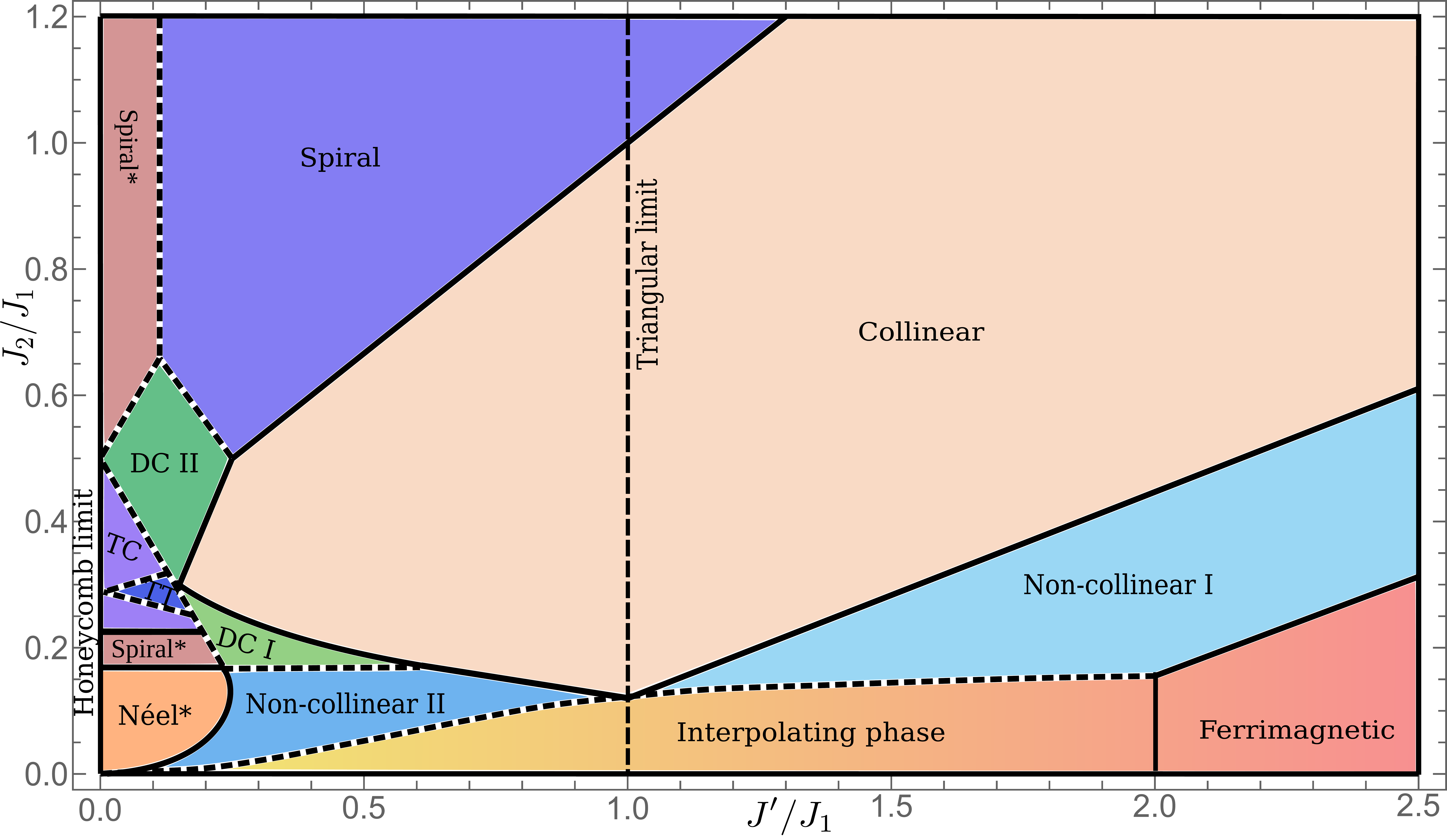}
\end{center}
\caption{\label{cpd}Classical ground state phase diagram as a function of $J_2/J_1$ versus $J'/J_1$.  This phase diagram interpolates between the honeycomb limit (far left) to the triangular limit (middle) and beyond. Thick dotted lines separating the phases indicate a first order phase transition while solid lines imply a continuous transition.  Details of each phase are given in the following sections: N\'{e}el*, spiral*, triple conical (TC) and ``triangle of triangles'' (TT) are described in Section \ref{hl}; the double conical phases, DC I and DC II are described in Section \ref{dcp}; the spiral, collinear, interpolating and ferrimagnetic phases are described in Section \ref{trilimit}; and the non-collinear phases I and II are discussed in Section \ref{ncp}.}
\end{figure*}

\section{Model}\label{mod}

The stuffed honeycomb lattice is shown in Fig.\ref{model}. It is a non-Bravais lattice with space group symmetry {\it p6m}.  The hexagonal lattice vectors are,
\begin{equation}\label{latticevec}
{\bf a}_1=\left(\frac{3}{2},\frac{\sqrt{3}}{2}\right); \;{\bf a}_2=\left(\frac{3}{2},-\frac{\sqrt{3}}{2}\right),
\end{equation}
where we take the nearest-neighbor distance between sites to be one. Two sites (A,B) are on the honeycomb lattice, while the C sites sit in the center of the hexagons; the basis vectors are,
\begin{equation}\label{basisvec}
\pmb{\delta}_A = (0,0);\,\, \pmb{\delta}_B = (\frac{1}{2\sqrt{3}},\frac{1}{2});\,\, \pmb{\delta}_C = (\frac{1}{\sqrt{3}},0).
\end{equation}

We consider Heisenberg spins with three different antiferromagnetic exchange interactions,
\begin{equation}\label{ham}
\mathcal{H} = J_1\sum_{\left<i,j\right>}\vec{S}_{i}^A\cdot \vec{S}_{j}^B+ J'\sum_{\mathclap{\substack{\left<i,j\right>\\ \eta=A,B}}}\vec{S}_{i}^{\eta}\cdot \vec{S}_{j}^{C} +  J_2\sum_{\mathclap{\substack{\left<\left<i,j\right>\right> \\ \eta = A,B,C}}}\vec{S}_{i}^{\eta}\cdot \vec{S}_{j}^{\eta} 
\end{equation}
$J_1$ and $J'$ both correspond to nearest-neighbor (NN) interactions. While $J_1$ couples the A and B sublattices, $J'$ couples the C sublattice with both A and B sublattices. $J_2$ is the next-nearest-neighbor (NNN) interaction, which couples spins in the same sublattice; we take $J_2$ on the honeycomb (AA,BB) and central spins (CC) to be identical for simplicity; although this identity is not required by symmetry, it is present in the triangular tri-layer. 

There are three limits of particular interest: $J_1 = J'$ gives the $J_1-J_2$ triangular lattice; $J' = 0$ yields a $J_1-J_2$ honeycomb lattice completely decoupled from a nearest-neighbor (here, $J_2$) triangular lattice; and finally $J_1 = 0$ gives the $J_1-J_2$ dice lattice, perhaps best known as the dual to the kagom\'{e} lattice.

\section{Methods}\label{methods}

While obtaining the ground state phase diagram for a Bravais lattice may be done by assuming a single $\bQ$ planar spiral variational ansatz, and minimizing $J(\bQ)$, non-Bravais lattices are generically more complicated and require a combination of numerical and analytical techniques.  Our goal is to obtain a variational ansatz for each phase, and to then find phase boundaries by comparing energies. As ansatz can be arbitrarily complicated, we first use iterative minimization to find the ground state configuration numerically at each point in the phase diagram, and then develop the corresponding variational ansatz that matches or beats the iterative minimization ground state energies.

Iterative minimization is a numerical technique that begins with a random spin configuration on a finite size lattice with periodic boundary conditions.  At each step in the algorithm, a spin is chosen randomly and aligned with the exchange field due to its neighbors. This exchange field can be seen by rewriting the Hamiltonian,
\begin{equation}
\mathcal{H} = \sum_i H_i, \quad \mathrm{where}\; H_i = \vec{S}_i\cdot(-\frac{1}{2}\sum_jJ_{ij}\vec{S}_j).
\end{equation}
The spin, $\vec{S}_i$ will then be set to,
\begin{equation}
\vec{S}_i \rarrow \frac{\sum_jJ_{ij}\vec{S}_j}{||\sum_jJ_{ij}\vec{S}_j||^2}.
\end{equation}
The algorithm is run until the energy converges.  In order to avoid finite size effects, and also to check that we avoid local minima, we ran the algorithm on lattices of all sizes from 4x4 to 30x30 unit cells, taking the minimum energy of these. 

We then used a variety of variational ansatz, each of which treats the classical spins as unit-vectors, setting $S = 1$.  Most of the phases fit into two classes of ansatz: a 3{\bf Q} ansatz we describe here, and a double conical ansatz described in section \ref{dcp}. When these two classes of ansatz failed, we developed new variational ansatze by examining the spin configurations given by iterative minimization. If our variational ansatz correctly describes the ground state, its energy is less than or equal to the minimum iterative minimization energy.  In principle, this process could miss states with unit cells larger than 30x30; here, we would expect the iterative minimization spin configurations to locally resemble the correct ground state, with topological defects or lock-in to nearby commensurate wave-vectors.  We have visually spot checked that the iterative minimization spin configurations locally match the configurations obtained from the variational ansatz.

The 3{\bf Q} ansatz allows each of the three sublattices to be treated independently. We define the sublattice spin, $\vec{S}^\eta(\bR_i)$, where $\bR_i$ denotes a Bravais lattice site and $\eta$ labels the sublattice. The most general form of this vector describes a conical spiral,
\begin{equation}
\vec{S}^\eta(\bR_i) = [\cos \theta \cos({\bf Q}\cdot{\bf R}_i),\cos \theta \sin({\bf Q}\cdot{\bf R}_i),\sin \theta],
\end{equation}
with the conical axis along the $\hat z$ direction and conical angle $\theta$.  The perpendicular spin components are determined by a planar spiral with ordering wave-vector $\bQ$.  Both $\theta$ and $\bQ$ are variational parameters.  We then require two sets of Euler angles to relate the three sublattices.  The A sublattice is chosen to be oriented as above, with the B axes rotated by Euler angles  $(\alpha,\beta,\gamma)$ and the C sublattice rotated by $(\alpha',\beta',\gamma')$.  Typically, most of these parameters are not needed to describe a phase; most phases are planar, with $\theta_\eta = 0$ and $\alpha = \alpha' = 0$, $\beta = \beta' = 0$. Once the relevant parameters are determined, and the classical energy minimized with respect to these parameters, the nature of the phases, and location and nature of the phase transitions can be determined. In particular, we can determine the first or second order nature of a phase transition by examining the derivatives of the energies at the phase boundaries. More complicated variational ansatz, like the double conical spiral and twelve-sublattice ansatze are described in the sections for each phase.

\section{Classical Phase Diagram}\label{phases}

\begin{figure}[H]
\begin{center}
\includegraphics[scale = 0.18]{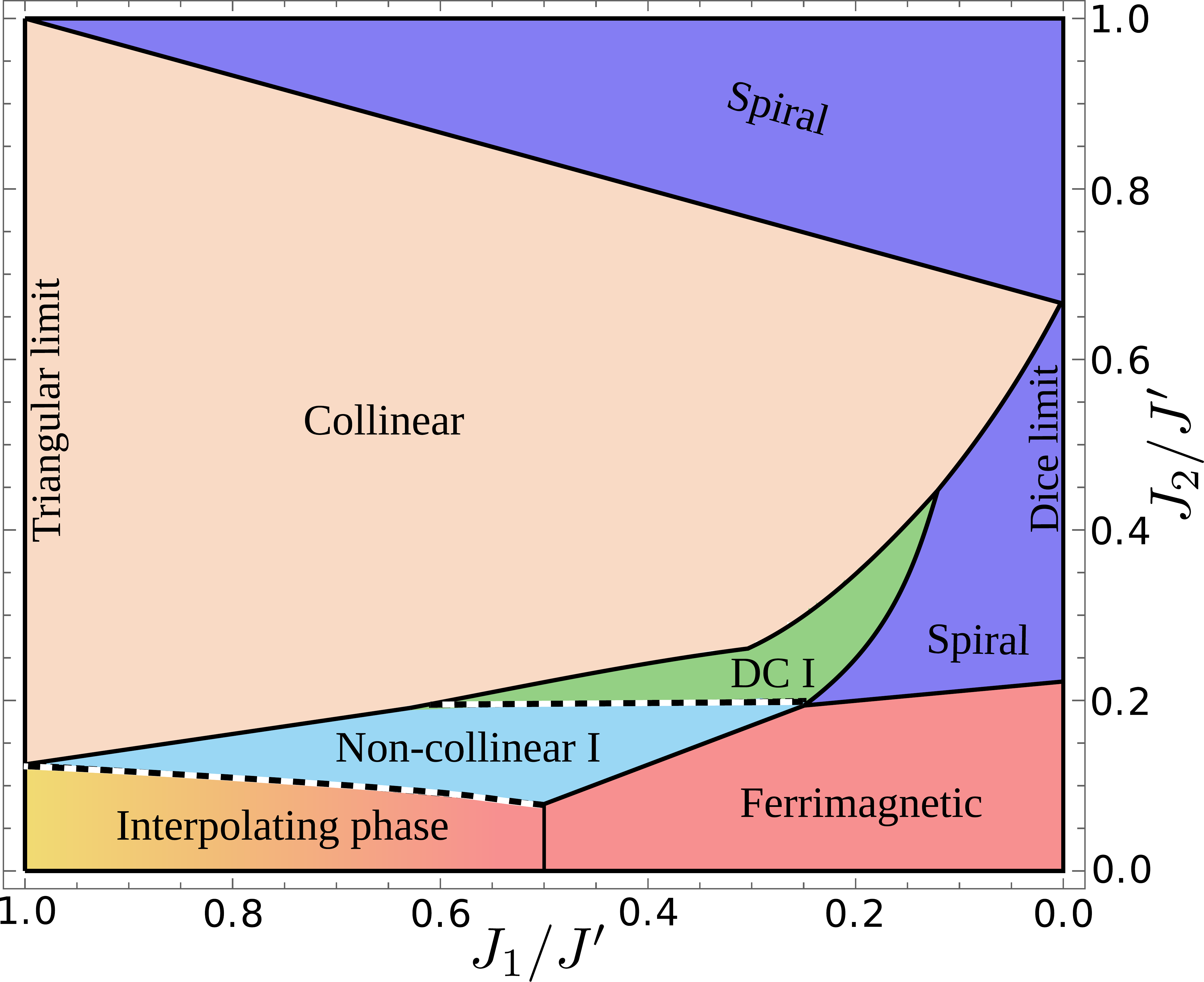}
\end{center}
\caption{\label{dpd} Classical ground state phase diagram as a function of $J_2/J'$ and $J_1/J'$; this phase diagram interpolates between the triangular (far left) and dice limits (far right). Dotted lines indicate a first order phase transition, while solid lines imply a continuous transition. Details of each phase are given in the following sections: the double conical phase, DC I is described in Section \ref{dcp}; the spiral, collinear, interpolating and ferrimagnetic phases are described in Section \ref{trilimit}; and non-collinear phase I is discussed in Section \ref{ncp}.}
\end{figure}

In this paper, we solve the classical, $S\rarrow \infty$ limit of this lattice for all values of $J'/J_1$ and $0 < J_2/J_1 < 1.2$; no new phases appear beyond this upper limit.  We show the phase diagram in two different figures in order to capture the relevant limits.  In Fig. \ref{cpd}, we plot the phase diagram as $J_2/J_1$ versus $J'/J_1$ in order to capture the interpolation from honeycomb to triangular lattice and beyond.  In Fig. \ref{dpd}, we instead plot the phase diagram as a function of $J_2/J'$ versus $J_1/J'$ in order to capture the evolution from the triangular to the dice limit.

%Classical ground state phase diagram as a function of $J_2/J_1$ versus $J'/J_1$.  This phase diagram interpolates between the honeycomb limit (far left) to the triangular limit (middle) and beyond. Thick dotted lines separating the phases indicate a first order phase transition while solid lines imply a continuous transition.  Details of each phase are given in the following sections: N\'{e}el*, spiral*, triple conical (TC) and ``triangle of triangles'' (TT) are described in Section \ref{hl}; the double conical phases, DC I and DC II are described in Section \ref{dcp}; the spiral, collinear, interpolating and ferrimagnetic phases are described in Section \ref{trilimit}; and the non-collinear phases I and II are discussed in Section \ref{ncp}.

First and second order transitions are indicated by dashed and solid lines, respectively. There are several multi-critical points.  We note that these naively seem to disobey the Gibbs phase rule, wherein we expect only three unrelated phases to meet at any given multicritical point in a two-dimensional phase diagram.  However, this constraint can be avoided when two or more of the phases are really different limits of the same ansatz.  For example, the collinear phase is a special case of non-collinear I and II, as well as double conical I and II; the N\'{e}el* phase is a special case of the spiral* phase; and the ferrimagnetic phase is a special case of both the interpolating and spiral phases.  These phases do, however, break different symmetries and are truly distinct.

\section{Phases near the honeycomb axis}\label{hl}

The classical ground state phase diagram of the honeycomb lattice itself is well known, with a N\'{e}el phase for $J_2/J_1 < 1/6$ and a spiral phase for $J_2/J_1 > 1/6$. For $J' = 0$, the central spins form 120$^\circ$ order on the C sublattice linked by $J_2$.  The small $J_2/J_1$ phases are unaffected by $J'$, but at larger $J_2/J_1$, the spiral is highly unstable.  With a small $J'/J_1$, the spiral phase distorts into one of three non-coplanar phases: the triple conical and triangle of triangles phases discussed below, which require many sublattices to describe, and a double conical phase, DC II discussed in section \ref{dc2}. This complexity suggests the fundamental instability of the spiral phase of the honeycomb lattice, and indeed that phase does not survive to $S = 1/2$, replaced by a VBS \cite{albuquerque11,clark11}.

\subsection{N\'{e}el* Phase}

\begin{figure}[H]
  \centering
  \includegraphics[width=.45\linewidth]{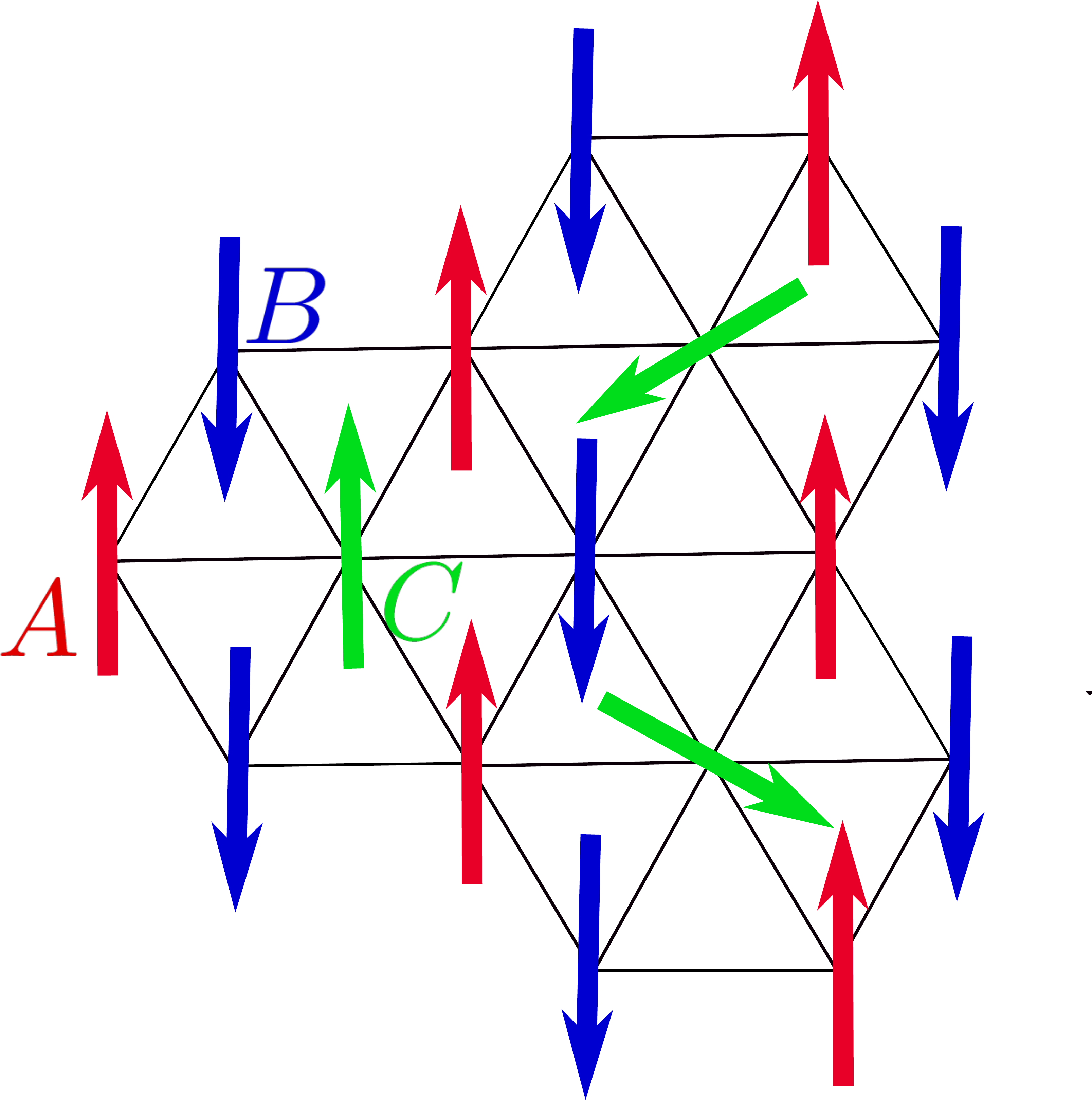}
\caption{N\'{e}el* phase. The A(red) and B(blue) spins are N\'{e}el ordered while the decoupled C (green) spins form $120^o$ order.  This particular arrangement is one of the six favored by thermal and quantum fluctuations.}
\label{neel}
\end{figure}

For $0<J_2/J_1<\frac{1}{6}$, the honeycomb spins (AB) order in the conventional N\'{e}el configuration while the C sublattice forms $120^o$ order, as shown in Fig. \ref{neel}.  In the classical, $T = 0$ limit, the C spins are completely decoupled from the AB spins, even for finite $J'$; we use the * suffix to indicate that the AB and C spins are decoupled, with the AB spins in their honeycomb limit phase, and the C spins forming 120$^\circ$ order. Thermal and quantum fluctuations will drive this phase into a coplanar order where one of the three C spin axes aligns with one of the AB spin axes.  This six-fold degeneracy leads to a $\mathds{Z}_6$ order driven by order by disorder \cite{orth12,orth14}. The classical energy for this phase is
\begin{equation}
E_{N\acute{e}el}[J_2] = -3 + 9/2 J_2,
\end{equation}
where for simplicity we set $J_1 = 1$ here, and in much of the rest of the paper. The spins are parametrized as 
\begin{align}
&\vec{S}_A = [0,0,1]; \;\vec{S}_B = [0,0,-1]\cr
&\vec{S}_C({\bf R}_i) = [\cos({\bf Q}_{tri}\cdot{\bf R}_i),0,\sin({\bf Q}_{tri}\cdot{\bf R}_i)],
\end{align}
where ${\bf Q}_{tri} = (\frac{2\pi}{3},\frac{2\pi}{3\sqrt{3}})$ is the $120^o$ ordering vector.

%%%%%%%%%%%%%%%%%%%%%%%%%%%%%%

\subsection{Spiral* phase}
\begin{figure}[t]
  \centering
\begin{subfigure}[t]{.2\textwidth}
\centering
  \includegraphics[width=\linewidth]{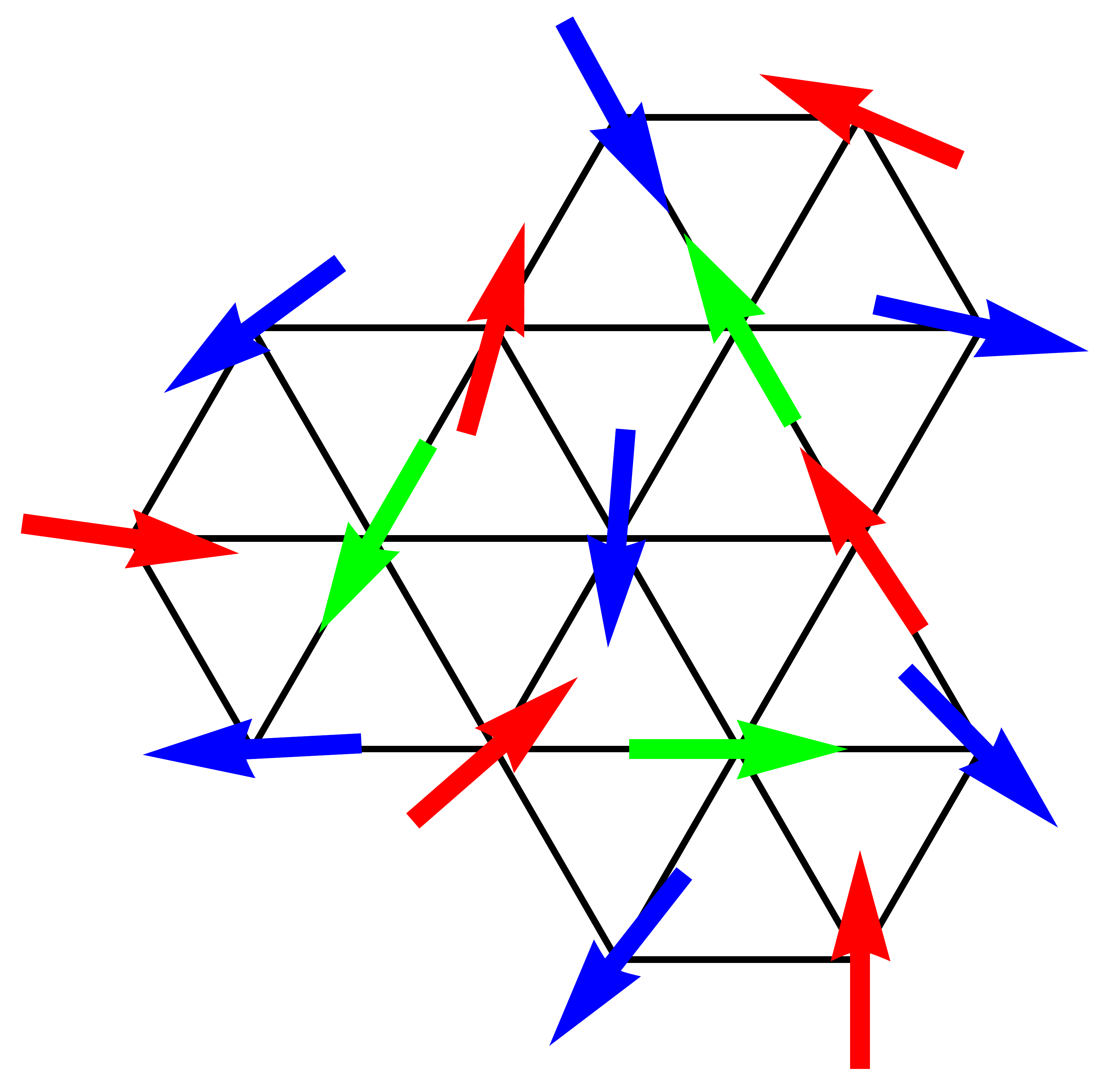}
\caption{\label{sp1}}
\end{subfigure}\hfill
\begin{subfigure}[t]{.24\textwidth}
  \centering
  \includegraphics[width=\linewidth]{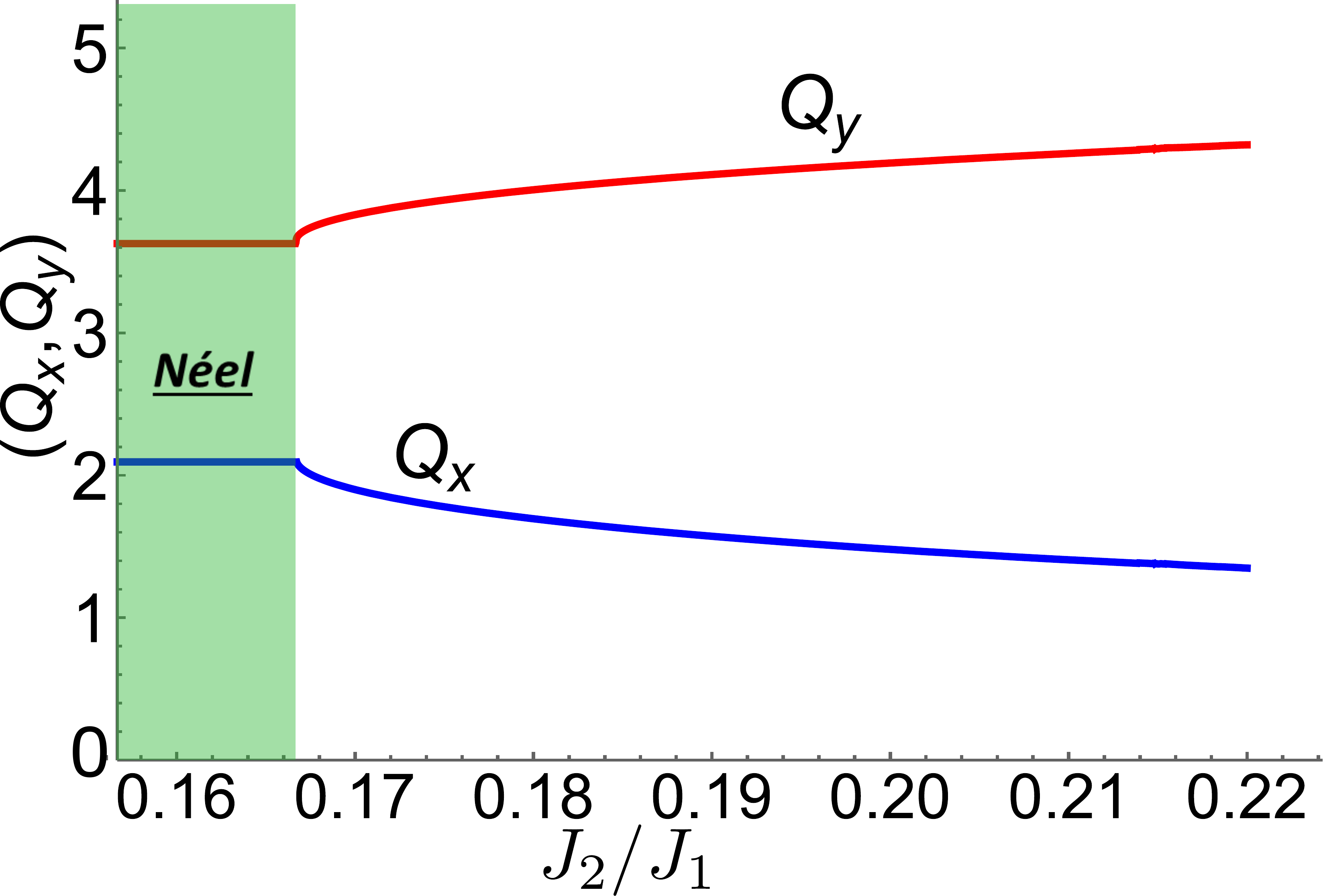}
\caption{\label{sp2}}
\end{subfigure}
\caption{\label{spiral}{Spiral* phase:  (a) There is an incommensurate coplanar spiral ordering on the A and B sublattices with {\bf $Q_{sp}$} being a function of $J_2$ only. The decoupled C spins have a $120^o$ order. (b) $Q_x(Q_y)$ plotted as a function of $J_2$. For $J_2/J_1 \leq 1/6$, {\bf $Q_{sp}$} limits to a constant value, and the N\'{e}el* phase is a special case of the spiral phase.  Note that all parameters of this phase are independent of $J'$.}}
\end{figure}

For $\frac{1}{6}<J_2 \lesssim 0.225$ and small $J'$, the honeycomb spins (AB) are driven into an incommensurate coplanar spiral order as shown in Fig. \ref{sp1}. The C sublattice remains decoupled and $120^o$ ordered even for finite $J'$, due to the cancellation of the overall exchange field at the C sites. In order to distinguish this phase from the planar spiral on all three sublattices, we call this phase spiral*, where * again indicates that the AB and C spins are decoupled. Quantum and thermal fluctuations again force the sublattices to be coplanar\cite{villain80,shendar,henley}.  The spin configuration is given by the variational ansatz,
\begin{align}
&\vec{S}_A({\bf R}_i) = [\cos({\bf Q}_{sp}\cdot{\bf R}_i),0,\sin({\bf Q}_{sp}\cdot{\bf R}_i)]\cr
&\vec{S}_B({\bf R}_i) = [\cos({\bf Q}_{sp}\cdot{\bf R}_i+\gamma),0,\sin({\bf Q}_{sp}\cdot{\bf R}_i+\gamma)]\cr
&\vec{S}_C({\bf R}_i) = [\cos({\bf Q}_{tri}\cdot{\bf R}_i),0,\sin({\bf Q}_{tri}\cdot{\bf R}_i)].
\end{align}
Here, the variational parameters are the spiral ordering wave-vector, ${\bf Q}_{sp}$ and the angle between the $A$ and $B$ spins in the same unit cell, $\gamma$. The variational energy of this phase is,
\begin{align}
\begin{split}
 E&_{spiral}[{\bf Q}_{sp}, \gamma, J_2] \\
&= \frac{J_2}{2} \Big[-3+ 4 \cos\{{\bf Q}_{sp}\cdot{({\bf a}_1 - {\bf a}_2)}\} + 4 \cos({\bf Q}_{sp}\cdot{{\bf a}_2})\\
&+ 4 \cos({\bf Q}_{sp}\cdot{{\bf a}_2})\Big] + \frac{1}{2} \Big[2 (1 + \cos\{{\bf Q}_{sp}\cdot{({\bf a}_1 - {\bf a}_2)}\} \\
&+ \cos\{{\bf Q}_{sp}\cdot{{\bf a}_1}\}) \cos(\gamma) - 2 (\sin\{{\bf Q}_{sp}\cdot{({\bf a}_1 - {\bf a}_2)}\} \\ 
& + \sin\{{\bf Q}_{sp}\cdot{{\bf a}_1}\}) \sin(\gamma)\Big].
\end{split}
\end{align}
Only the first term $-3J_2/2$ comes from the C spins.  Minimization of this function shows that ${\bf Q}_{sp} $ and $\gamma$ are independent of $J'$, as indeed is the entire energy of this phase.  We are also only finding one of a classically degenerate manifold of ${\bf Q}_{sp}$, which cause this phase to be strongly affected by quantum fluctuations\cite{okamura10}. The $J_2$ dependence of  ${\bf Q}_{sp} $ is shown in Fig. \ref{sp2}; note that for $J_2/J_1 = 1/6$,  ${\bf Q}_{sp} \rightarrow (0,0)$, and thus the N\'{e}el* phase is a special case of the spiral* phase, and the transition between the two is second order.  For $J_2/J_1 > 1/2$, the spiral* phase is again the lowest energy phase; it persists out to $J_2/J_1 = \infty$, where $\bQ_{sp}$ limits to $\bQ_{tri}$.

%%%%%%%%%%%%%%%%%%%%%%%%%%%%%%%%%%%%%%%%%%%%%%%%%%%%%%%%%%

\subsection{Triple conical phase}

\begin{figure}[t]
  \centering
\begin{subfigure}[t]{.15\textwidth}
\centering
  \includegraphics[width=\linewidth]{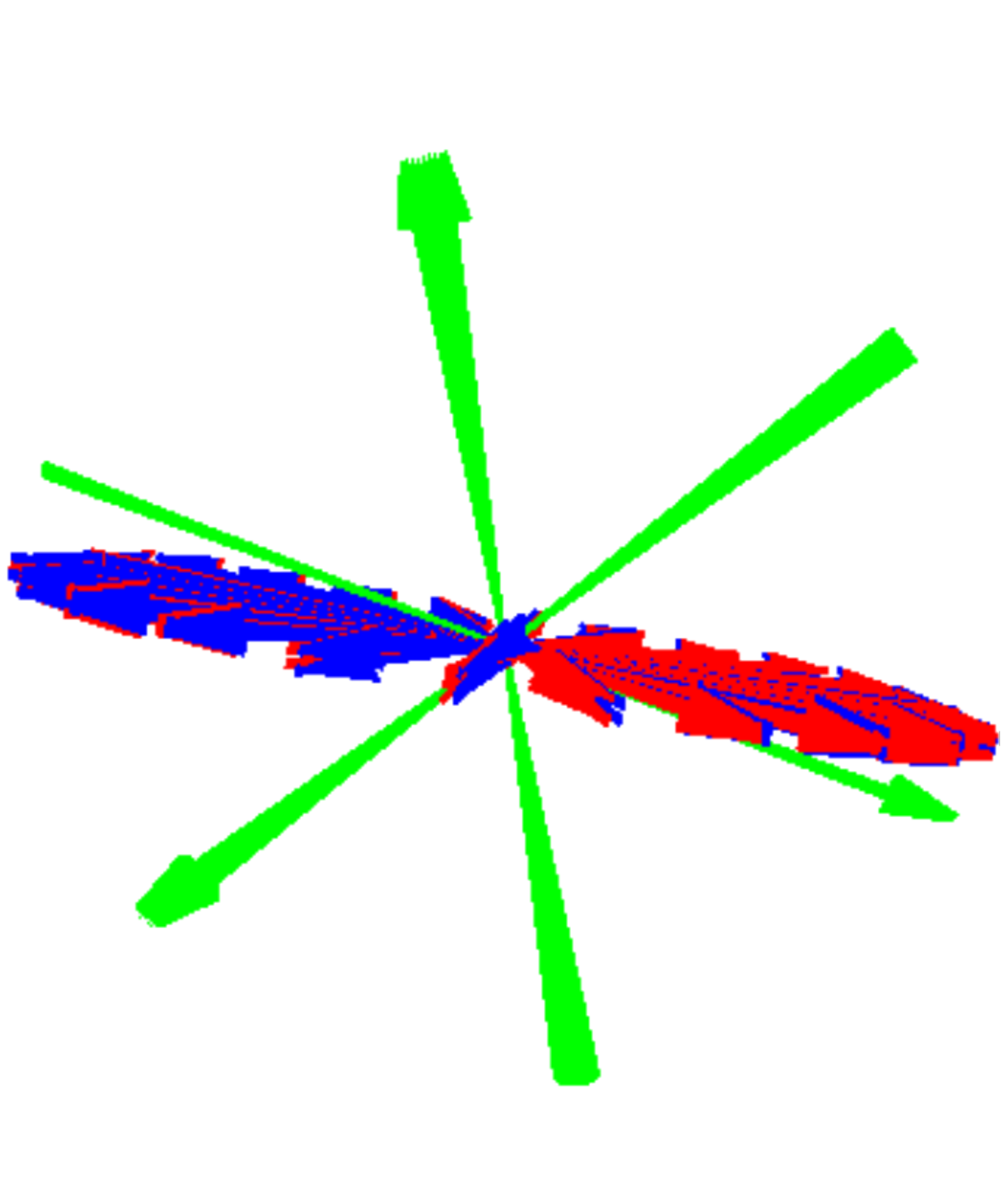}
\caption{\label{tcs1}$J'/J_1 = 0.05$}
\end{subfigure}\hfill
\begin{subfigure}[t]{.15\textwidth}
  \centering
  \includegraphics[width=\linewidth]{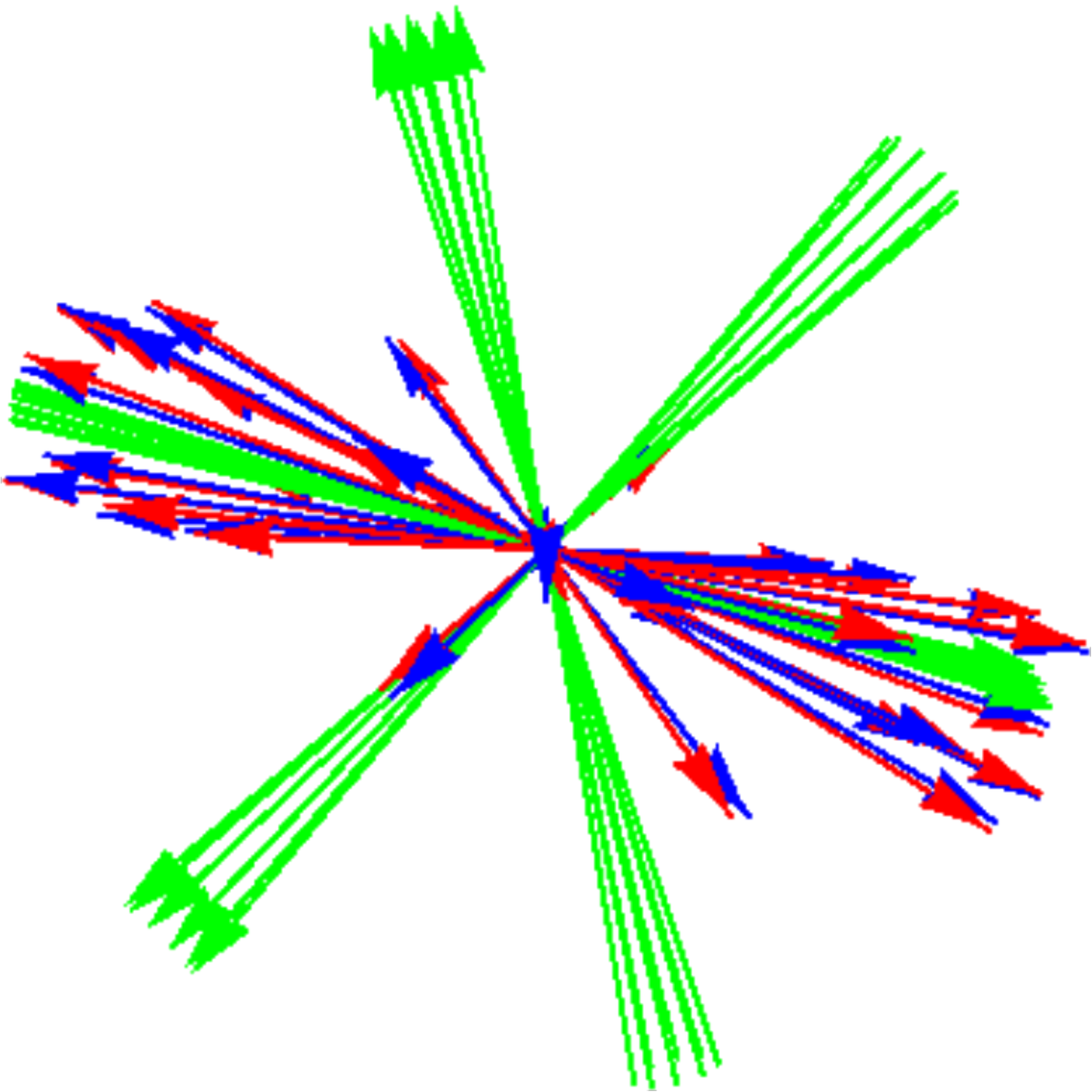}
\caption{\label{tcs2}$J'/J_1 = 0.075$}
\end{subfigure}
\begin{subfigure}[t]{.15\textwidth}
  \centering
  \includegraphics[width=\linewidth]{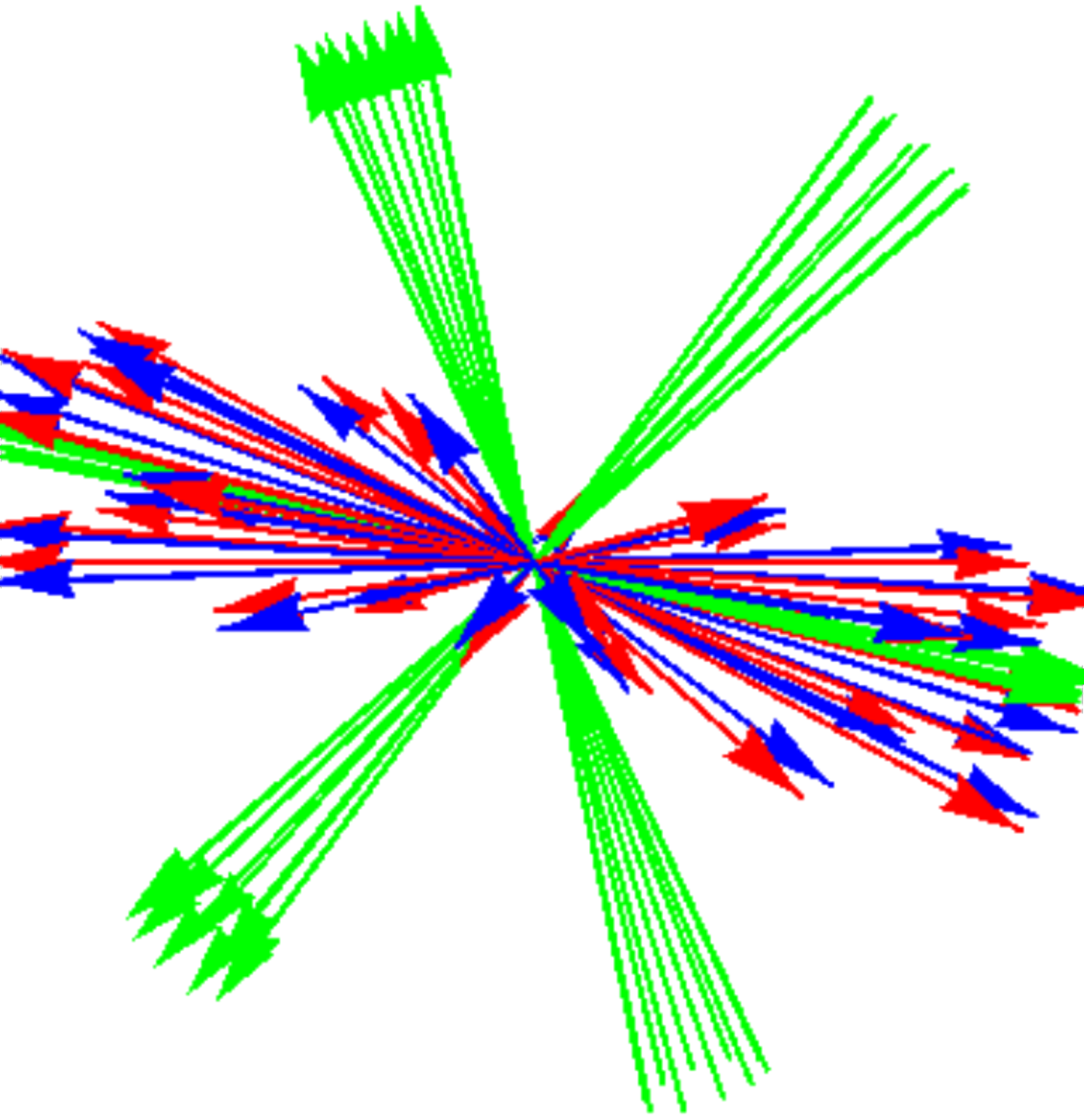}
\caption{\label{tcs3}$J'/J_1 = 0.1$}
\end{subfigure}
\caption{\label{tc} Spin configurations in the triple conical phase plotted using a common origin plot, with A, B, C spins shown in red, blue, green, respectively.  Three representative configurations are plotted for $J_2/J_1 = .3$, with increasing $J'/J_1$.  While the configuration in (a) is very close to the spiral* phase, the AB spirals wave out of the plane with increasing $J'$, and the C spins distort into cones around the original 120$^\circ$ axes.}
\end{figure}

For $J_2/J_1 \gtrsim 0.225$ and $J'/J_1 > 0$, the spiral* phase distorts into a ``triple conical'' phase.  While for $J_2/J_1 < .22$, the AB spirals are flat and decoupled from the C spins, with larger $J_2$ these spirals begin to wave out of the plane in order to couple to the C spins and take advantage of the $J'$ exchange coupling. The C spins are only slightly distorted from their 120$^\circ$ order, and now align such that their ordering plane is perpendicular to the initial AB ordering plane.  The case for small $J'/J_1$ is shown in Fig. \ref{tc} (a), where we plot all of the spins obtained in iterative minimization with a common origin.  As $J'/J_1$ increases, the AB spirals wave more and more out of the plane, and the C spins form three cones around the original 120$^\circ$ axes, as shown for Fig. \ref{tc} (b, c).  Note that one of these conical axes is in the AB plane, and that cone flattens out with larger $J'$ to better align with the AB spins.  This phase is quite complicated, and we were unfortunately unable to find a variational parameterization for it.  The phase boundaries were determined by comparing iterative minimization energies to the analytical energies of the surrounding phases, and the spin configuration of each phase point was checked to ensure that no additional phases were present. While the transition from the spiral to the triple conical phase appears to be smooth, it may instead be weakly first order; our data could not resolve this difference. Due to its non-coplanar nature, this phase is unlikely to survive substantial quantum fluctuations.  For sufficiently large $J'$, this phase undergoes a first order phase transition to the DC II double conical phase.

%%%%%%%%%%%%%%%%%%%%%%%%%%%%%%%%%%%%%%%%%%%%%%%%%%%%%%%%%%%%

\subsection{Triangle of triangles phase}

\begin{figure}[t]
  \centering
\begin{subfigure}[t]{.28\textwidth}
  \includegraphics[width=1.05\linewidth]{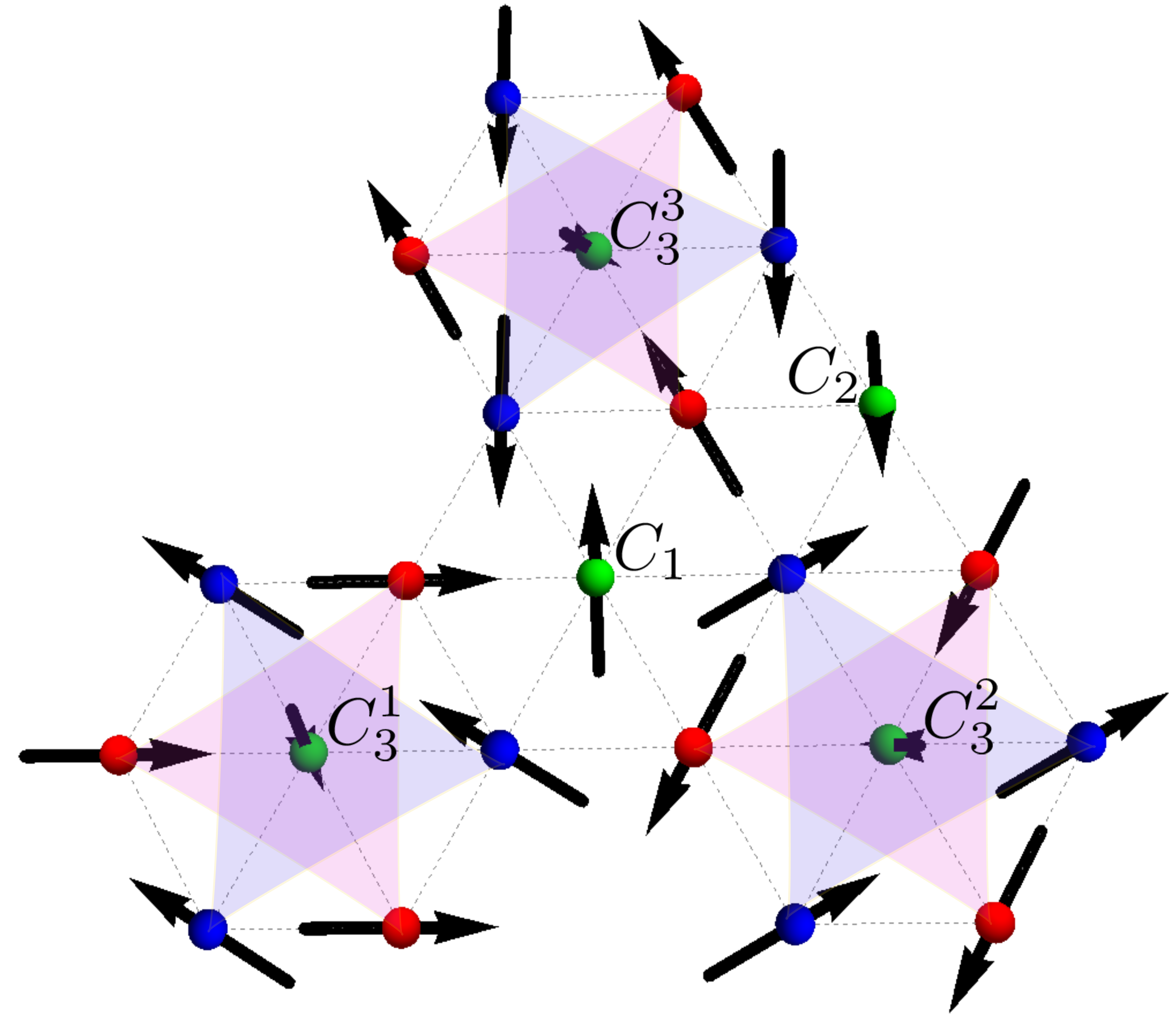}
\caption{\label{TT1}}
\end{subfigure}%
~
\begin{subfigure}[t]{.25\textwidth}
  \centering
  \includegraphics[width=.75\linewidth]{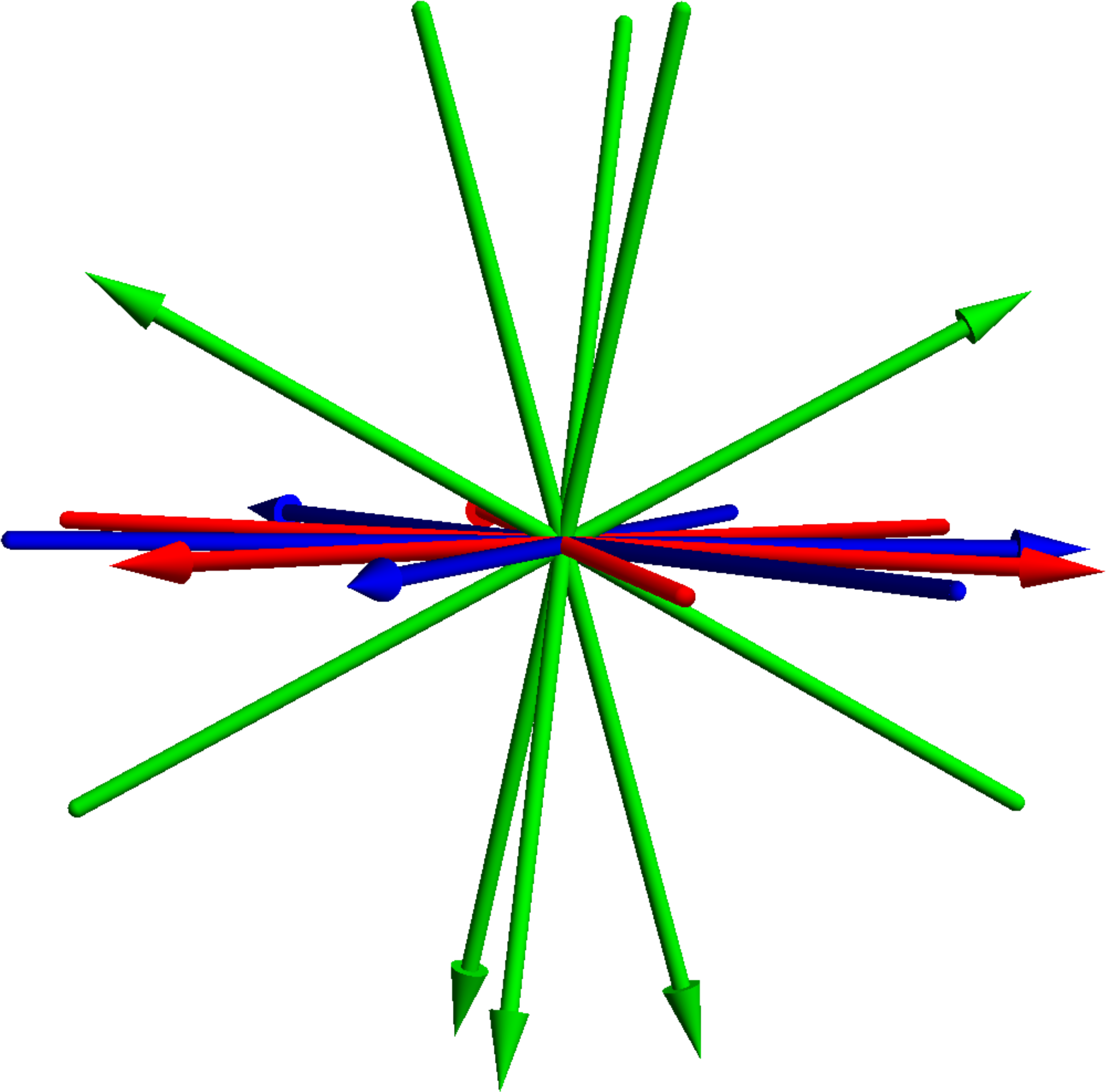}
\caption{\label{TT2}}
\end{subfigure}
\caption{\label{tt}{The triangle of triangles phase exists in a wedge near the honeycomb axis. The unit cell is shown in (a)  with A, B, C spins in red, blue and green respectively. The triangle of triangles feature is particularly emphasized by the solid blue and red triangles on the three distinct types of ferromagnetic hexagons.  (b) While the AB spins lie in a plane, the C spins are oriented out of the plane; their orientations are shown via a common origin plot. These plots are for $J_2/J_1$ = 0.275 and $J'/J_1$ = 0.15.}}
\end{figure}

Right in the middle of the triple conical phase, there is wedge of another unusual non-coplanar phase that almost touches the $J' = 0$ axis at $J_2/J_1 \sim 0.29$.  This phase is best described as consisting of ``triangles of triangles'' on the A and B sublattices, as shown in Fig. \ref{TT1}; it cannot be simply described using ordering wave-vectors.  Here, on some subset of the hexagons, all three A (B) spins will be ferromagnetically aligned with each other, with a relative angle, $\gamma$ between the coplanar A and B spins.  These hexagons are then arranged as if they were single spins forming 120$^\circ$ order.  There are five types of C spin sites: three sites located within the different types of ferromagnetic hexagons, $C_3^{(1,2,3)}$, and two sites located in the two different types of intermediate hexagons, $C_1$ and $C_2$.  $C_1$, $C_2$ and the average of $C_3^{(1,2,3)}$ form 120$^\circ$ order in a plane perpendicular to the AB spins.  The $C_3$ spins have a conical structure.  The actual variational parameterization is slightly more complicated, 
\begin{align}
&\vec{S}_A^{\Delta1} = [1,0,0]\cr
&\vec{S}_A^{\Delta2} = [-1/2,0,\sqrt{3}/2]\cr
&\vec{S}_A^{\Delta3} = [-1/2,0,-\sqrt{3}/2]\cr
&\vec{S}_B^{\Delta1} = [\cos(\gamma),0,\sin(\gamma)]\cr
&\vec{S}_B^{\Delta2} = [\cos(\gamma+2\pi/3),0,\sin(\gamma+2\pi/3)]\cr
&\vec{S}_B^{\Delta3} = [\cos(\gamma+4\pi/3),0,\sin(\gamma+4\pi/3)]\cr
&\vec{S}_{C_1} = [\cos(\eta),\sin(\eta),0]\cr
&\vec{S}_{C_2} = [\cos(\eta+\lambda),\sin(\eta+\lambda),0]\cr
&\vec{S}_{C_3^1} = [\cos(\eta+\theta+\lambda+\nu),\sin(\eta+\theta+\lambda+\nu),0]\cr
\begin{split}
&\vec{S}_{C_3^2} = \Big[\frac{1}{4}(3\cos(\eta-\theta+\lambda+\nu)+\cos(\eta+\theta+\lambda+\nu)),\\
&\,\,\,\,\,\,\,\,\,\,\,\,\,\,\,\,\,\,\,\,\,\frac{1}{4}(3\sin(\eta-\theta+\lambda+\nu)+\sin(\eta+\theta+\lambda+\nu)),\\
&\,\,\,\,\,\,\,\,\,\,\,\,\,\,\,\,\,\,\,\,\,-\frac{\sqrt{3}}{2}\sin(\theta)\Big]
\end{split}\cr
\begin{split}
&\vec{S}_{C_3^3} = \Big[\frac{1}{4}(3\cos(\eta-\theta+\lambda+\nu)+\cos(\eta+\theta+\lambda+\nu)),\\
&\,\,\,\,\,\,\,\,\,\,\,\,\,\,\,\,\,\,\,\,\,\frac{1}{4}(3\sin(\eta-\theta+\lambda+\nu)+\sin(\eta+\theta+\lambda+\nu)),\\
&\,\,\,\,\,\,\,\,\,\,\,\,\,\,\,\,\,\,\,\,\,-\frac{\sqrt{3}}{2}\sin(\theta)\Big]
\end{split}\cr
\end{align}
where $S_{A,B}^{\Delta i}$ describes the A,B spins on the three types of ferromagnetic hexagons, as shown in Fig.\ref{TT1}. There are five variational parameters: $\gamma$ is the angle between the A and B spins on a given ferromagnetic hexagon;  $\theta$ is the conical angle for the $C_3$ spins; $\lambda$ is the angle between $C_1$ and $C_2$; $\nu$ is the angle between $C_2$ and the axis of the $C_3$ cone; and $\eta$ is the angle by which $C_1$ is rotated with respect to the projection of $S_A^{\Delta 1}$ onto the $C_1$, $C_2$, $C_3$ plane.  The variational energy is,
\begin{align}\label{TTeq}
\begin{split}
E_{TT}&[J_2, J', \theta, \lambda, \eta, \nu, \gamma] = \frac{1}{2} \Big\{ 3 \cos \gamma + 2 J_2 \cos \lambda\\
 &\!\!\!\!\!\! + 4 J_2 \cos \theta \cos \frac{\lambda}{2} \cos\left(\frac{\lambda}{2}+ \nu\right)\\ 
 &\!\!\!\!\!\!- \sin \gamma \left[\sqrt{3} + J' \sin \theta \left(1 + \sin[\eta + \lambda + \nu]\right)\right]\Big\}
\end{split}
\end{align}
This phase is sandwiched in the middle of the triple conical phase, separated by what we believe must be first order transitions. While in Fig. \ref{cpd}, it appears to touch the $J' = 0$ axis, the spiral* phase does extend for a small, but finite $J'$.

%%%%%%%%%%%%%%%%%%%%%%%%%%%%%%%%%%%%%%%%%%%%%%%%%%%%%%%%%%%%%
%%%%%%%%%%%%%%%%%%%%%%%%%%%%%%%%%%%%%%%%%%%%%%%%%%%%%%%%%%%

\section{Phases on the triangular and $J_2 = 0$ axes}\label{trilimit}

Next, we turn to the phases on the triangular axis ($J' = J_1$), and discuss their evolution off-axis; we will additionally discuss the $J_2 = 0$ axis phases, as these have substantial overlap with the triangular axis phases.  
On the triangular axis, there are only two phases for $J_2/J_1 < 1$, the 120$^\circ$ phase for $J_2/J_1 < 1/8$, which evolves smoothly off axis, and the collinear phase for $1 > J_2/J_1 > 1/8$, which remains unchanged off-axis.  Beyond $J_2/J_1 = 1$, there is a planar spiral phase that evolves smoothly to three independent triangular lattices for $J_1 = 0$, and extends out to the dice lattice limit.  

On the $J_2 = 0$ line, a single phase interpolates from the N\'{e}el order of the honeycomb limit ($\vec{S}_A = -\vec{S}_B$) to the 120$^\circ$ order of the triangular limit, and out to a ferrimagnetic limit at $J' = 2J_1$ ($\vec{S}_A = \vec{S}_B = -\vec{S}_C$).  Beyond $J' = 2J_1$, this ferrimagnetic phase does not evolve further, and is the ground state out to the dice limit.

\subsection{Interpolating phase}\label{ipp}

\begin{figure}[H]
\captionsetup[subfigure]{justification=centering}
\centering
\begin{subfigure}[t]{.2\textwidth}
  \centering
  \includegraphics[width=\linewidth]{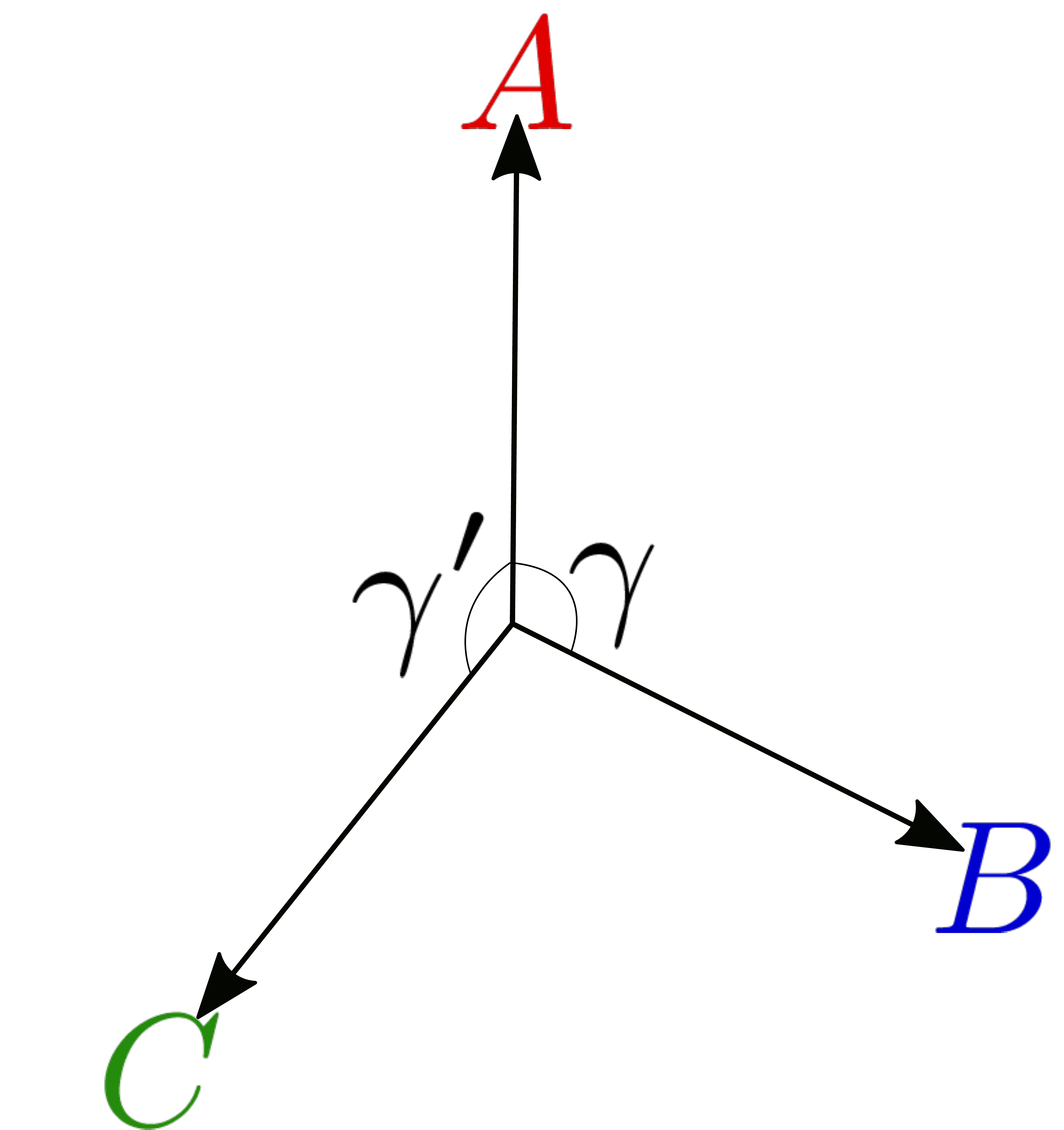}
\caption{}  
\label{intAngle}
\end{subfigure}\hfill
\begin{subfigure}[t]{.18\textwidth}
  \centering
\includegraphics[width=\linewidth]{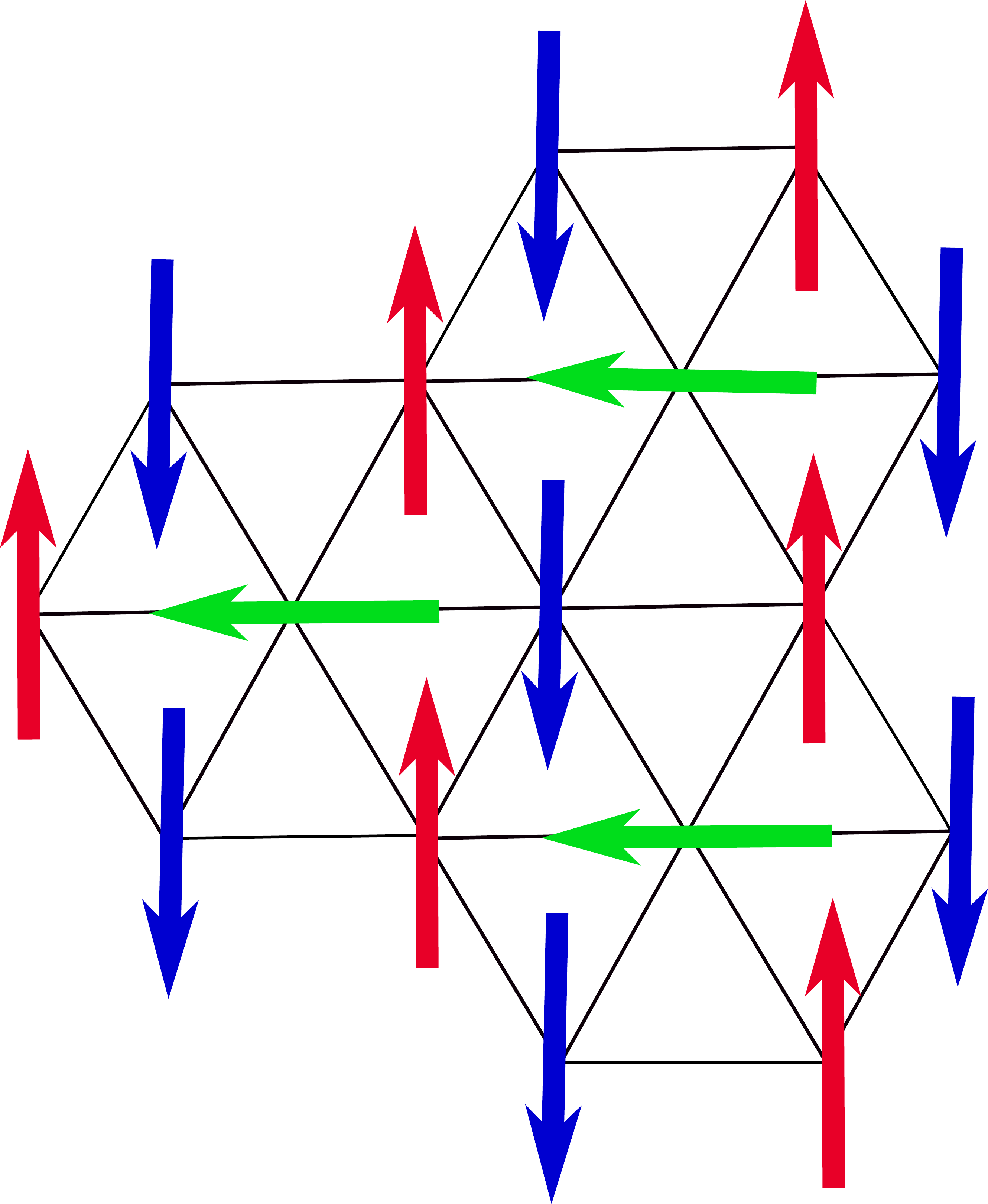}
  \caption{N\'{e}el}
  \label{int1}
\end{subfigure}
\begin{subfigure}[t]{.19\textwidth}
  \centering
\includegraphics[width=\linewidth]{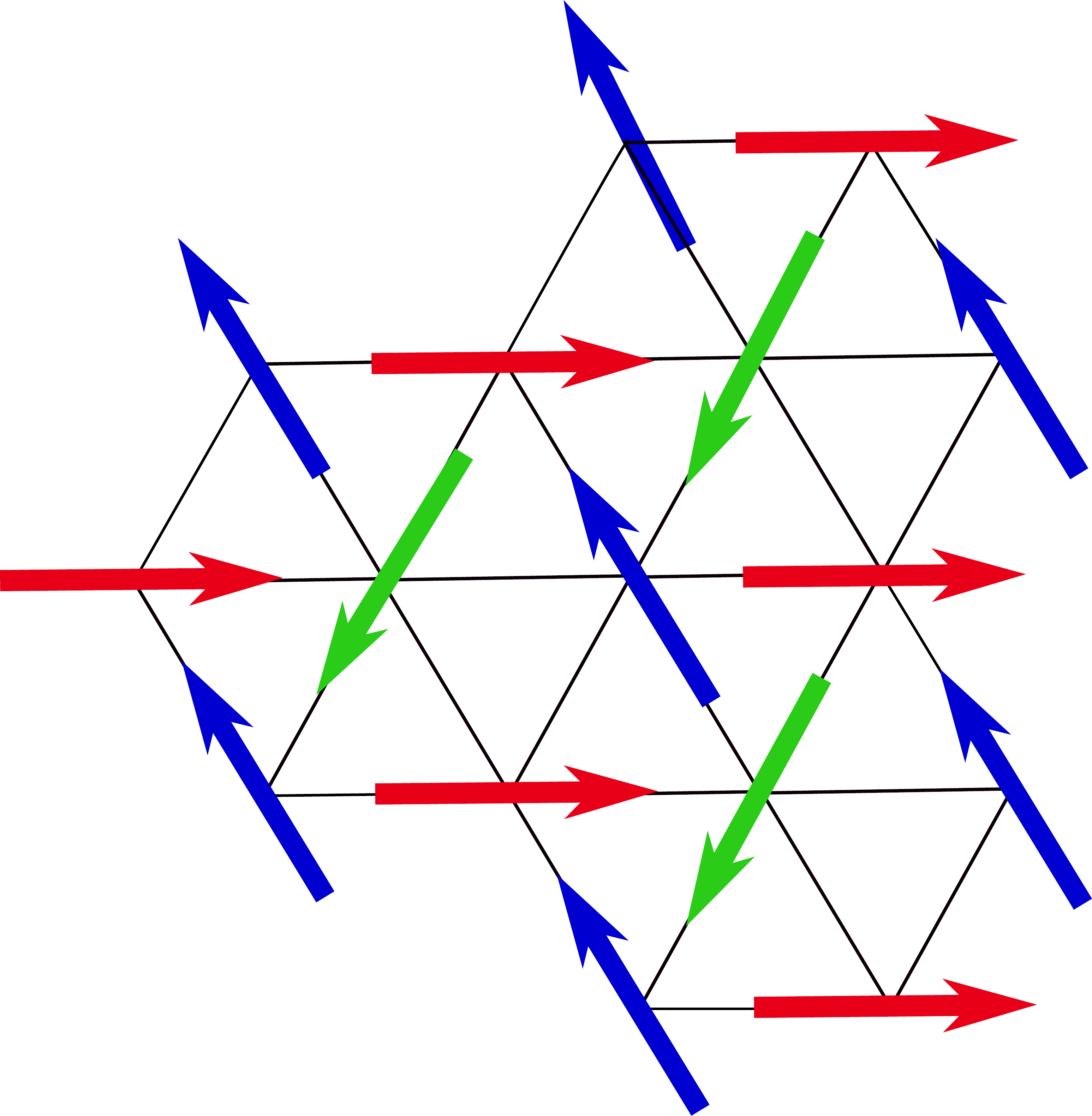}
  \caption{$120^o$}
  \label{int2}
\end{subfigure}\hfill
\begin{subfigure}[t]{.17\textwidth}
  \centering
\includegraphics[width=\linewidth]{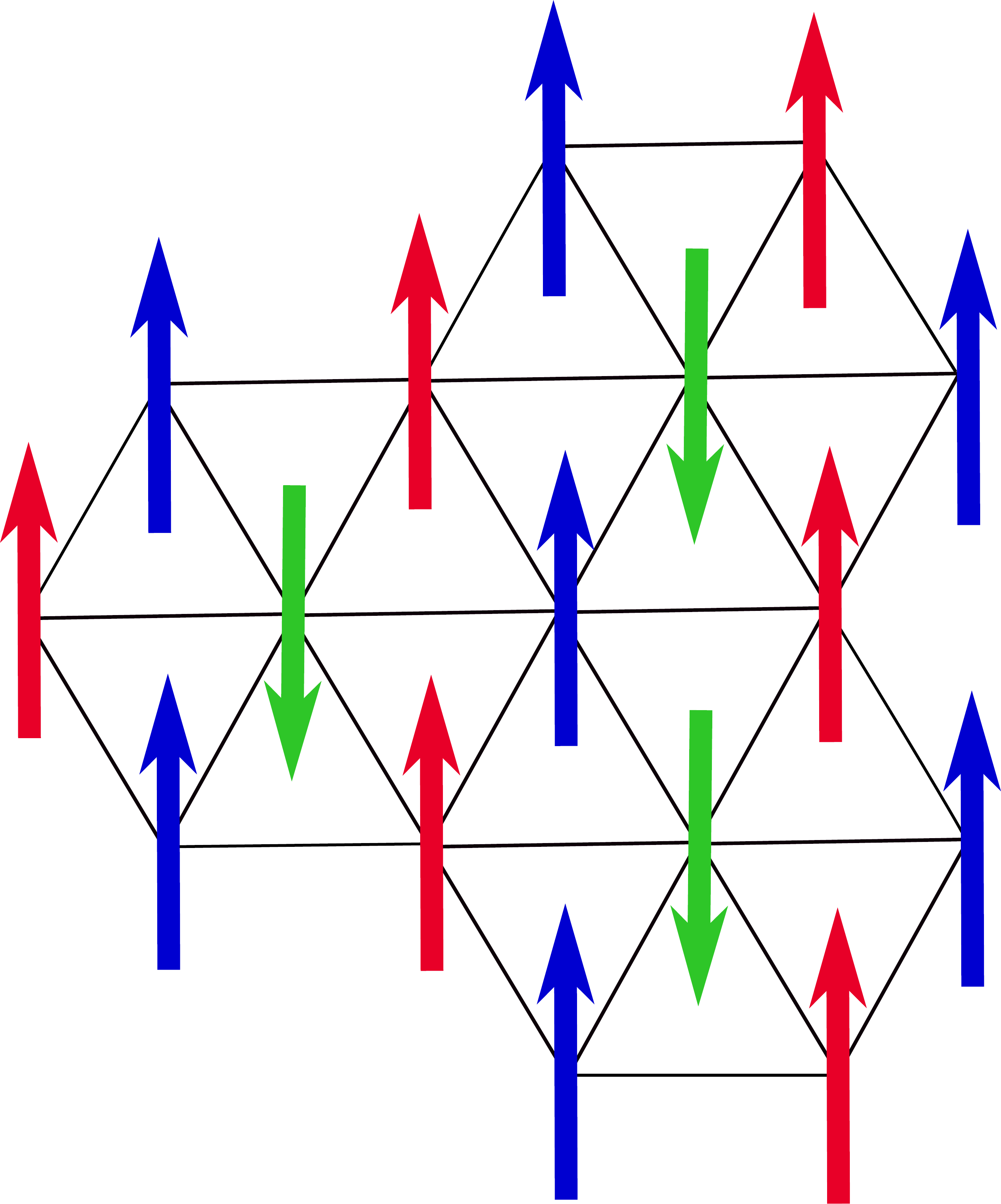}
  \caption{Ferri}
  \label{int3}
\end{subfigure}
\caption{The interpolating phase occupies the entire $J_2 = 0$ axis out to $J_2/J_1 = 2$, and is described by a single spin orientation on each sublattice.  (a) depicts the angle between the sublattice spins, while (b)-(c)-(d) highlight the three key limits for (b) $J' = 0$ (N\'{e}el order on the honeycomb), (c) $J' = J_1$ (120$^\circ$ order) and (d) $J' > 2 J_1$ (the ferrimagnetic order of the dice lattice).}
\label{ip}
\end{figure}

\begin{figure}[t]
\centering
  \includegraphics[width=0.8\linewidth]{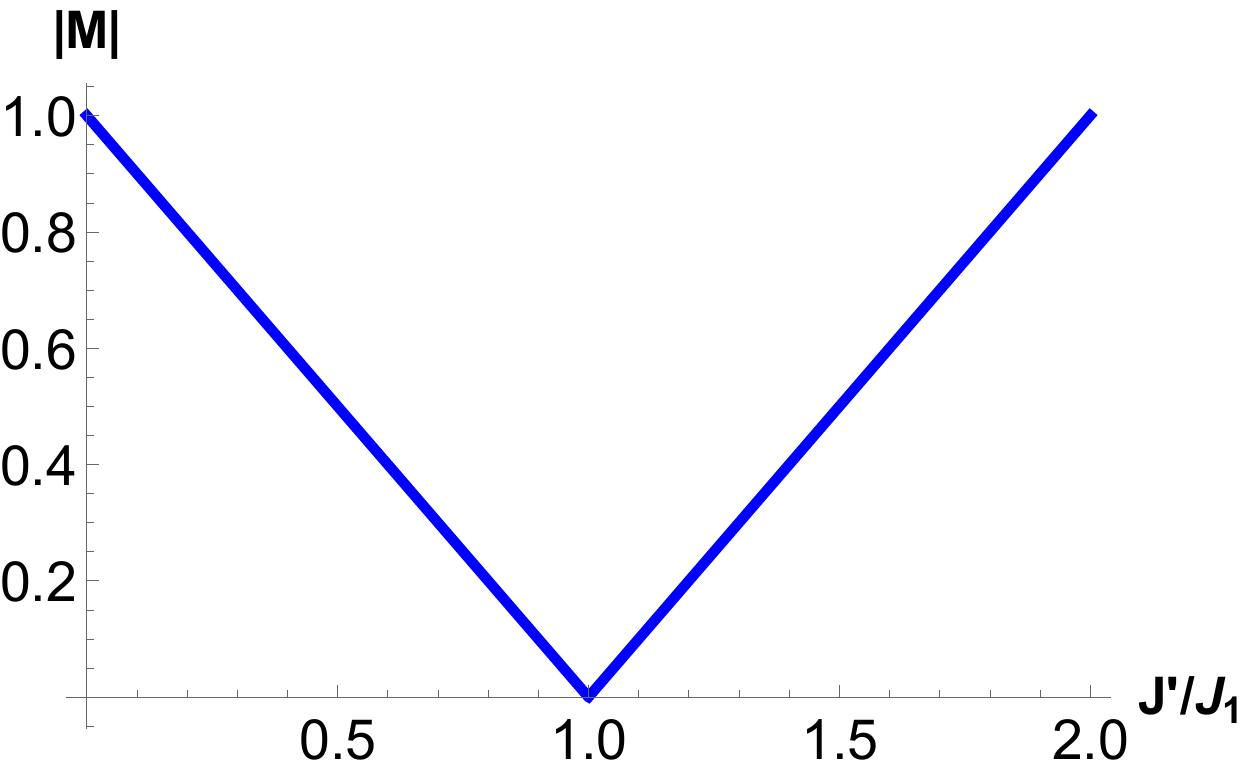}
\caption{Magnetization of the interpolating phase as a function of $J'/J_1$.  We plot the magnitude of the net moment per unit cell  in units of $S$. The moment is maximal for $J' = 0$, and again for $J' = 2 J_1$, only vanishing on the triangular lattice for $J' = J_1$.}
\label{interp}
\end{figure}

The interpolating phase extends along the $J_2 = 0$ axis, interpolating between N\'{e}el order on the honeycomb lattice ($J' = 0$) to 120$^\circ$ order on the triangular lattice ($J' = J_1$) to ferrimagnetic order ($J' = 2 J_1$). It can be captured by a simple variational ansatz where each sublattice is ferromagnetic within itself, and the interpolating is captured by the angles $\gamma$ and $\gamma'$ between the AB and AC sublattices, as shown in Fig. \ref{ip}.  These angles are,
\begin{equation}
\gamma = 2\cos^{-1}\frac{J'}{2};\quad \gamma' = \gamma/2 + \pi.
\end{equation}
For $J' = 0$, $\gamma = \pi$ captures the N\'{e}el order on the honeycomb lattice.  In contrast to the N\'{e}el* phase, where the C spins form 120$^\circ$ order, they are ferromagnetic here.  For $J' = 0$, the relative angle is free, but we choose $\gamma' = 3\pi/2$, for consistency with the finite $J'$ results; see Fig. \ref{int1}.  Note that this phase generically has a net moment, except at the 120$^\circ$ point, as shown in Fig. \ref{interp}.  As $J'$ increases, $\gamma$ smoothly trends towards zero and $\gamma'$ trends towards $\pi$, reaching that point at $J'/J_1 = 2$.  The classical energy has a simple analytic form and is given by: 
\begin{align}\label{int}
E_{int}[J_2,J'] = &[9 J_2 - \frac{3}{2} (2 + J'^2)].
\end{align}

\subsection{Ferrimagnetic phase}

At $J'/J_1 = 2$, the interpolating phase become fully ferrimagnetic and maximizes the gain from $J'$ bonds, with the AB sublattices ferromagnetically aligned, and the C spins anti-aligned to both, as shown in Fig. \ref{int3}. The moment per spin is $S/3$. This phase extends out to the dice lattice limit, and has the classical energy,
\begin{equation}
E_{ferri}[J_2,J'] = 3 + 9 J_2 - 6 J'.
\end{equation}
The ferrimagnetic phase is a limit of the interpolating phase, much as the N\'{e}el* phase is a limit of the spiral* phase; being collinear it has a higher symmetry than the interpolating phase, and is a distinct phase. 

\subsection{Collinear phase}

\begin{figure}[h]
  \centering
  \includegraphics[width=0.5\linewidth]{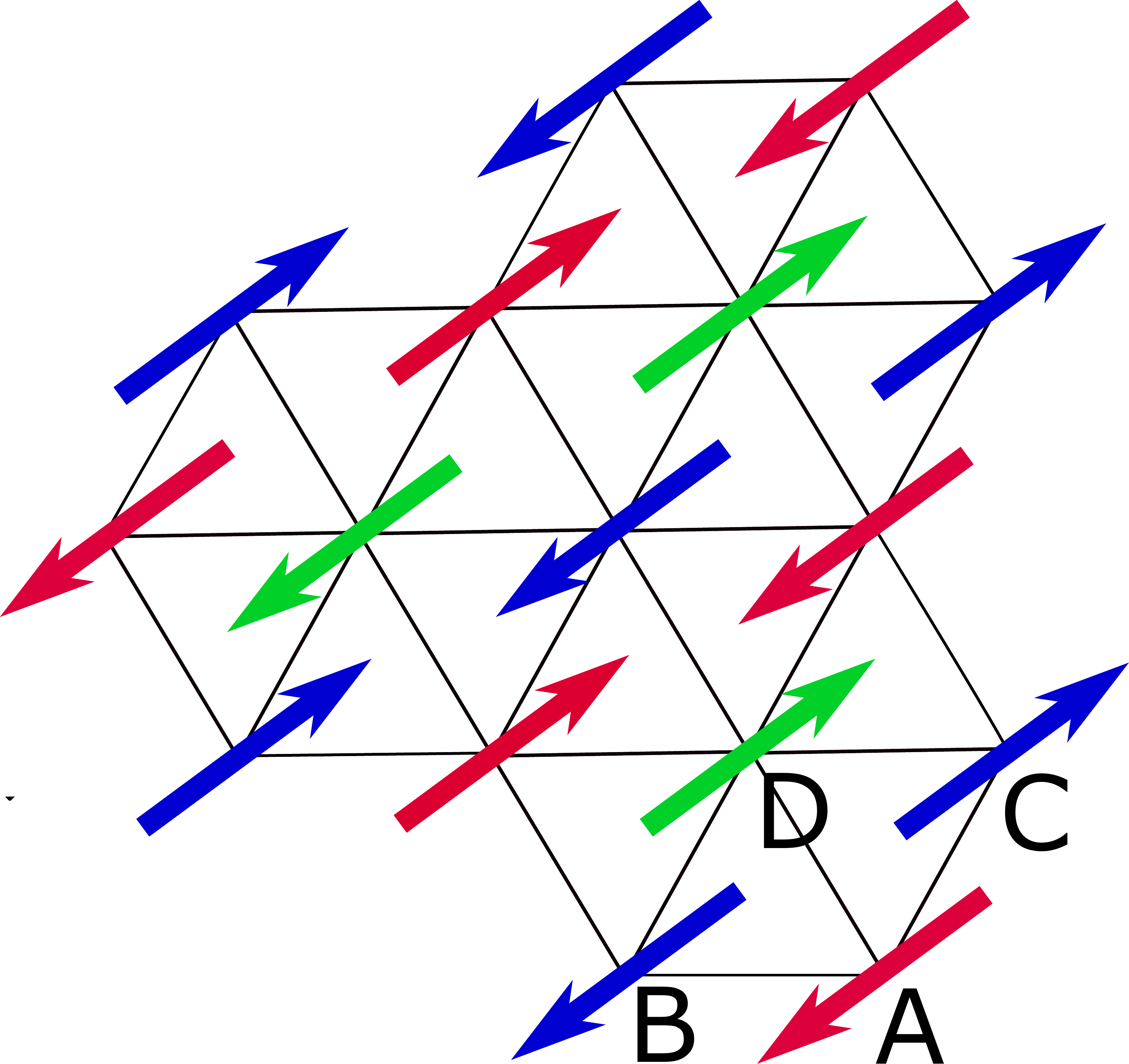}
 \caption{The collinear phase consists of ferromagnetic stripes along one triangular axis (here, the $\hat x$ axis), and antiferromagnetic stripes along the other two axes. This particular arrangement is selected via order by disorder.  More generally, it is a four sublattice phase labeled by the sites $ABCD$ in the figure, with $\vec{S}_A + \vec{S}_B+\vec{S}_C+\vec{S}_D = 0$ the only ground state condition.}
\label{coll}
\end{figure}

The bulk of the phase diagram is occupied by the \emph{collinear} phase, and many of its neighboring phases borrow some of its features.  Fig. \ref{coll} shows the typical collinear arrangement that gives this phase its name, where all of the spins align ferromagnetically along one of the three triangular axes, and alternate antiferromagnetically along the other two; this phase therefore breaks the six-fold rotational symmetry and allows a $\mathbb{Z}_3$ nematic order parameter, as we discuss further in the related non-collinear phases.  However, this phase is only one of a set of classical ground states, which may be more generically described by a four-sublattice arrangement around a rhombus, where the sum of spins, $\vec{S}_A + \vec{S}_B+\vec{S}_C+\vec{S}_D = 0$.  This four-sublattice arrangement can be taken on the triangular lattice, as shown in Fig. \ref{coll}, or may be taken on each of our three sublattices individually, where the same four spins must be taken for each of the A,B,C sublattices.  The Hamiltonian, (\ref{ham}) can be rewritten as,
\begin{equation}
H = \frac{J_1 + 2 J' + 3 J_2}{4}\left[(\vec{S}_A + \vec{S}_B+\vec{S}_C+\vec{S}_D)^2-4S^2\right],
\end{equation}
with the overall classical energy,
\begin{equation}
E_{coll}[J_2,J'] = -1 - 3J_2 - 2J'.
\end{equation}
There is thus a continuous manifold of classical ground states, including non-coplanar phases like those where the four spins point along the vertices of a tetrahedra.  One particular such state has a non-zero uniform scalar chirality, $\kappa = \vec{S}_i \cdot (\vec{S}_j \times \vec{S}_k)$ around each triangle\cite{kurz01,martin08}.  However, quantum and thermal fluctuations select the collinear states via order by disorder \cite{chubukov92}.  This particular state can also be captured by a single $\bQ$, here given for each of our three sublattices,  $\bQ_{coll} = (2\pi/3,0)$, with
\begin{equation}
\vec{S}({\bf R}_i) = [\cos({\bf Q}_{coll}\cdot {\bf R}_i), 0, \sin({\bf Q}_{coll}\cdot {\bf R}_i)].
\end{equation}

\subsection{Spiral phase}

\begin{figure}[t]
  \centering
\begin{subfigure}[t]{.2\textwidth}
\centering
  \includegraphics[width=\linewidth]{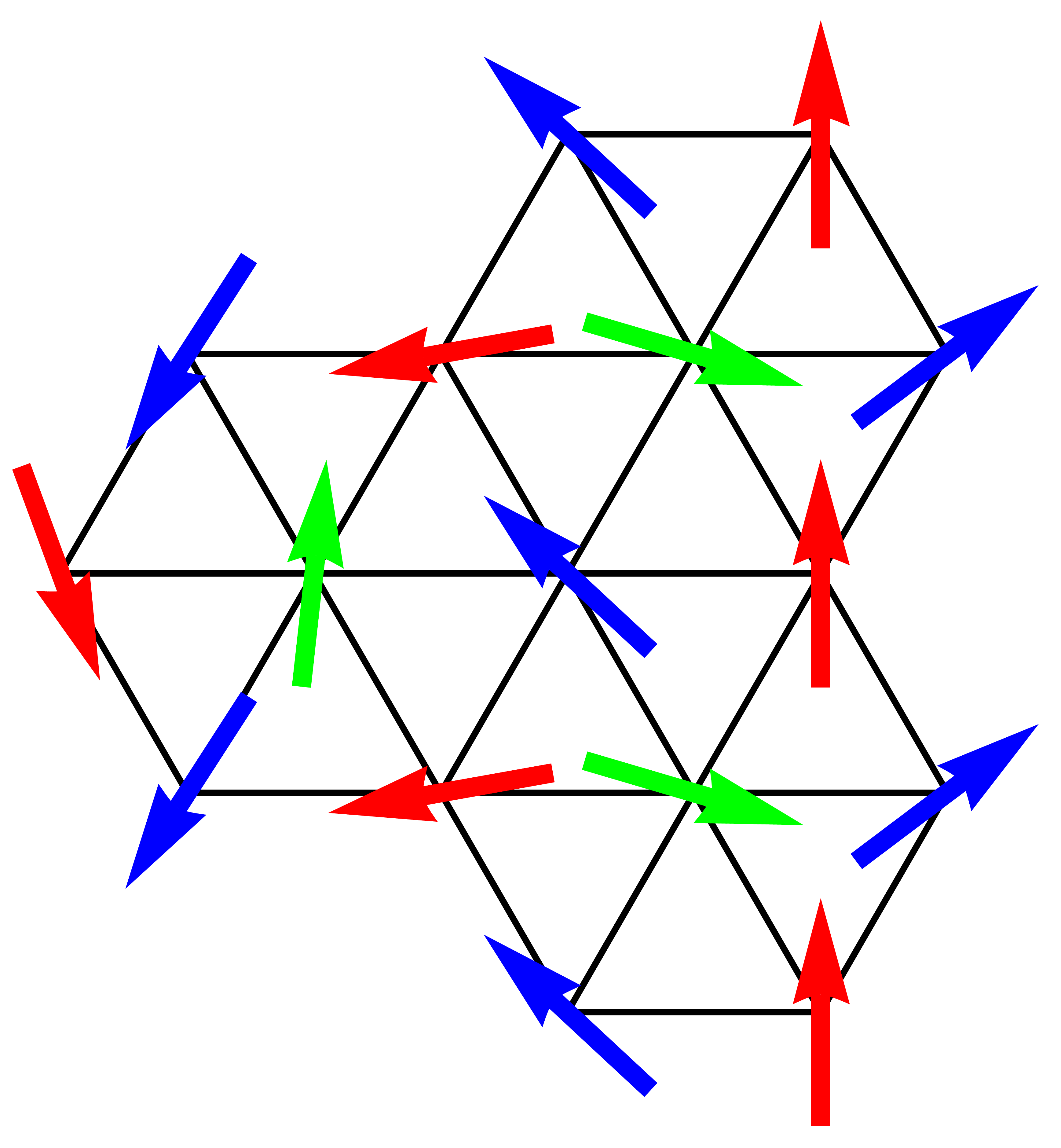}
\caption{\label{spd1}}
\end{subfigure}\hfill
\begin{subfigure}[t]{.26\textwidth}
  \centering
  \includegraphics[width=\linewidth]{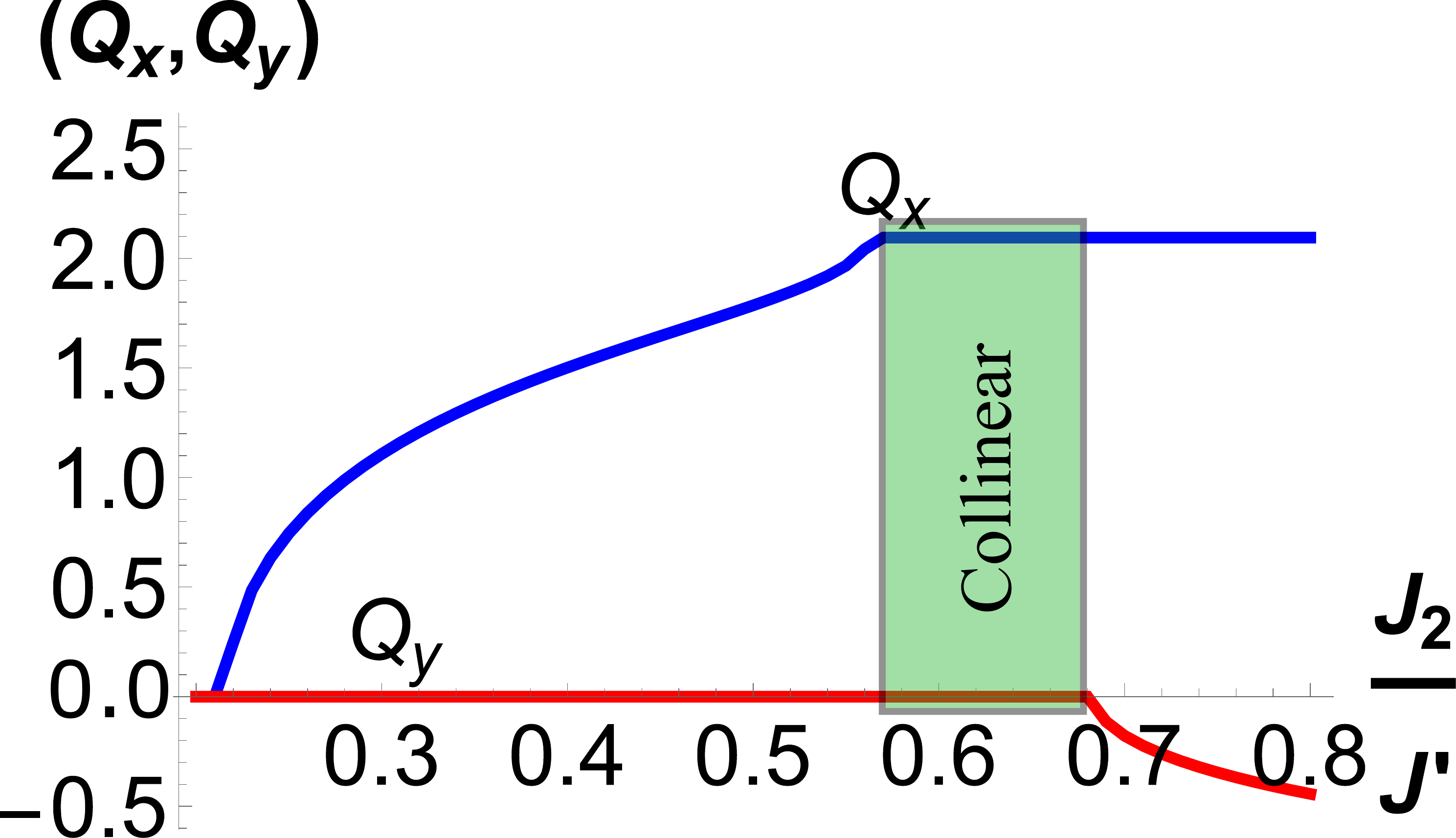}
\caption{\label{spd2}}
\end{subfigure}
\caption{\label{spdice} In the spiral phase, all three sublattices spiral with a generically incommensurate wave-vector, $\bQ_{sp}$ that is a function of both $J_2/J'$ and $J_1/J'$.  (a) shows the real space ordering for $J_2/J' = 0.3 $ and $J_1/J' = 0.1$. (b) shows $(Q_x,Q_y)$ as a function of $J_2$ for $J_1 = 0.1$; the plateau in $Q_x$ begins at the transition to the collinear phase, while $Q_y$ becomes non-zero when the spiral phase is reentered.}
\end{figure}

In the triangular limit, at $J_2 > J_1$, the collinear state gives way to a planar spiral phase that encompasses most of the large $J_2$ region; it extends for $J' < J_1$ down to meet the honeycomb spiral* phase, where only the AB spins spiral, and out to the dice lattice limit. Here, each of the three sublattices forms a planar spiral, with the same ordering wave-vector $\bQ_A = \bQ_B = \bQ_C$, and relative angles $\gamma$ and $\gamma'$ between AB and AC sublattices, as before.  This wave-vector is $\bQ_{coll}$ at the boundary with the collinear phase, and $\bQ = (0,0)$ at the boundary with the ferrimagnetic phase; it asymptotes smoothly to $\bQ_{tri}$ for large $J_2$. Indeed the phase can be described by three variational parameters: $\xi$, the fraction $\bQ = \xi \bQ_{tri}$ of the triangular lattice ordering vector, $\gamma$ and $\gamma'$.  All three parameters depend on both $J_2/J'$ and $J_1/J'$.  This planar spiral connects smoothly with the planar spiral on the triangular lattice axis for $J_2/J_1 > 1$, however the honeycomb spiral* is distinct until $\bQ_{sp} \rightarrow \bQ_{tri}$, as the C spins always have $\bQ_{tri}$. As the ferrimagnetic and collinear phases are special cases of the spiral, the transition between them and the spiral is second order, as seen in Fig. \ref{spd2}.

%%%%%%%%%%%%%%%%%%%%%%%%%%%%%%%%%%%%%%%%%%%%%%%%%%%%%%%%%%%%

\section{Double conical phases}\label{dcp}

At intermediate $J'$, there are two distinct ``double conical'' phases, where one or more sublattices can be described variationally by a double conical structure. Here, one wave-vector $\bQ_1$ controls the in-plane ordering perpendicular to the conical axis, while another, $\bQ_2$ controls the out-of-plane ordering.  The spins on a single sublattice are parameterized by the unit vector,
\begin{align}\label{dcP} 
\begin{split}
\vec{S}(\theta,\bQ_1,&\bQ_2,\bR_i) = [\sin\theta\cos \bQ_1\cdot\bR_i,\\ &\sin\theta\sin \bQ_1\cdot\bR_i,  \cos\theta\cos \bQ_2\cdot\bR_i],\end{split}
\end{align}
where $\theta$ is the conical angle, and the conical axis is $\hat S_z$.  See Fig. \ref{dc1} for an example.  As $\vec{S}$ must be a unit vector, $\cos \bQ_2 \cdot \bR_i = \pm 1$, which limits the possible values of $\bQ_2$ to different collinear configurations; we always find $\bQ_{coll} = (2\pi/3,0)$.  $\bQ_1$ is not so limited and can be incommensurate.  Different sublattices may have non-trivial relative cone orientations, in which case the relative angles will also be variational parameters.

\subsection{Double conical phase I}

The first double conical phase, DC I occurs twice in the phase diagram: in a wedge between the spiral*, triple conical, triangle of triangles and the collinear phases, shown in Fig. \ref{cpd}, and in a wedge for $J' > J_1$ between the spiral, non-collinear I and collinear phases, shown in Fig. \ref{dpd}.  

All three sublattices form double cones, with $\theta_B = \pi-\theta_A$, and $\theta_C$ distinct; all three sublattices share the same $\bQ$'s. The out-of-plane components form a collinear structure, $\bQ_2 = \bQ_{coll}$, while the in-plane $\bQ_1 = (Q_{1x},0) = \xi \bQ_{coll}$ is generally incommensurate.  The classical energy is,
\begin{align}\label{EDCI}
E_{DCI} & [\xi,\theta_A,\theta_C,\gamma,\gamma',J_2,J'] = - \cos^2\theta_A \cr
& \; + \left[2 \cos \gamma + \cos (\gamma - \pi \xi)\right]\sin^2 \theta_A \cr
&\!\!\!\!\! + J_2\left[-2 \cos^2 \theta_A -\cos^2 \theta_C \right. \cr
& \left. \; + (1+ 2 \cos \pi \xi)(2 \sin^2 \theta_A + \sin^2 \theta_C)\right]\cr
& \!\!\!\!\!+ J'\left[-2 \cos \theta_A \cos \theta_C + \{2 \cos (\gamma-\gamma') + \cos \gamma' \right.\cr
& \left.+ 2 \cos (\gamma'-\pi \xi)+\cos (\gamma-\gamma'+ \pi \xi)\} \sin \theta_A \sin \theta_C \right]\cr
\end{align}

\subsubsection{$J' < J_1$ occurrence}

\begin{figure}[h]
\captionsetup[subfigure]{justification=centering}
  \centering
\begin{subfigure}[h]{.44\textwidth}
  \centering
  \includegraphics[width=.8\linewidth]{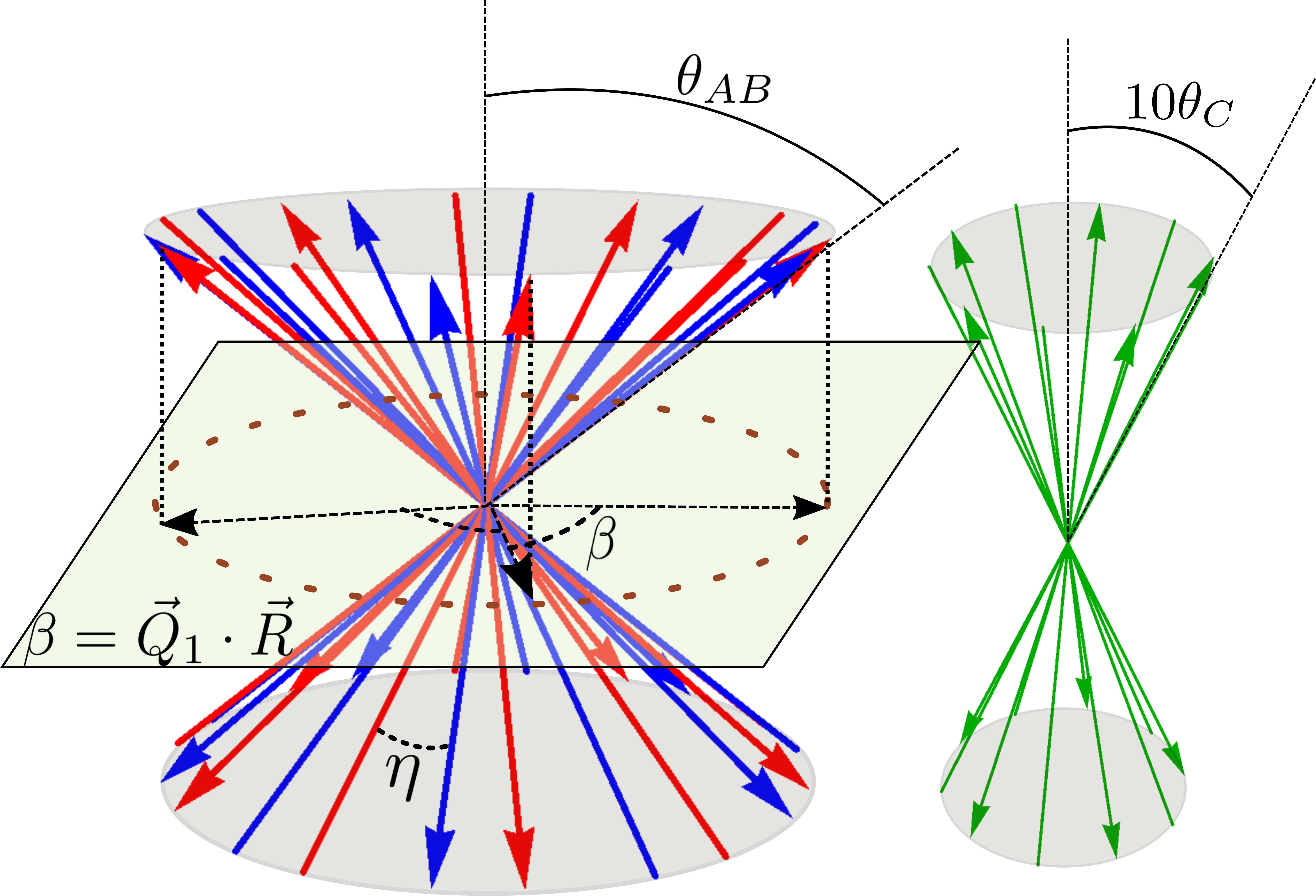}
  \caption{}
  \label{dc1}
\end{subfigure}\\[1ex]
\begin{subfigure}[h]{.44\textwidth}
  \centering
  \includegraphics[width=.9\linewidth]{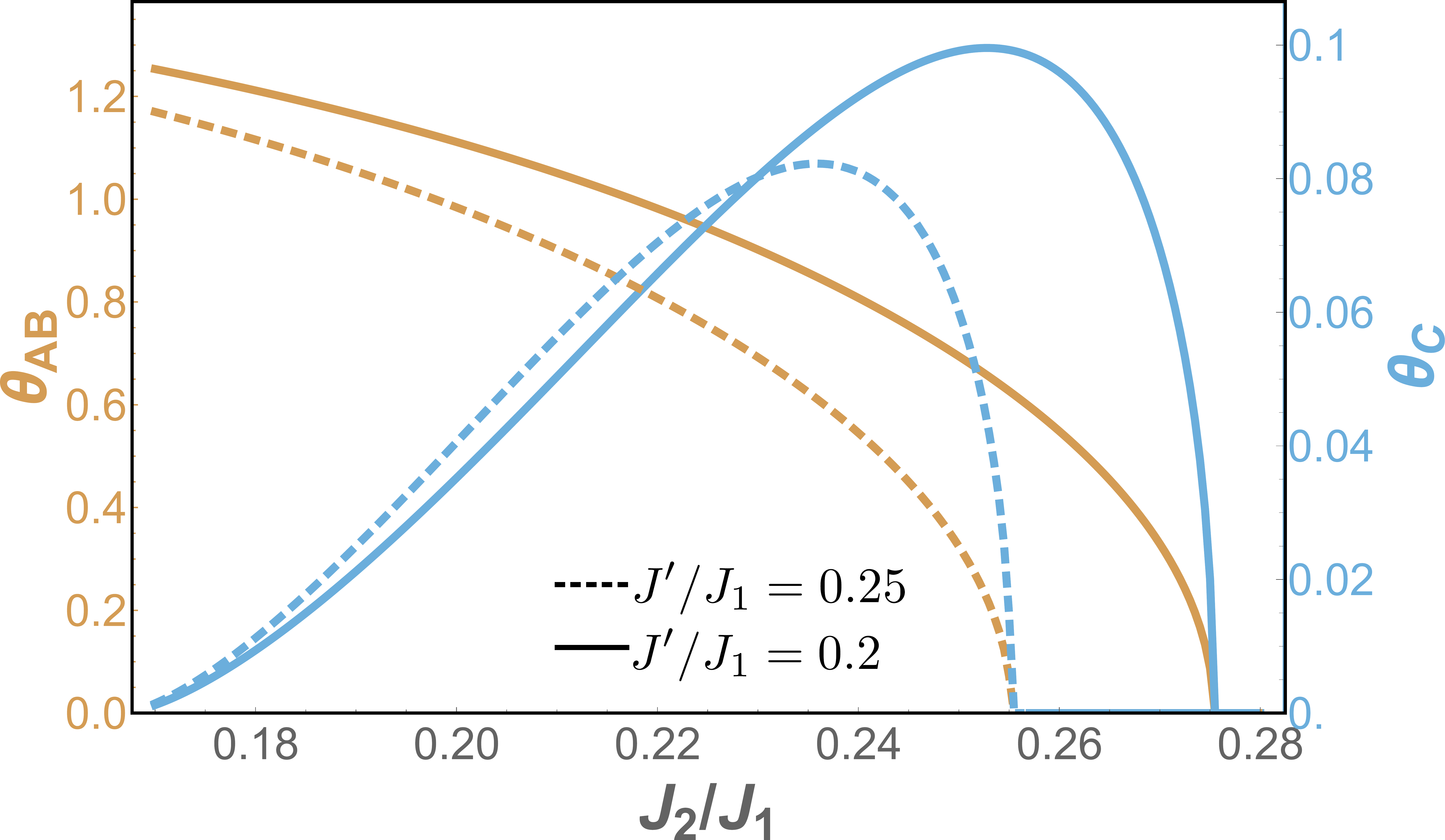}
  \caption{}
  \label{dcdetail}
\end{subfigure}
\caption{\label{fig:dc1} Double conical I for $J' < J_1$. (a) This phase consists of double conical spirals on each of the A, B and C sublattices, with conical angles $\theta_A$, $\theta_B = \pi-\theta_A$ and $\theta_{C} \ll \theta_A$. The AB and C spins are shown in common origin plots on the left and right, respectively. The ordering wave-vectors are generally incommensurate in the plane, $\bQ_1$, although here we plot the special commensurate case where $\bQ_1 = \xi \bQ_{coll}$ with $\xi = 2/5$. The corresponding $J_2/J_1$ and $J'/J_1$ values are 0.198 and 0.29575 respectively. There is a small angle, $\eta$ between the A,B spins, as well as a small angle $\eta_C$ between the A and C spins (not shown). Note: as $\theta_C$ is very small, we plot a conical angle of $10\theta_C$ for clarity. (b)  DC I parameters as functions of $J_2/J_1$ and $J'/J_1$. The variation of the conical angles $\theta_{AB}$ (yellow) and $\theta_C$ (blue) is shown from the boundary of DC I shared with the non-collinear II phase to the critical $J_2$ beyond which it becomes collinear.}
\end{figure}

 Here we discuss its first appearance; the DC I phase occurs for larger $J'$ beyond $J_2/J_1 = 1/6$, above the spiral phase; an example is shown in Fig. \ref{dc1}. 

The conical angles vary strongly with both parameters, as is shown in Fig. \ref{dcdetail}, with $\theta_C \ll \theta_A$.  As the border to the collinear phase is approached, $\theta_A, \theta_C \rightarrow 0$, indicating that the collinear phase is a special case of DC I; as such, the transition is second order.  The transitions to other neighboring phases are all first order.

\subsubsection{$J' > J_1$ occurrence}\label{diceDC}

The second appearance is near the dice lattice limit, between the spiral and collinear phases.  As $J_1/J'$ increases, the planar spiral phase continuously tilts out of the plane to form a recurrence of the double conical DC I phase. In contrast to the small $J'$ version, here the conical angles $\theta_A$ and $\theta_C$ have similar orders of magnitude, as shown in Fig. \ref{dc1}.  Otherwise, the two phases are quite similar.  Again, the in-plane $\bQ_1$ is generically incommensurate, and is smoothly connected to $\bQ_{sp}$ across the second order phase boundary separating the planar spiral and DCI phases.  Note that we have a multicritical point with three second order lines where DC I, spiral and ferrimagnetic phases all join along with a first order line between DC I and the non-collinear I phase.

\begin{figure}[h]
\captionsetup[subfigure]{justification=centering}
  \centering
\begin{subfigure}[h]{.44\textwidth}
  \centering
  \includegraphics[width=.8\linewidth]{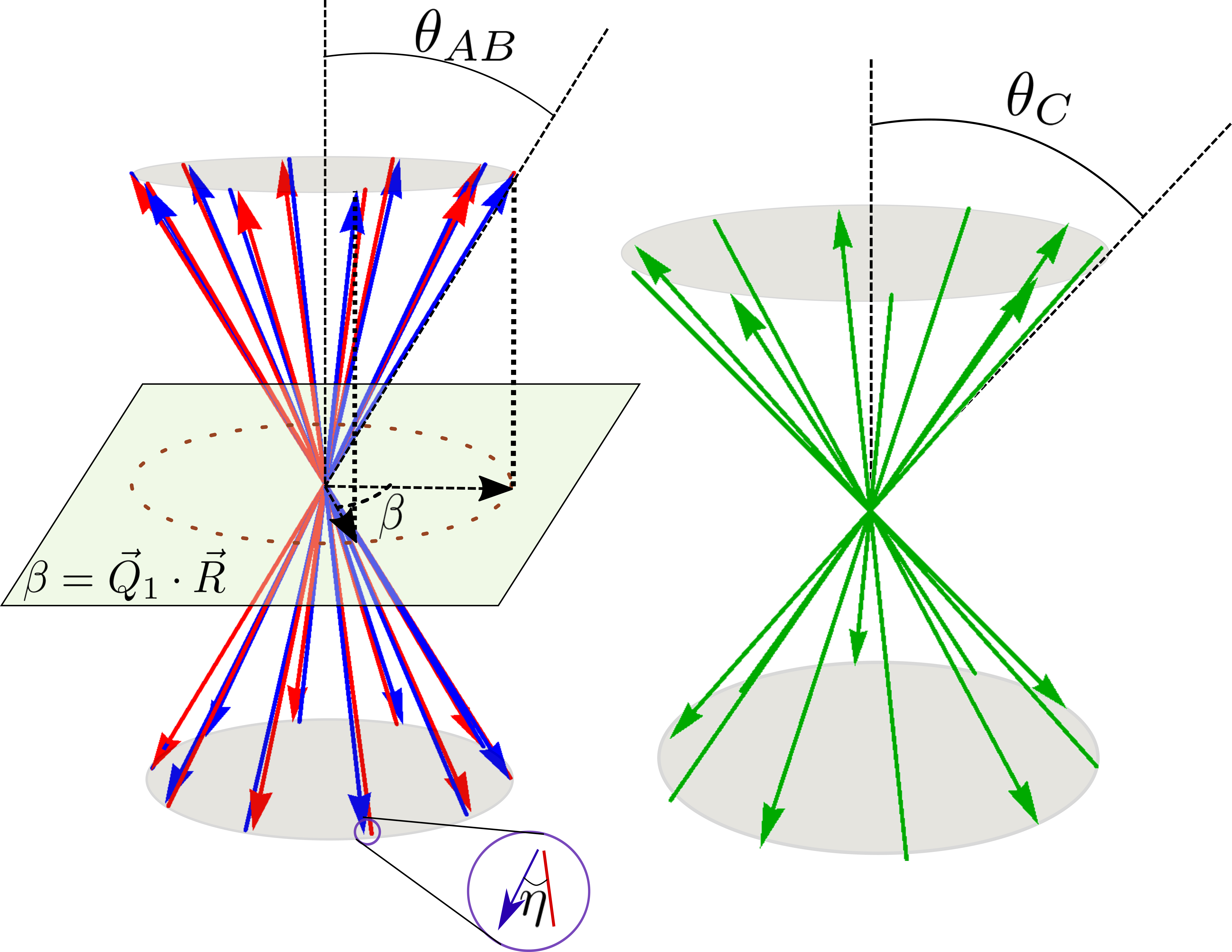}
  \caption{}
  \label{dcdice}
\end{subfigure}\\[1ex]
\begin{subfigure}[h]{.44\textwidth}
  \centering
  \includegraphics[width=.95\linewidth]{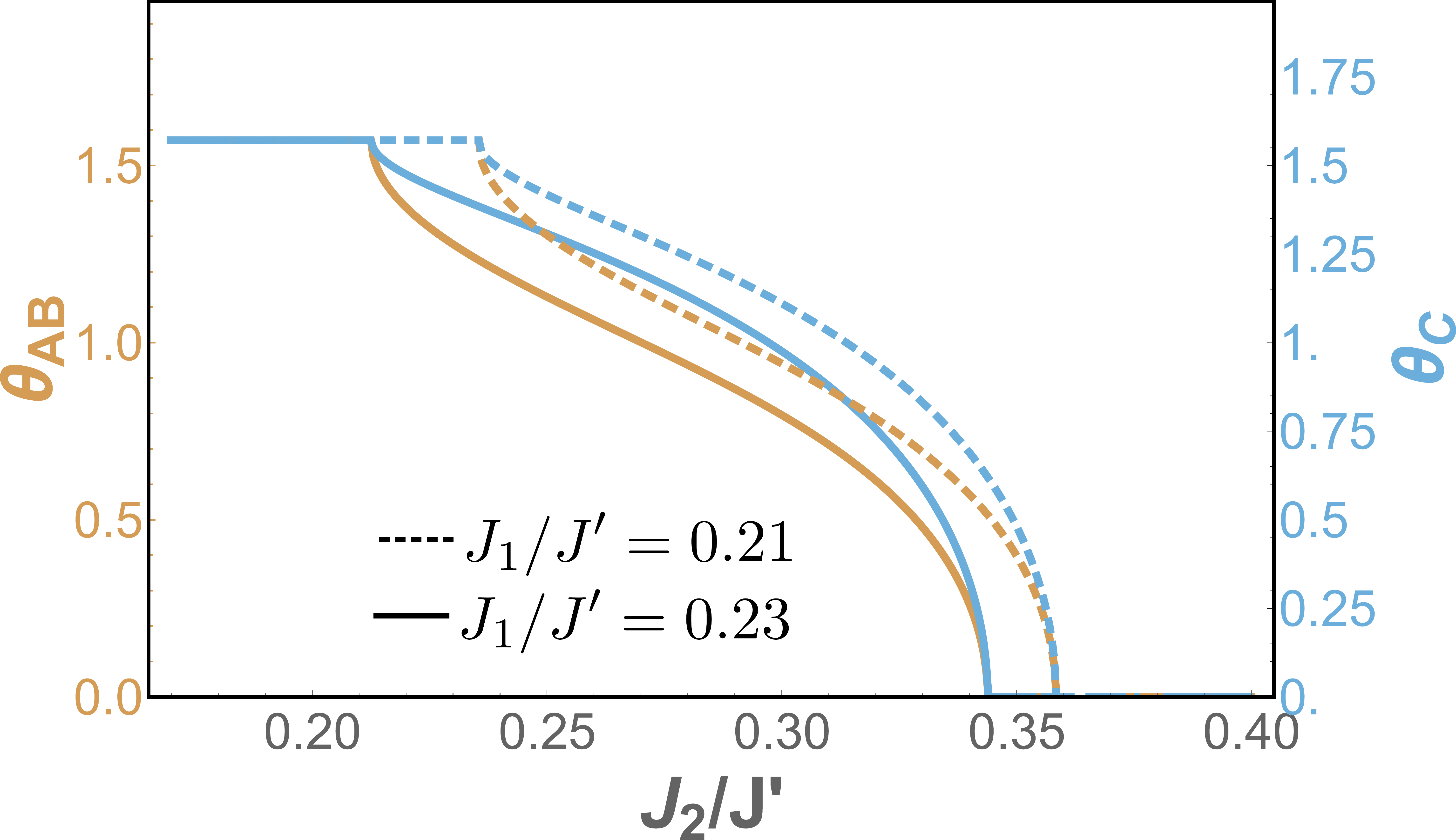}
  \caption{}
  \label{dcdicedetail}
\end{subfigure}
\caption{\label{fig:dcdice} The dice lattice version of DC I. (a) The AB and C spins are shown in common origin plots on the left and right, respectively.  Each sublattice forms a double conical spiral with conical angles $\theta_A$, $\theta_B = \pi-\theta_A$ and $\theta_{C}$ respectively. Here, $\bQ_1$ takes the special commensurate value, $\xi \bQ_{coll}$ with $\xi = 2/5$, which occurs for $J_1/J' = .4$ and $J_2/J'=.24$. (b)  The variation of the double conical I parameters as a function of $J_2$ for two values of $J_1/J'$ values is shown. The conical angles, $\theta_{AB}$ (yellow) and $\theta_C$ (blue) are plotted from the boundary of DC I and non-collinear II out to the critical $J_2$ beyond which it becomes collinear.}
\end{figure}

%%%%%%%%%%%%%%%%%%%%%%%%%%%%%%%%%%%%%%%%%%%%%%%%%%%%%%%%%%%%

\subsection{Double conical phase II}\label{dc2}

\begin{figure}[h]
\captionsetup[subfigure]{justification=centering}
  \centering
\begin{subfigure}[t]{.23\textwidth}
  \includegraphics[width=\linewidth]{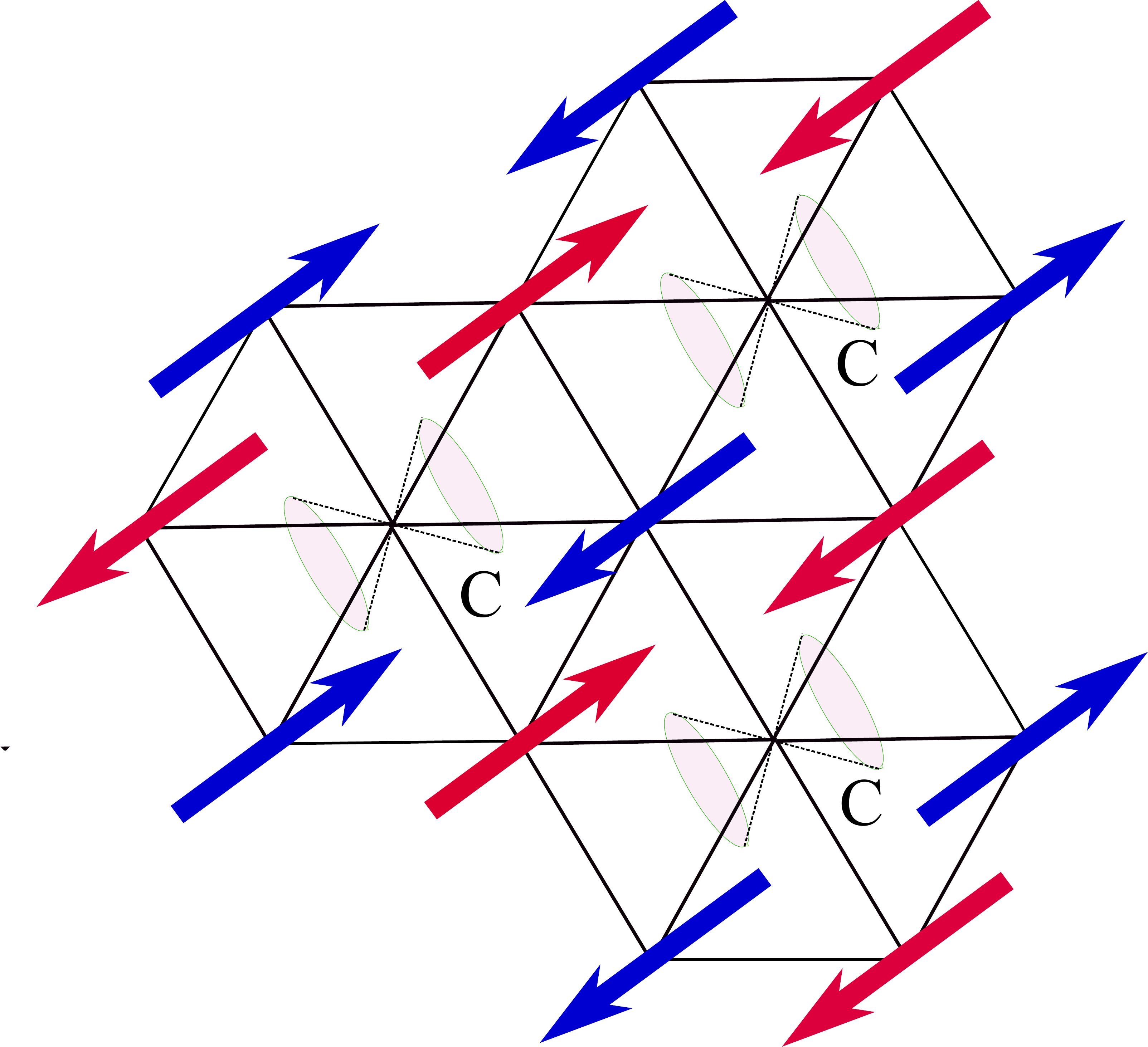}
\caption{}
\label{dcII1}
\end{subfigure}\hfill
\begin{subfigure}[t]{.23\textwidth}
  \centering
  \includegraphics[width=\linewidth]{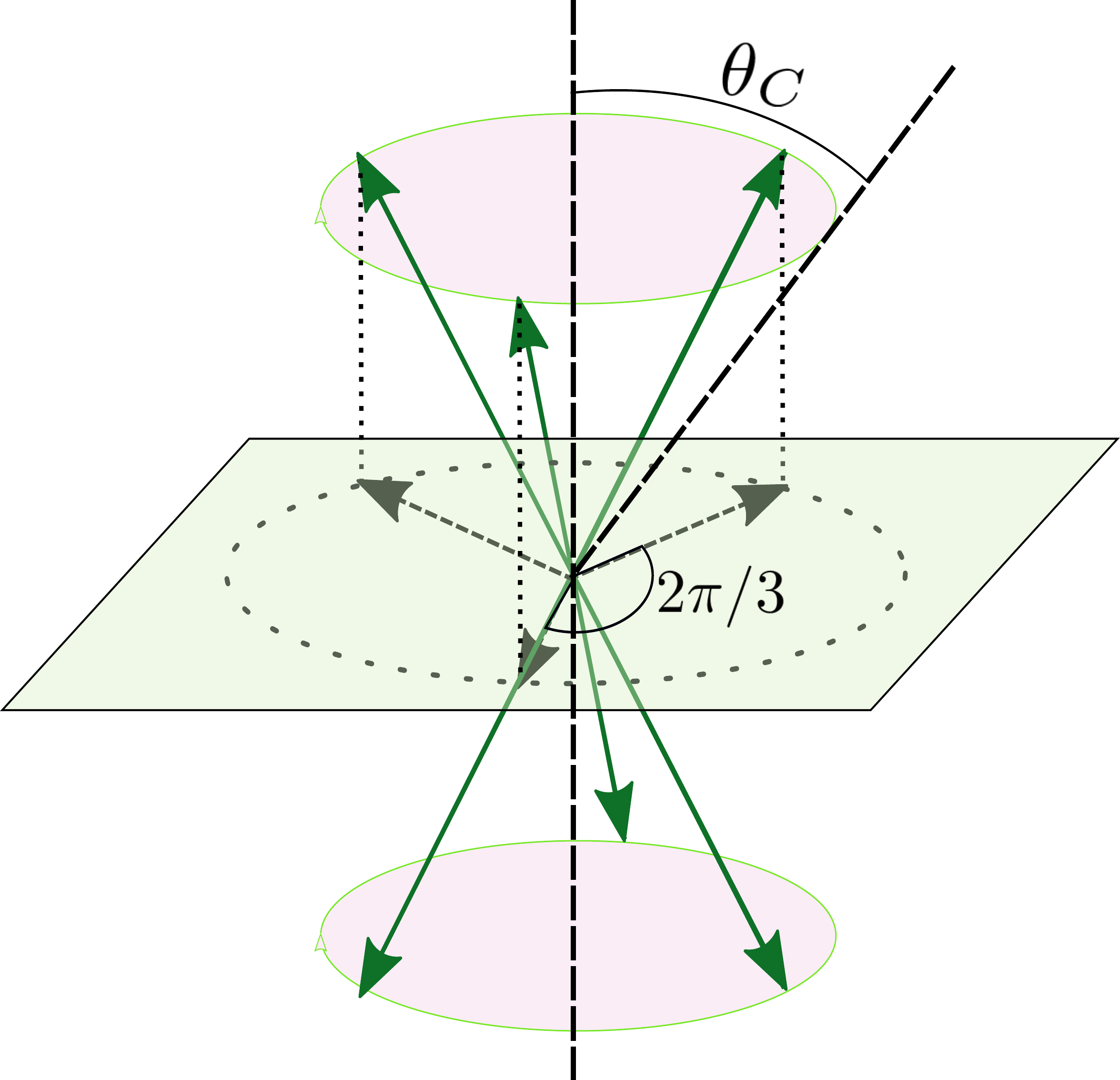}
\caption{}
\label{dcII2}
\end{subfigure}
\caption{\label{fig:dc2} Double Conical II:  The A and B sublattices are each in a collinear configuration, while the C sublattice forms a double conical structure, with the conical axis oriented along the A/B spins. (a) shows the real space lattice configuration.  (b) shows the double conical structure of the C spins in a common-origin plot, where $\bQ_1 = \bQ_{tri}$ and $\bQ_2 = \bQ_{coll}$.  The conical angle $\theta_C$ is a function of both $J_2$ and $J'$.}
\end{figure}

Sandwiched between the triple conical, spiral*, spiral and collinear phases is a second double conical phase, double conical II. The A and B spins remain in a collinear structure, with $\bQ_{coll} = (2\pi/3,0)$, while the C spins form a double conical spiral, as shown in Fig. \ref{fig:dc2}, with the conical axis collinear with the AB spins.  This double conical spiral has a single free parameter,  the conical angle, $\theta_C$, while $\bQ_1 = \bQ_{tri} = (\frac{2\pi}{3},\frac{2\pi}{\sqrt{3}})$ and $\bQ_2 = \bQ_{coll}$ are all fixed. The classical energy for this phase is,
\begin{equation}
E_{DCII}[J_2, J',\theta_C] = -1 + \frac{13 J_2}{4} - 2 J' \cos \theta_C - \frac{J_2}{4} \cos 2\theta_C
\end{equation}
This phase smoothly evolves into the collinear phase, as $\theta_C \rightarrow 0$, but all other phase boundaries are first order.  While the wedge of DC II appears to touch the honeycomb axis, as with the triangle of triangles phase, it merely approaches closely.

%\begin{table}[H]
%\begin{center}
%\begin{tabular}{|r|l|}

%\hline
%          Collinear &  -1 - 3$J_2$ - 2$J'$ \\
%\hline
    %   N\'{e}el &  -3 + 9/2 $J_2$ \\
%\hline
    %  Spiral & $\begin{aligned} &Min[1/2 J_2 (-3 + 4 \cos(\sqrt{3} q_y) + 4 \cos((3 q_x)/2 - (\sqrt{3} qy)/2) + 4 \cos((3 q_x)/2 + \\ & (\sqrt{3} q_y)/2))  + 1/2 (2 (1 + \cos(\sqrt{3} q_y) + \cos((3 q_x)/2 +   (\sqrt{3} q_y)/2)) \cos(\gamma) + \\ & 2 (-\sin(\sqrt{3} q_y) - sin((3 q_x)/2 + (\sqrt{3} q_y)/2)) \sin(\gamma))]\,\,w.r.t\,\,{q_x, q_y, \gamma}\end{aligned}$ \\
%\hline
  %     Interpolating & $(9 J_2 - 3/2 (2 + J'^2)) \theta(2 - J') + (3 + 9 J_2 - 6 J') \theta(J' - 2)$ \\
%\hline
    %   Non-collinear II &  $\begin{aligned} J_2 - 3 \cos^2(\delta)  +  4 J_2 \cos(2 \delta) - & 2 J'  \sin(\delta) -  \sin^2(\delta) \\ & with\,\, \delta = Re\left[-Sin^{-1}\left(\frac{J'}{8 J_2 - 2}\right)\right]\end{aligned}$\\
%\hline
 %      Non-collinear I &  $\begin{aligned}  Min[2 J_2 (1 & + 2 \cos(2 \theta_{AB})) + J_2 (1 + 2 \cos(2 \theta_C)) +  (1 + 2 \cos(2 \theta_{AB})) 
%\\ & + J' (-6 \cos(\theta_{AB}) \cos(\theta_C) - 2 \sin(\theta_{AB})\sin(\theta_C))]\,\,\, w.r.t\,\,\, (\theta_{AB}\,\, and \,\, \theta_C) \end{aligned}$ \\
%\hline
  %    TT & $\begin{aligned}Min[1/2 (3 \cos(\gamma) & + 2 J_2 \cos(\lambda) + 
  % 4 J_2 \cos(\theta) \cos(\lambda/2) \cos(\lambda/2 + \nu) \\ & - \sin(\gamma) (\sqrt{3} + J' \sin(\theta) (1 + \sin(\eta + \lambda + \nu))))]\,\,w.r.t\,\,(\gamma, \lambda, \theta, \eta, \nu)\end{aligned}$\\
%\hline
%\end{tabular}
%\end{center}
%\caption{Analytic expressions of energy for the phases}
%\label{table:table1}
%\end{table}  

\section{Non-collinear phases}\label{ncp}

Just off of the triangular lattice axis near the triangular lattice critical point, we find two interesting phases whose width vanishes as the critical point is approached, as seen in Fig. \ref{cpd}.  These are each separated from the collinear phase by second order transition lines, and share a number of common characteristics. As the fluctuations of these phases may strongly influence the spin liquid found on the triangular axis, we study these phases and their fluctuations in more detail. In particular, both phases have a $\mathbb{Z}_3$ nematic order parameter, and a free classical angle that allows them to be non-coplanar in principle, although order by disorder naturally selects the coplanar configuration.

\begin{figure}[h]
  \centering
  \includegraphics[width=.8\linewidth]{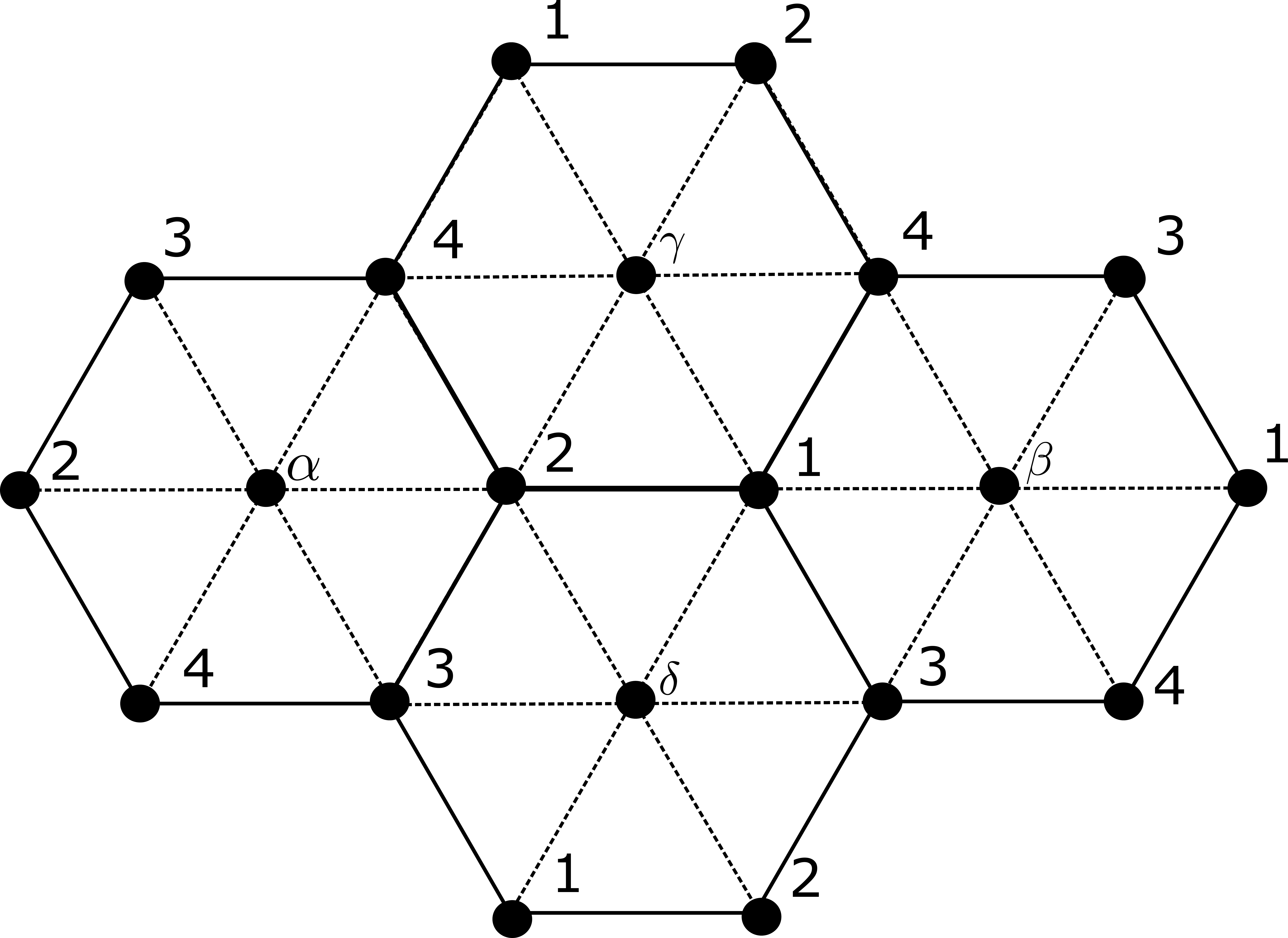}
  \caption{The sublattice configuration of the non-collinear phases. Each non-collinear phase has a 12 sublattice structure, wherein each of the A,B, and C sublattices has four different spin configurations, arranged as shown in the figure, where the AB and C sublattices are labeled with numbers and greek letters respectively.}
  \label{nc3}
\end{figure}

Both non-collinear phases are most generally described in a twelve-sublattice basis, where each of the three A, B and C sublattices has four sublattices; these four sublattices are the same ABCD sublattices from the collinear phase, although the collinear condition is of course not satisfied.  These are shown in Fig. \ref{nc3}, where the AB spins are labeled with (1,2,3,4) and the C spins with ($\alpha,\beta,\gamma,\delta$).  Generically these A,B,C spins all sit on different cones forming pairs of spins, as shown in Fig. \ref{nc1} and Fig. \ref{nc2}.

\subsection{Non-collinear phase I}\label{NCI}

\begin{figure}[h]
  \centering
  \includegraphics[width=.4\linewidth]{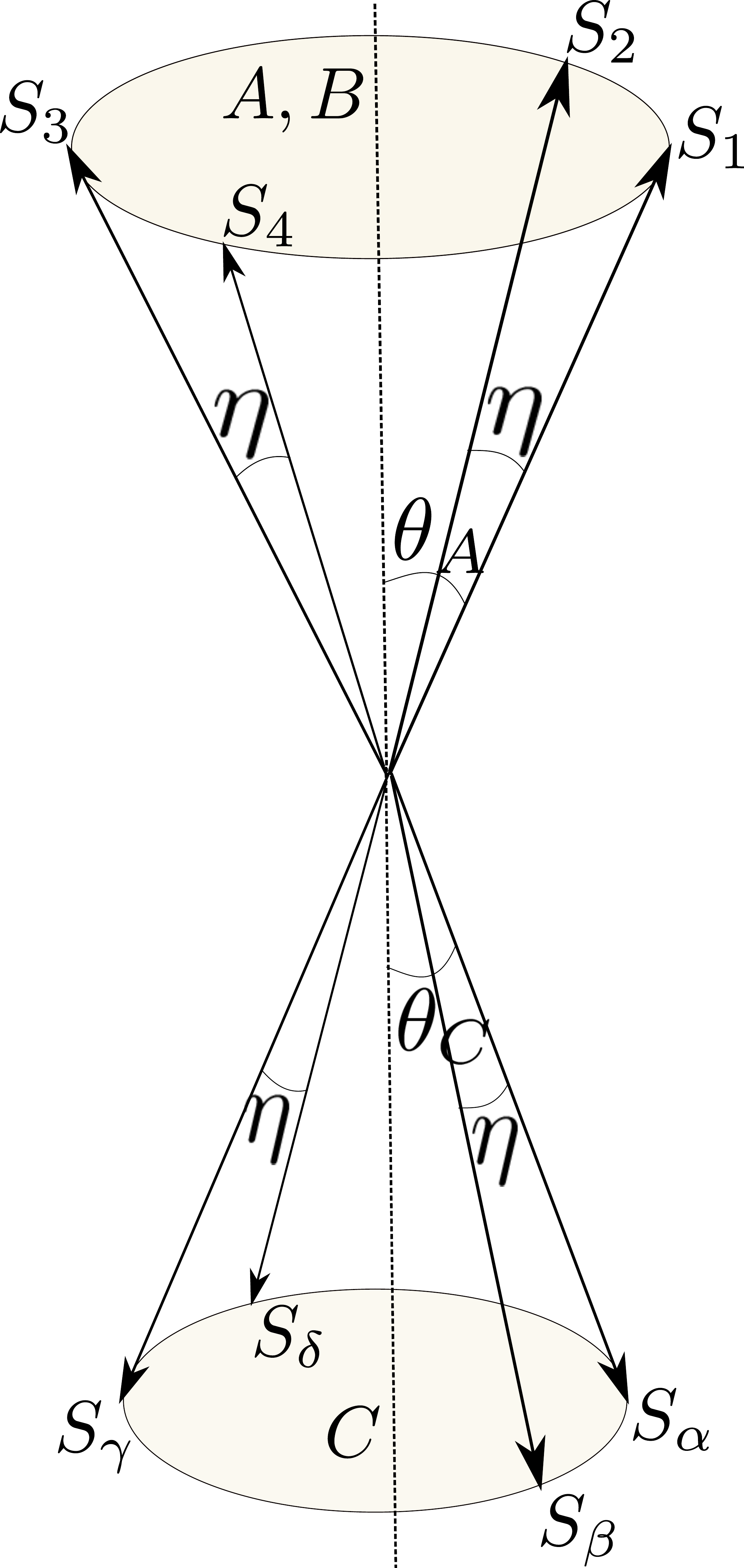}
  \caption{Non-collinear phase I consists of spins on two cones: one for the AB spins (up, with conical angle $\theta_A$) and one for the C spins (inverted, with $\theta_C$). The spin components perpendicular to the conical axis form opposing pairs, (1,3) and (2,4); ($\alpha$, $\gamma$) and ($\beta$, $\delta$).}
  \label{nc1}
\end{figure}

\begin{figure}[h]
  \centering
  \begin{subfigure}[h]{.44\textwidth}
  \centering
  \includegraphics[width=.9\linewidth]{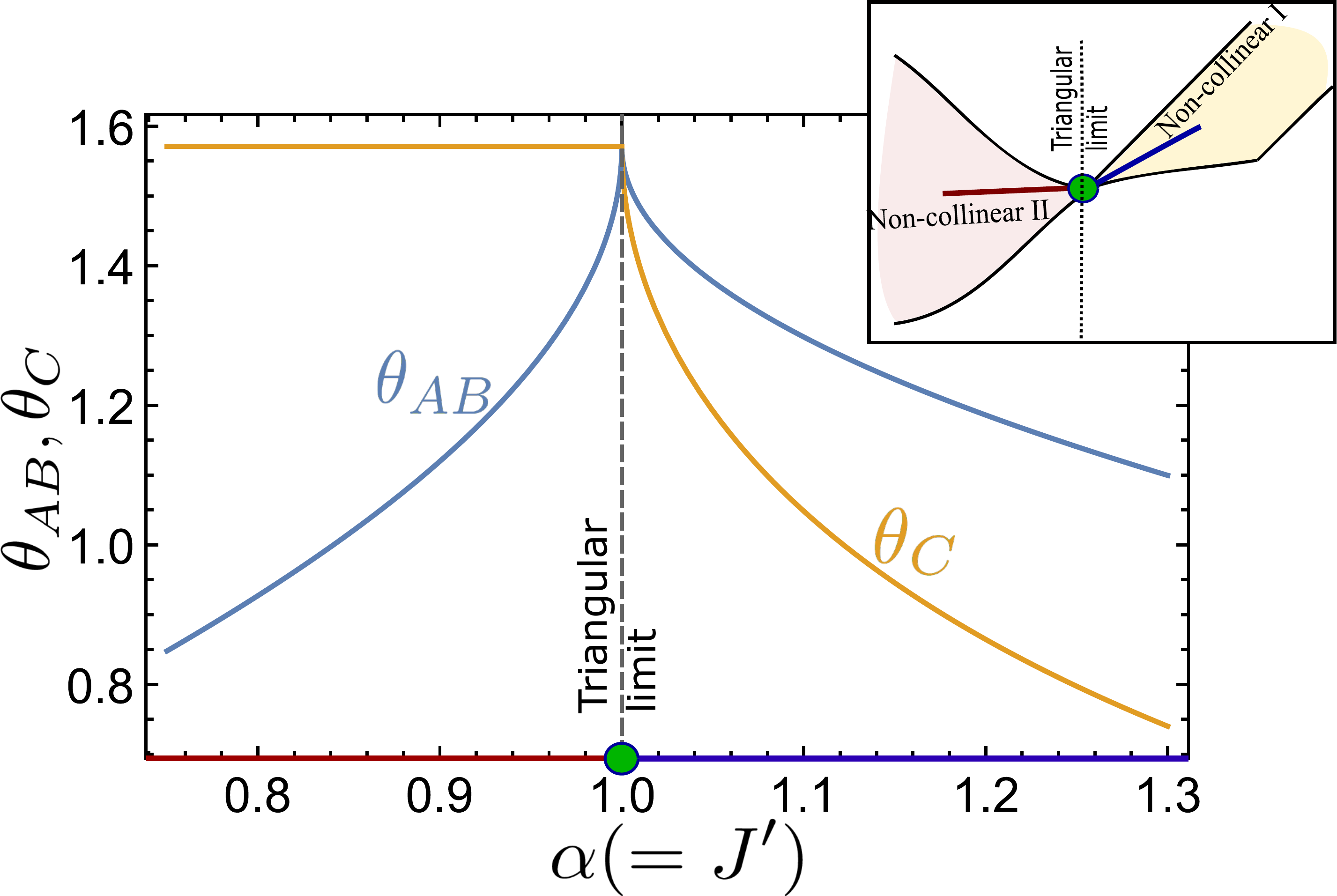}
  \caption{Conical angles}
  \label{thtnc}
\end{subfigure}\\[1ex]
\begin{subfigure}[h]{.44\textwidth}
  \centering
  \includegraphics[width=.9\linewidth]{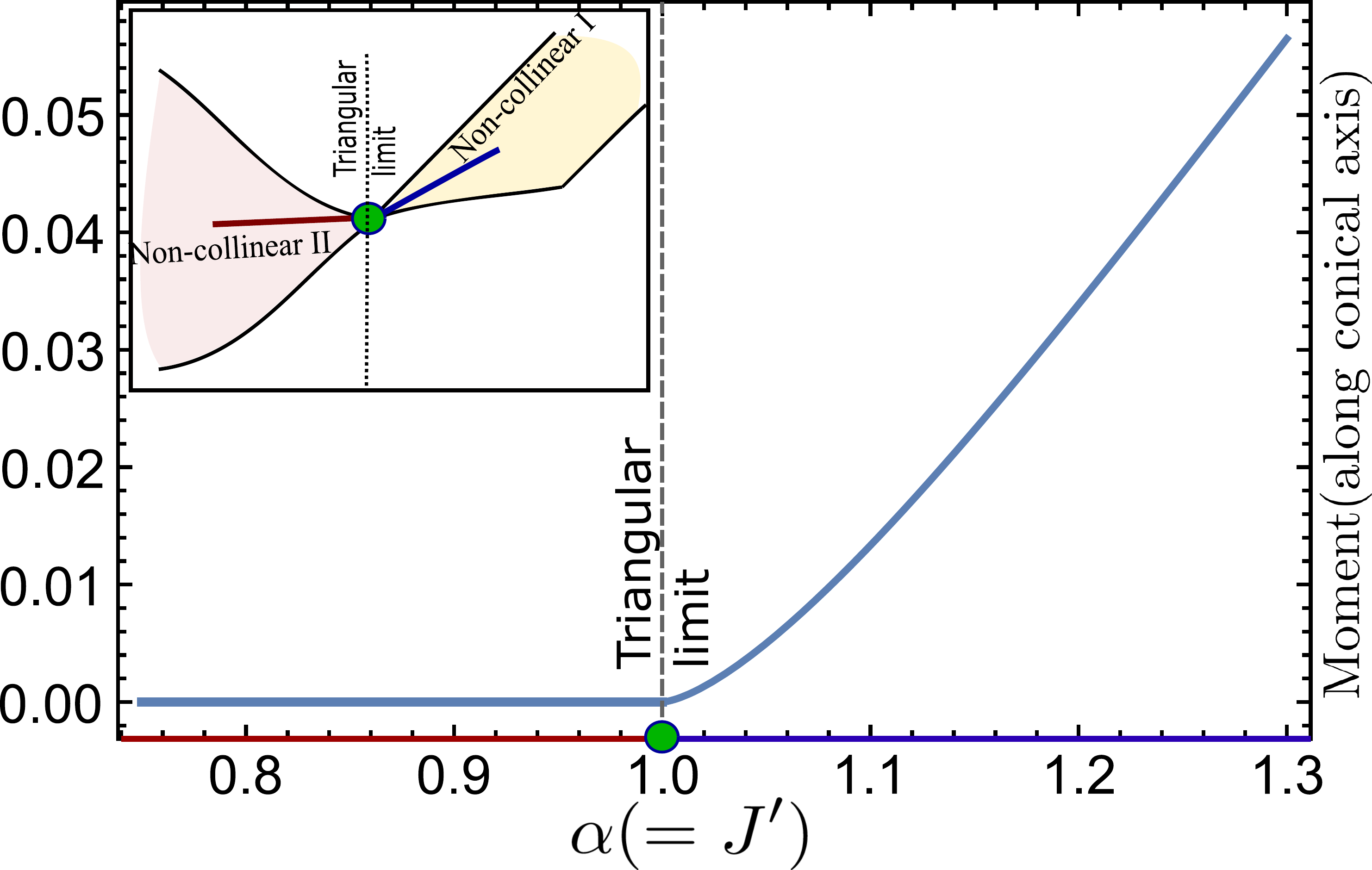}
  \caption{Magnetic moment}
  \label{mom}
\end{subfigure}
  \caption{(a) Variation of conical angles for the non-collinear phases along a path $J' = \alpha$, $J_2 = \frac{3}{20}(\alpha-1)\theta(\alpha-1)+\frac{1}{8}$ (shown in inset). The C spins are planar for $\alpha < 1$, with $\theta_C = \pi/2$; $\theta_C$ then decreases with increasing $\alpha$ as the spins tilt down into an increasingly narrowing cone.  The AB spins are similarly planar exactly at $\alpha = 1$, with decreasing conical angles away from the triangular axis in both phases. (b) Ferromagnetic moment in the non-collinear phases. The inset at the left corner shows the path with the same parametrization as before, along which the variation of the moment is plotted. The path and the x axis are also color matched to further elucidate this fact. }
\end{figure}

For $J' > J_1$, the four A and B spin configurations overlap, and exist on a cone with fixed angle $\theta_A = \theta_B$, while the C spins are on an inverted cone with angle $\theta_C \neq \theta_A$; these spin configurations are shown on a common origin plot in Fig. \ref{nc1}. This phase generically has a net moment along the common conical axis, which is zero on the triangular axis, and increases smoothly with increasing $J'$, as shown in Fig. \ref{mom}.  The AB spin components perpendicular to the conical axis form opposing pairs, ($\vec{S}_1$, $\vec{S}_3$) and ($\vec{S}_2$, $\vec{S}_4$) with a free angle $\eta$ between the two pairs. Similarly, the perpendicular C spin components form pairs, ($\vec{S}_\alpha$, $\vec{S}_\gamma$) and ($\vec{S}_\beta$, $\vec{S}_\delta$), separated by the same free classical angle. The AB spins are,
\begin{align}\label{AB}
\begin{split}
&\vec{S}_1 = [\sin\theta_{A},0,\cos\theta_{A}]\\
&\vec{S}_2 = [\sin\theta_{A}\cos\eta,\sin\theta_{A}\sin\eta,\cos\theta_{A}]\\
&\vec{S}_3 = [-\sin\theta_{A},0,\cos\theta_{A}]\\
&\vec{S}_4 = [-\sin\theta_{A}\cos\eta,-\sin\theta_{A}\sin\eta,\cos\theta_{A}],
\end{split}
\end{align}
while the C spins are,
\begin{align}\label{RR}
\begin{split}
&\vec{S}_\alpha = [\sin\theta_{C},0,-\cos\theta_{C}]\\
&\vec{S}_\beta = [\sin\theta_{C} \cos \eta,\sin\theta_{C}\sin\eta,-\cos\theta_{C}]\\
&\vec{S}_\gamma = [-\sin\theta_{C},0,-\cos\theta_{C}]\\
&\vec{S}_\delta = [-\sin\theta_{C}\cos\eta,-\sin\theta_{C}\sin\eta,-\cos\theta_{C}]
\end{split}
\end{align}
These spins are arranged as shown in Fig. \ref{nc3}. The resulting classical energy is given by
\begin{align}
E_{nc1}[J_2, &J']  = (1+2J_2)(1+2\cos 2\theta_{A})\cr
& +J_2(1+2\cos 2\theta_C) \cr
& -2J'(\sin\theta_{A}\sin\theta_C + 3\cos\theta_{A}\cos\theta_C),
\end{align}
where $\theta_A$ and $\theta_C$ are variational parameters, and the energy is independent of $\eta$.
The variation of the conical angles for both non-collinear phases along a particular parametric path $J_2(J')$ is shown in Fig. \ref{thtnc}.

In section \ref{obd}, we shall show that order by disorder selects $\eta = 0$, favoring the coplanar spin configuration, as expected. Nevertheless, the relatively low energy competing states may affect the nature of the spin liquid. In particular, the non-coplanar configurations will generically have nonzero \emph{scalar spin chirality}, defined on a triangle of spins (1,2,3),
\begin{equation}
\kappa_{\Delta} = \vec{S}_1\cdot \vec{S}_2\times \vec{S}_3.
\end{equation}
In Fig. \ref{c1}, we show the pattern of striped spin chirality for the nearest-neighbor triangles, with $\eta \neq 0$.  There are four chiralities $\pm \kappa_{1,2}$, given by,
\begin{align}
\label{kappa}
&\kappa_1 = \sin\eta \sin\theta_A \sin(\theta_{A}-\theta_C)\cr
&\kappa_2 = \sin\eta \sin 2 \theta_{A} \sin \theta_{C},
\end{align}
which lead to two types of hexagons, with positive and negative chiralities arranged in a striped pattern. Note that there is no uniform chirality.
%\begin{figure}[h]
%  \centering
%  \includegraphics[width=.9\linewidth]{chiral1final.pdf}
%  \caption{For $\eta \neq 0$, non-collinear I has a striped pattern of chirality, as shown by the four values of chirality on nearest-neighbor triangles, indicated with four colors.  Each hexagon has a distinct sign of chirality, which is arranged in a collinear pattern.}
%  \label{c1}
%\end{figure}

In addition, we also see that this magnetic order breaks the three-fold lattice rotational symmetry by developing a $\mathbb{Z}_3$ nematic bond order parameter,
\begin{equation}
\mathcal{N} = \frac{1}{N}\sum_i \langle \vec{S}_i \cdot \vec{S}_{i+\mathbf{a}_1}\rangle + \langle \vec{S}_i \cdot \vec{S}_{i+\mathbf{a}_2}\rangle \mathrm{e}^{\frac{2\pi i}{3}}+ \langle \vec{S}_i \cdot \vec{S}_{i+\mathbf{a}_2-\mathbf{a}_1}\rangle \mathrm{e}^{\frac{4\pi i}{3}},
\end{equation}
where $i$ sums over all spins, in all sublattices and $N$ is the total number of sites. As this $\mathbb{Z}_3$ order parameter breaks a discrete symmetry, it can, and will develop at finite temperatures before the magnetic ordering sets in. This finite temperature phase transition also occurs in the neighboring collinear phase, and is not fundamentally different here.  Essentially, in the ground state, spins along one of the three lattice directions are ferromagnetically aligned: for $\eta = 0$, this is the $\hat x$ direction which connects $\vec{S}_1$ to $\vec{S}_2$, and $\vec{S}_3$ to $\vec{S}_4$. The particular direction of such correlations is selected at this nematic transition, even though the spins themselves do not order until $T = 0$; this is a \emph{$\mathbb{Z}_3$ bond order}.

\subsection{Non-collinear phase II}\label{NCII}

\begin{figure}[t]
  \centering
  \includegraphics[width=.5\linewidth]{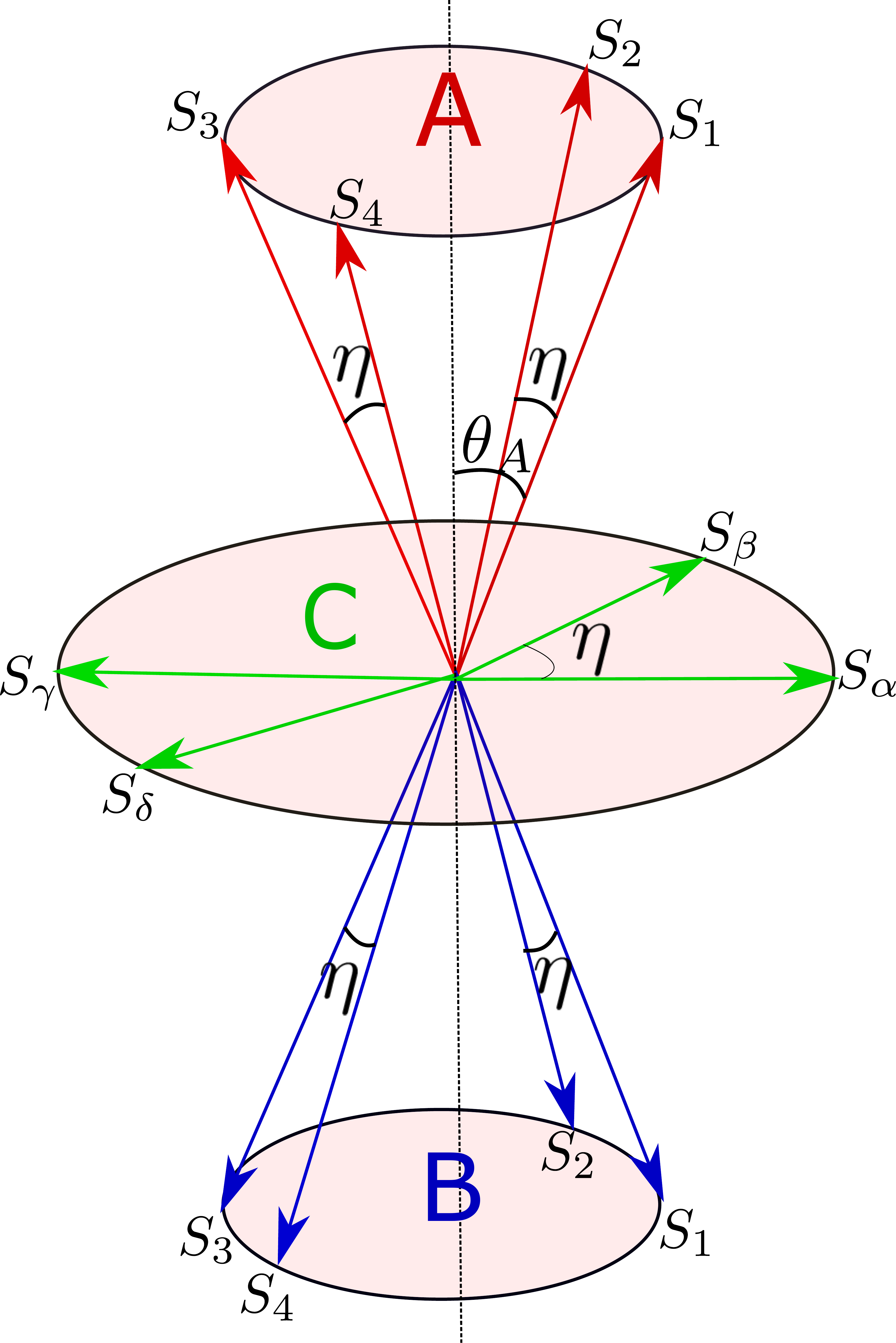}
\caption{ Non-collinear II is similar to non-collinear I, but here, the A and B cones are inverted with respect to one another and the C spins lie in the plane.  This phase has the same classical free angle $\eta$ between pairs of spins.}
 \label{nc2}
\end{figure}
\begin{figure}[t]
\centering
\begin{subfigure}[h]{.44\textwidth}
  \centering
  \includegraphics[width=.9\linewidth]{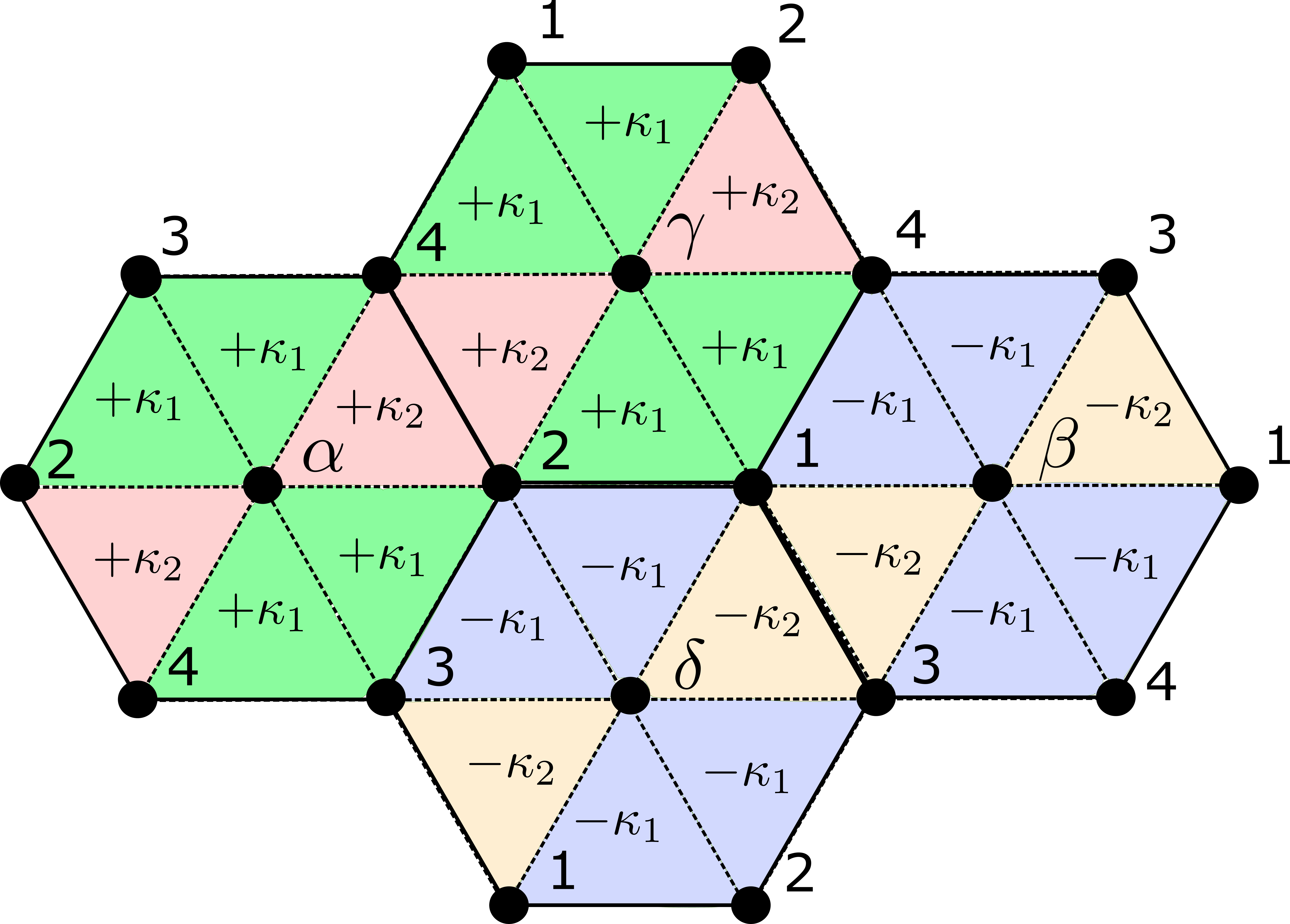}
  \caption{Chirality for non-collinear I}
  \label{c1}
\end{subfigure}\\[1ex]
\begin{subfigure}[h]{.44\textwidth}
  \centering
  \includegraphics[width=.9\linewidth]{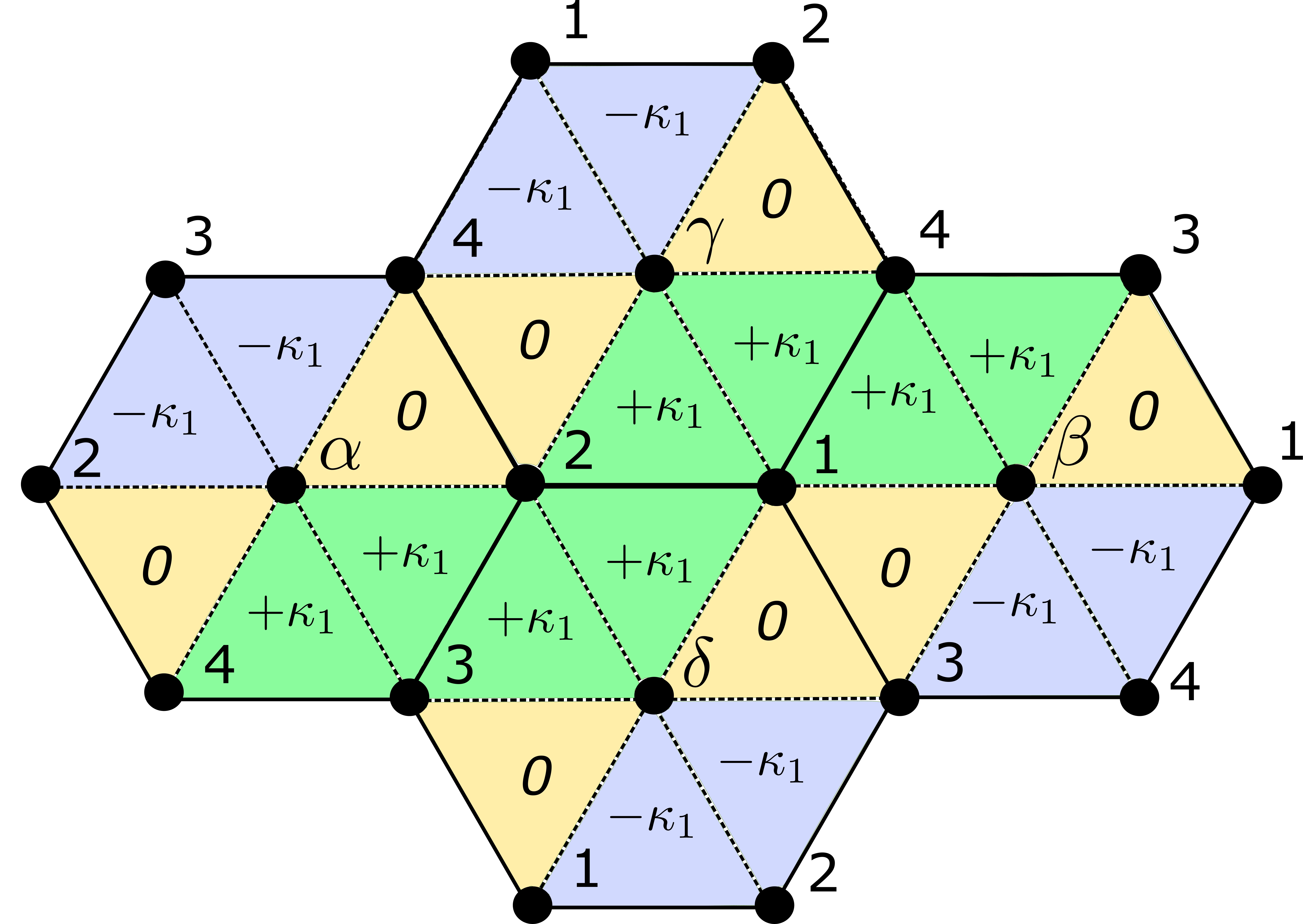}
  \caption{Chirality for non-collinear II}
  \label{c2}
\end{subfigure}
\caption{(a) For $\eta \neq 0$, non-collinear I has a striped pattern of chirality, as shown by the four values of chirality ($\pm \kappa_{1,2}$) on nearest-neighbor triangles, indicated with four colors.  Each hexagon has a distinct sign of chirality, which is arranged in a collinear pattern. (b) Non-collinear II similarly has a striped pattern of chirality, with $\kappa_1$ corresponding to $\theta_C = \pi/2$ in eqn. \ref{kappa} while the zero chirality comes from the  antiparallel orientation of spins $A_1$ and $B_3$ and also $A_2$ and $B_4$ as shown in fig.\ref{nc2}. The flipping of the signs of $\kappa_1$ is also due to the opposite orientation of the A and B cones.}
\label{zeroeta}
\end{figure}

For $J' < J_1$, the B sublattice cone flips, with $\theta_B = \pi - \theta_A$ and consequentially, the C spins become planar, as shown in Fig. \ref{nc2}, with $\theta_C = \pi/2$.  There is no longer any net moment.  Otherwise, the structure of this phase is identical, with the same pairs of spins on each sublattice with a free classical angle $\eta$ between them.  The classical energy is,
\begin{align} 
\begin{split}
E_{nc2}[J_2,J'] & =\! J_2(1\!+\!4\cos 2 \theta_A)\! -\! 2 \cos^2\theta_A\! -\! 2 J'  \sin\theta_A\! - \!1\\
& \text{where}\,\,\,\theta_A = \sin^{-1}\left[\frac{J'}{8J_2-2}\right]
\end{split}
\end{align}
There is still a striped pattern of chirality for nonzero $\eta$, as shown in Fig. \ref{c2}; here, the form of the chiralities given in eqn. (\ref{kappa}) is identical, with $\theta_C = \pi/2$ and some sign changes due to the inversion of the B cone, as indicated in the figure.  Again, there is no uniform chirality, and this phase also possesses an identical $\mathbb{Z}_3$ nematic order.

%\subsection{Multi-critical point}

%\noindent The triangular lattice critical point is revealed as a multi-critical point in our phase diagram which implies lots of classical degeneracy making it an ideal place to look for a spin liquid. The non-collinear phases appear off the triangular axis as illustrated in fig.\ref{tetra}.

%In general, multicritical points are seen as points in the parameter space with a lot of fluctuations (requires second order phase transitions) providing a possible region to harbor a spin liquid phase. There are several multicritical points in the classical phase diagram. Of them, the ones around the triangular axis and the other closer to the triple conical phase with five phases around. The later consists of phases which are likely to be washed off by fluctuations and hence we consider only the first one. Interestingly, this point reveals what is a first order transition in the triangular lattice to have a possible second order transition ($120^o$ to collinear). Also, it includes a lot of fluctuations because of the infinite degeneracy from the free classical angle ($\eta$) in the non-collinear phases. 

%\begin{figure}[H]
%\centering
%  \includegraphics[width=0.8\linewidth]{tetrapoint.pdf}
%\caption{{\bf Multi-critical point}: Illustration of the multi-critical point right on the triangular limit line formed at the junction of 2 first order and 2 second order lines.}
%\label{tetra}
%\end{figure}

\subsection{Spin wave calculations and order by disorder}\label{obd}

In this section, we consider the effect of both quantum and thermal fluctuations on the two non-collinear phases.  These two phases behave quite similarly, and so we mostly focus on non-collinear I.  First, we develop the linear spin-wave theory for the non-collinear phases with $\eta = 0$.  We also run classical Monte Carlo to show that thermal fluctuations select the coplanar state. While there is no magnetic order at finite temperatures, there is a nematic phase transition where the stripes of ferromagnetic spins choose to run along one of the three lattice directions.

%%%%%%%%%%%%%%%%%%%%%%%%%%%%%%%%%%%%%%%%%%%%%%%%%

\begin{figure}[t]
\centering
\begin{subfigure}[h]{.5\textwidth}
  \centering
  \includegraphics[width=.5\linewidth]{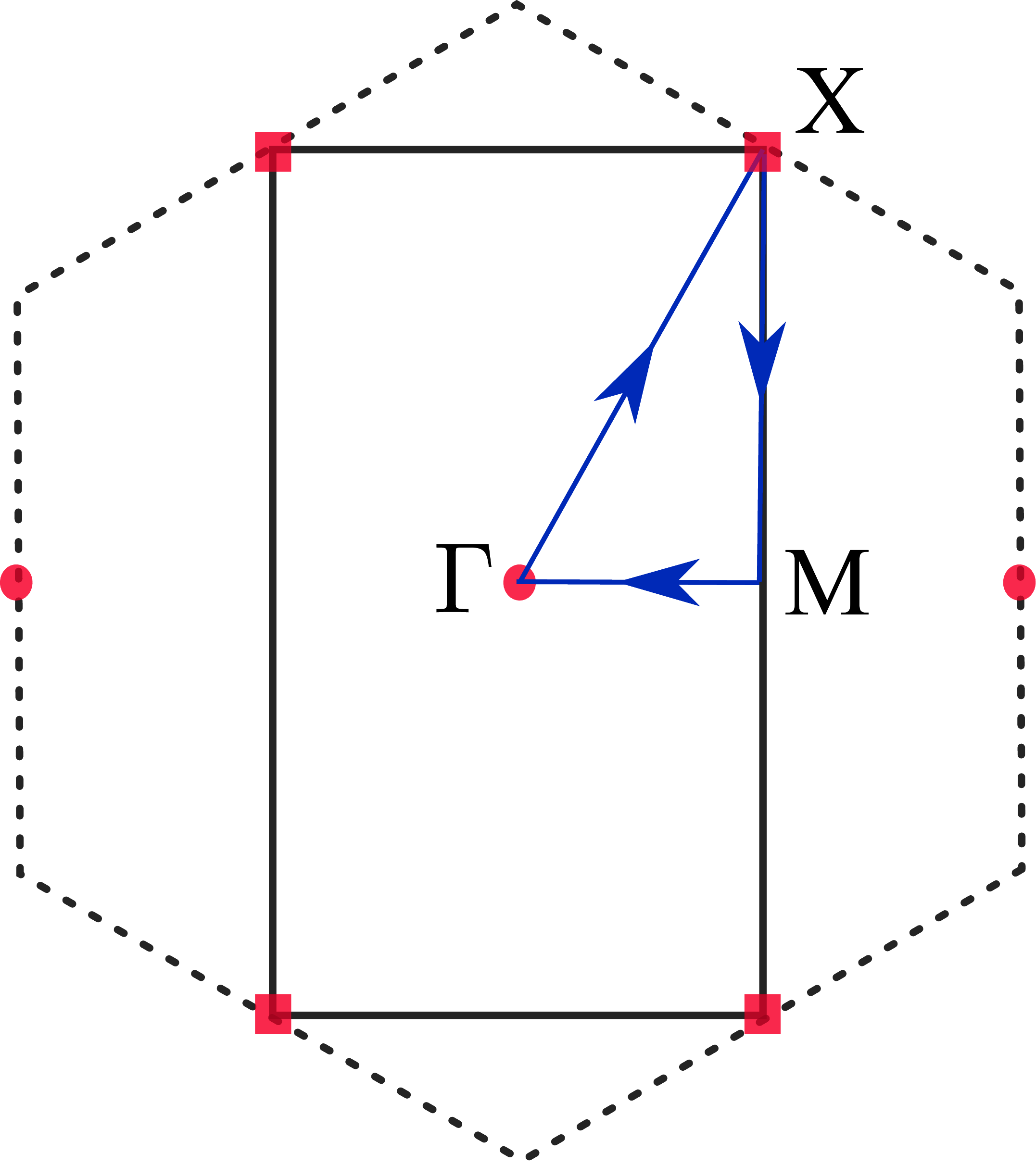}
  \caption{New and old Brillouin zones.}
  \label{eta01}
\end{subfigure}\\[1ex]
\begin{subfigure}[h]{.5\textwidth}
  \centering
  \includegraphics[width=.95\linewidth]{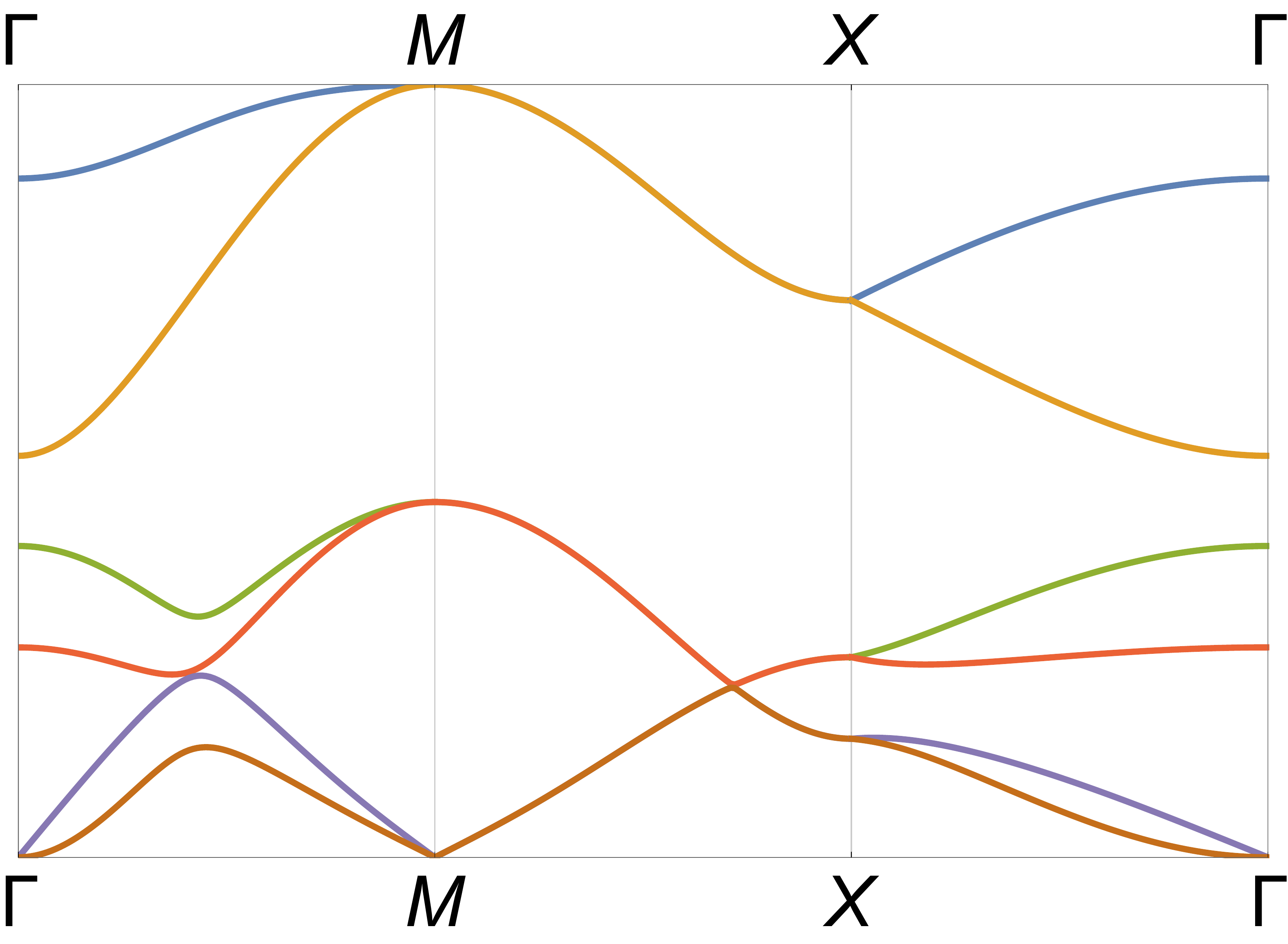}
  \caption{Spin wave dispersion.}
  \label{eta02}
\end{subfigure}
\caption{Spin wave dispersion for non-collinear phase I for $\eta =0$, which is selected via order by disorder. (a) shows both the original hexagonal Brillouin zone with the three sublattices, and the new rectangular Brillouin zone for this six site unit cell. As this six site unit cell derives from the collinear phase, there are linear modes at each of the M points of the original Brillouin zone: one at the reciprocal lattice vectors of the new Brillouin zone ($\bQ_{coll}$, indicated by red circles) and two at the corners (indicated by red squares).  (b) shows the spin wave dispersion.  In total, there are three Goldstone modes, as expected for noncollinear orders.  There is also a quadratic mode at $\Gamma$ associated with the accidental classical degeneracy in $\eta$ that will be gapped out by $1/S$ corrections \cite{chubukov92}.}
\label{zeroeta}
\end{figure}

%%%%%%%%%%%%%%%%%%%%%%%%%%%%%%%%%%%%%%%%%%%%%%%%%

%%%%%%%%%%%%%%%%%%%%%%%%%%%%%%%%%%%%%%%%%%%%%%%%%

%%%%%%%%%%%%%%%%%%%%%%%%%%%%%%%%%%%%%%%%%%%%%%%%%

%%%%%%%%%%%%%%%%%%%%%   The FIGURE for band dispersion  %%%%%%%%%%%%%%%%%%%%%%%%%%%%%%%

\subsubsection{Spin wave theory}

In this section, we give the details of our linear spin-wave calculation for $\eta =0$.  In this simplified case, two sets of spins are equivalent: $\vec{S}_1 = \vec{S}_2$, $\vec{S}_3 = \vec{S}_4$, $\vec{S}_\alpha = \vec{S}_\beta$, and $\vec{S}_\gamma = \vec{S}_\delta$, and so we can use six sublattices, instead of twelve. As we have six sublattices, we require six Holstein-Primakoff (HP) bosons.  We define a local triad of orthonormal vectors for each sublattice; these triads are related by rotations in the conical space defined by the angles $\theta_{A}$($\theta_C$) (Figs. \ref{nc1} and \ref{nc2}) for these non-collinear phases. A given spin operator can be expressed as 
\begin{equation}
\vec{S}_m = \sum_{i}\tilde{S}_{m,i}{\bf t}_{i,m}
\end{equation}
where the `tilde' on the spin components emphasizes the fact that these are defined in the local basis {\bf t} and $m$ is a sublattice index. The local bases are defined as,
\begin{equation}
\mathbf{t}_{i,m} = \mathcal{R}(\theta_m){\bf e}_i,
\end{equation}
with ${\bf e} = (e_x,e_y,e_z)$ and $\mathcal{R}$ the appropriate rotation matrix; for example, for $\vec{S}_1$, we have
\begin{equation}
\mathcal{R} = \begin{pmatrix}
    \cos\theta_{m}       & 0 & \sin\theta_{m} \\
    0     & 1 & 0\\
-\sin\theta_{m} &0 &\cos\theta_{m}
\end{pmatrix}.
\end{equation}
The spin components are then Fourier transformed via
\begin{equation}
\tilde{S}_{m,i}({\bf r}) = \frac{1}{\sqrt{N}}\sum_{\bf q} \tilde{S}_{m,i}({\bf q})e^{i{\bf q}\cdot{\bf r}}
\end{equation}
The Hamiltonian in Fourier space then becomes,
\begin{align}\label{hamfr}
\begin{split}
\mathcal{H} =& J_1\sum_{\mathclap{\substack{i,m,n, \\ \pmb{\delta}_{AB},{\bf q},{\bf q'}}}}\tilde{S}_{m,i}^{A}({\bf q})\tilde{S}_{n,i}^{B}({\bf q'}){\bf t}_{i,m}{\bf t}_{i,n}e^{i ({\bf q}+ {\bf q'}).{\bf r}}e^{i {\bf q'}.\pmb{\delta}_{AB}}\\
&+ J_2\sum_{\mathclap{\substack{i,m,n,\eta \\ \pmb{\delta}_{2},{\bf q},{\bf q'}}}}\tilde{S}_{m,i}^{\eta}({\bf q})\tilde{S}_{n,i}^{\eta}({\bf q'}){\bf t}_{i,m}{\bf t}_{i,n}e^{i ({\bf q}+ {\bf q'}).{\bf r}}e^{i {\bf q'}.\pmb{\delta}_2}\\
&+J'\sum_{\mathclap{\substack{i,m,n,\eta' \\ \pmb{\delta}_{C},{\bf q},{\bf q'}}}}\tilde{S}_{m,i}^{\eta'}({\bf q})\tilde{S}_{n,i}^{C}({\bf q'}){\bf t}_{i,m}{\bf t}_{i,n}e^{i ({\bf q}+ {\bf q'}).{\bf r}}e^{i {\bf q'}.\pmb{\delta}_C}
\end{split}
\end{align}
where $i = (x,y,z)$; $m$, $n$ run over the sublattice indices: $\{1,2\}$ for AB and $\{\alpha,\beta\}$ for C; $\pmb{\delta}_{AB}$, $\pmb{\delta}_2$ and $\pmb{\delta}_C$ represent the nearest neighbors of type $J_1$, the second nearest neighbors and the nearest neighbors of type $J'$, respectively. Finally, $\eta$ labels the original sublattice indices: A, B and C, while $\eta'$ only includes A and B. 

We then use a HP representation in real space
\begin{align}
&\tilde{S}_m^+({\bf r}) = \sqrt{2S - b_m^{\dagger}({\bf r})b_m({\bf r})}b_m({\bf r})\cr
&\tilde{S}_m^-({\bf r}) = b_m^{\dagger}({\bf r})\sqrt{2S - b_m^{\dagger}({\bf r})b_m({\bf r})}\cr
&\tilde{S}_m^z({\bf r}) = S - b_m^{\dagger}({\bf r})b_m({\bf r}),
\end{align}
expand for $S \gg 1$ and Fourier transform,
\begin{align}
&\tilde{S}_m^{x}[q] = \sqrt{\frac{S}{2}} (b_m^{\dagger}[-q] + b_m[q]) + O\left(\frac{1}{S^2}\right)\cr
&\tilde{S}_m^{y}[q] = i\sqrt{\frac{S}{2}} (b_m^{\dagger}[-q] - b_m[q]) + O\left(\frac{1}{S^2}\right) \cr
&\tilde{S}_m^{z}[q] =  \sum_k -b_m^{\dagger}[k-q] b_m[q]/\sqrt{N} + \sqrt{N}S \delta_{q,0},
\end{align}
where $N$ is the number of sites.
This representation is then substituted into the above Hamiltonian, and the $O(S)$ terms extracted. The resulting quadratic Hamiltonian is then diagonalized via a Bogoliubov transformation.  This transformation is most straightforwardly done by doubling the size of the matrix using the basis,
\begin{align}
\begin{split}
X[q] = (b_1[q], b_2[q],\hdots, b_{6}[q], b_1^{\dagger}[-q], \hdots, b_{6}^{\dagger}[-q]).
\end{split}
\end{align}
The form of the Hamiltonian in the Fourier space is, 
\begin{equation}
\mathcal{H} = \sum_q X^{\dagger}[q]H(q)X[q], 
\end{equation}
\begin{equation*}
\text{where}\,\,\, H(q) =   \begin{pmatrix}F[q] & G[q]\\ G^*[-q] & F^*[-q]\end{pmatrix}
\end{equation*}
$F[q]$ defines the coefficients of terms of the form $b_m^{\dagger}[q]b_n[q]$, while $G[q]$ collects coefficients of terms like $b_m^{\dagger}[q]b_n^{\dagger}[-q]$. Due to the large size of the magnetic unit cell, we do this extraction with the aid of a non-commutative algebra package in Mathematica \cite{NCAlgebra}. The Bogoliubov transformation is then done by diagonalizing $g\mathcal{H}$ instead of $\mathcal{H}$, where $g$ is the 12x12 matrix,
\begin{equation}
g =  \begin{pmatrix}
    \mathds{1}_{6}       &  0 \\
    0      & -\mathds{1}_{6}  
\end{pmatrix}.
\end{equation}
The eigenvalues of $g H(q)$ give us six bands with dispersion $\pm \omega_{q \lambda}$, $\lambda = 1,\ldots 6$. In order to calculate the critical spin, we also need the transformation matrix to convert between the original bosons and the emergent spin waves.  These satisfy,
\begin{align}
\mathcal{T}g\mathcal{T}^{\dagger} & = g\cr
\mathcal{T}^{-1}g H\mathcal{T} & = g D,
\end{align}
where the first condition ensures bosonic commutation relations are always satisfied, and the second condition ensures that $\mathcal{T}$ diagonalizes $gH$ to obtain the diagonal matrix $g D$ of $\omega_{q\lambda}$.  If there are degenerate eigenvalues, one must be more careful \cite{delmaestro04}, but these transformation matrices can still be obtained.

A representative spin wave dispersion for non-collinear I is shown in Fig. \ref{zeroeta}, plotted in the new rectangular Brillouin zone from $\Gamma$ to $M = (\pi/\sqrt{3}a,\pi/a)$ to $X = (0,\pi/a)$.  There are six distinct bands, with four zero modes: one linear and one quadratic mode at the $\Gamma$ point and two linear modes at the corner of the Brillouin zone, $M$. The three linear modes are Goldstone modes associated with the complete breaking of the $SO(3)$ continuous symmetry by non-collinear order.  The quadratic mode is a zero mode associated with the classical degeneracy in $\eta$; as this degeneracy is lifted with $1/S$ corrections, this is an ``accidental'' classical degeneracy. Such a mode is also found in the collinear phase \cite{chubukov92}, and is an artifact of the $O(S)$ expansion; further $1/S$ corrections are expected to gap it out. 

\subsubsection{Critical spin}

\begin{figure}[h]
\centering
  \includegraphics[width=1\linewidth]{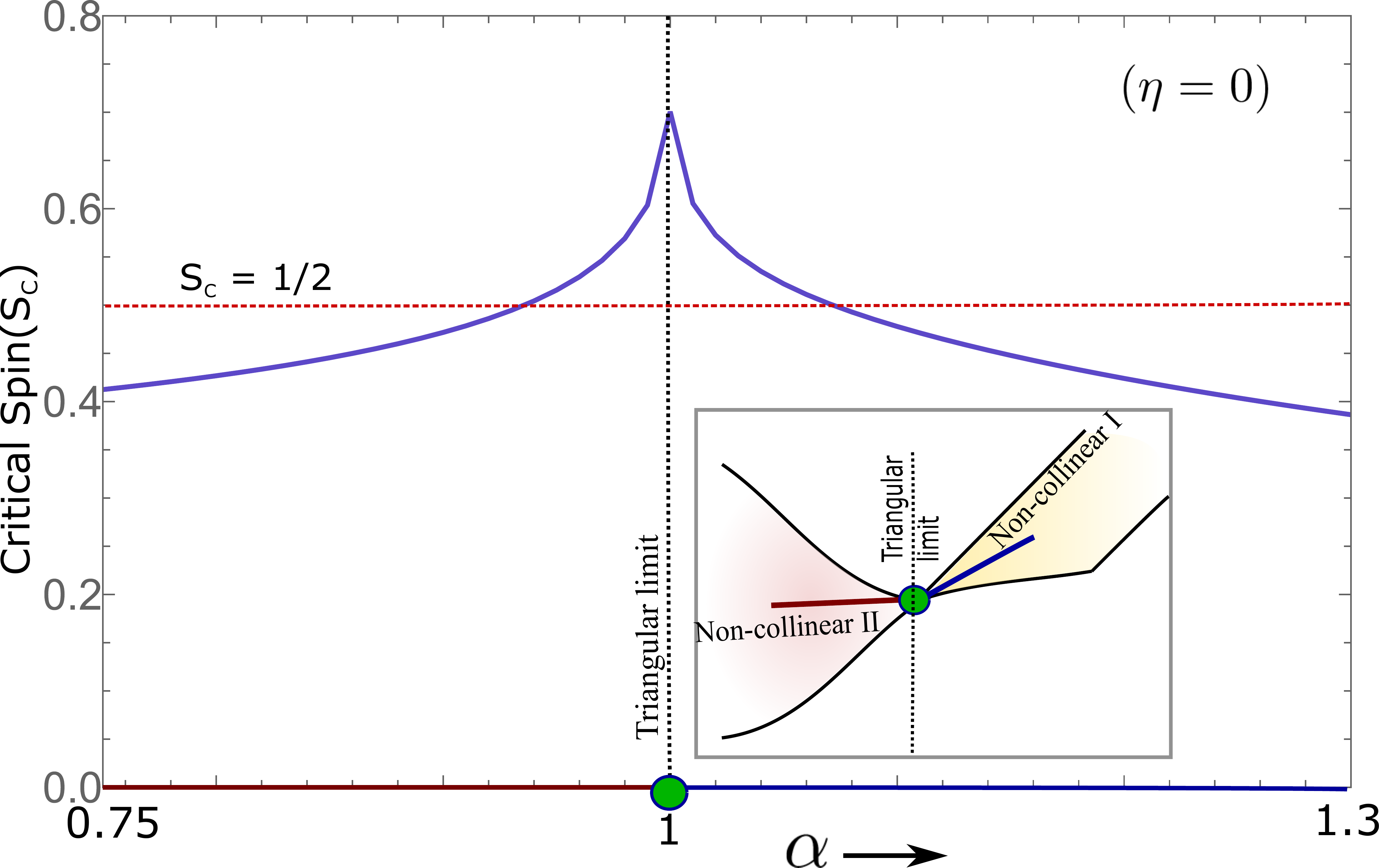}
\caption{Critical spin for the non-collinear phases, along the parametric path defined in Fig. \ref{mom} and shown in the inset. There is a substantial region around the triangular lattice critical point where $S_c > 1/2$ and we naively expect a quantum disordered state in this region, as is found on the triangular lattice line.}
 \label{c3}
\end{figure}

We can examine the reduction of the ordered moment due to quantum and thermal fluctuations for $\eta = 0$.  This reduction is given by,
\begin{align}
\begin{split}
\delta S &= \frac{1}{6N}\sum_{{\bf r},m}\left<b_m^{\dagger}({\bf r})b_m({\bf r})\right>\\
& = \frac{1}{6N}\sum_{{\bf q},m}\left<b_m^{\dagger}[{\bf q}]b_m[{\bf q}]\right>.
\end{split}
\end{align}
As these are the original bosons, $b$, we use the transformation matrices, $\mathcal{T}$ to rewrite,
\begin{equation}
\delta S = \frac{1}{2}\left(\frac{1}{6N}\sum_{\bf q}\sum_{n}[\mathcal{T}^{\dagger}\mathcal{T}]_{nn}-1\right)
\end{equation}
When $\delta S \geq S$, the ordered moment $\langle S \rangle$ is completely suppressed, and we define the critical spin, $S_c = \delta S$ that separates magnetic order for $S > S_c$ from an unknown quantum disordered phase.  In Fig. \ref{c3}, we plot the critical spin along a path that traverses both non-collinear phases.  There is a sizable region near the triangular limit where $S_c > 1/2$, and a quantum disordered phase is expected, at least in linear spin wave theory.  Thus we expect the spin liquid found surrounding the triangular lattice critical point to extend into a substantial region away from the triangular limit. 

\subsubsection{Finite temperatures: order by disorder and nematicity}
\label{monte}

We next turn to a classical Monte Carlo analysis at finite temperature, where we see that thermal fluctuations select the coplanar value, $\eta = 0$ of the free angle, and also see that a nematic order parameter develops at a finite temperature. In order to evaluate $\eta$ straightforwardly, we consider only the nematic order parameter for the AB sublattices,
\begin{align}
\mathcal{N}_{ab} = \frac{3}{2N}\sum_{i\in AB} & \langle \vec{S}_i \cdot \vec{S}_{i+\mathbf{a}_1}\rangle + \langle \vec{S}_i \cdot \vec{S}_{i+\mathbf{a}_2}\rangle \mathrm{e}^{\frac{2\pi i}{3}}\cr
& + \langle \vec{S}_i \cdot \vec{S}_{i+\mathbf{a}_2-\mathbf{a}_1}\rangle \mathrm{e}^{\frac{4\pi i}{3}},
\end{align}
where $i$ now sums over only the AB spins, and $N$ is the total number of sites.  Here, $\langle \cdots \rangle$ is the usual thermal average.

\begin{figure}[h]
 \label{fig:MonteCarlo}
\centering
  \includegraphics[width=1\linewidth]{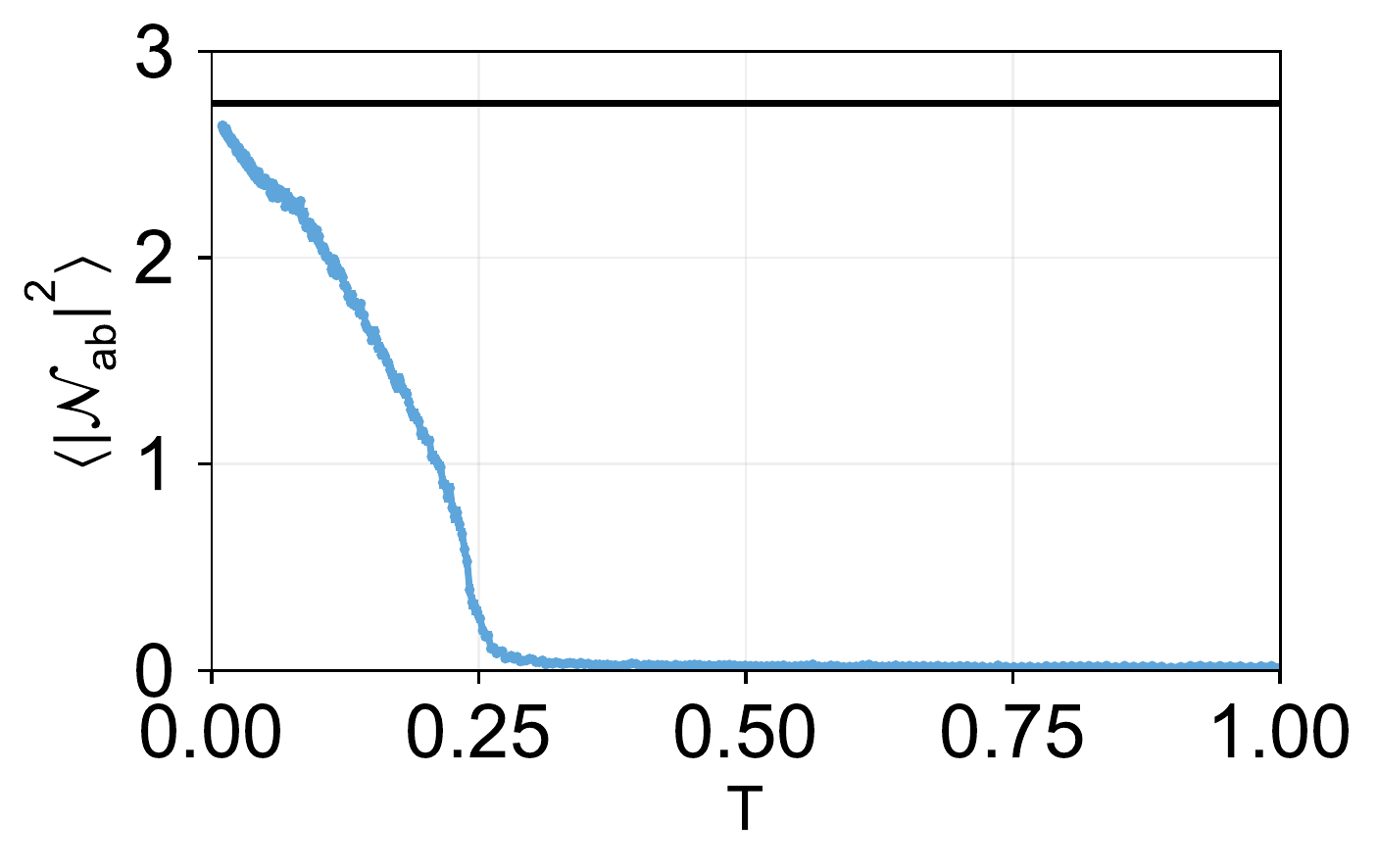}%{OPvsT.pdf}
%\caption{$D(T)$ vs $T$ for $10 \times 10$ unit cells at  $\{J_2/J=0.28, J'/J=1.6\}$ }
\vspace{-0.5cm}
\caption{$\langle|\mathcal{N}_{ab}|^2\rangle$ as a function of temperature for a lattice of $10 \times 10$ unit cells at $J_2/J_1=0.28, J'/J_1=1.6$. The nematic phase transition is seen at $T_N = J_1/4$, with some broadening due to the finite system size.  As $T \rightarrow 0$, the $\langle|\mathcal{N}_{ab}|^2\rangle$ approaches the line $4 \sin^4 \theta_A =  2.747$, indicating that the coplanar, $\eta = 0$ classical angle is selected.}
 \label{c4}
\end{figure}

We focus on a single point in the middle of the non-collinear I phase, $(J_2/J_1=0.28, J'/J_1=1.6)$, but expect the results to be generic to both phases. We use a lattice of 
$10 \times 10$ unit cells at a temperature $T_\textrm{min}=10^{-2}$.  To avoid problems with equilibration, we use parallel tempering with 3200 replicas with a temperature schedule such that replica $i$ has temperature $T_\textrm{min} (1.01)^i$. As the data will contain snapshots with all three values of the nematic order parameter, we calculate the Monte Carlo average of its modulus squared, $\langle|\mathcal{N}_{ab}|^2\rangle$. In the thermodynamic limit, this quantity is zero above the nematic transition, and at $T = 0$ becomes,
\begin{equation}
\langle|\mathcal{N}_{ab}|^2\rangle(T=0) = \frac{1}{2}(5+ 3 \cos 2 \eta) \sin^4 \theta_A.
\end{equation} 
For this point, $\theta_A = 1.144$ rad, and for $\eta = 0$, $\langle|\mathcal{N}_{ab}|^2\rangle \rightarrow 2.747$ as $T$ approaches zero.  $\langle|\mathcal{N}_{ab}|^2\rangle$ is shown as a function of temperature in Fig. \ref{c4}, where it turns on at $T_\mathcal{N} = J_1/4$ and clearly limits to 2.747 as $T \rightarrow 0$.

As expected, thermal fluctuations force both non-collinear phases to be coplanar.  In addition, while thermal fluctuations immediately melt the continuous magnetic order parameter, the nematic order parameter survives.  One might expect such a nematic order to also survive when quantum fluctuations melt these phases into a quantum spin liquid, and some signatures of such a nematic spin liquid have been seen on the triangular lattice limit\cite{zhu15,hu15,saadatmand15}.

\section{Conclusion}\label{con}

We have established the complete ground state classical phase diagram of the stuffed honeycomb lattice, which interpolates between the known honeycomb, triangular and dice lattice limits.  We find a wide variety of non-collinear and even non-coplanar phases, and reveal the transition between 120$^\circ$ and collinear order on the triangular lattice to be a multicritical point where four phases intersect. We have examined the structure and fluctuations of the two additional phases, and propose that the triangular lattice spin liquid extends into a substantial region in the stuffed honeycomb phase diagram. Future work will address the possible existence and nature of this spin liquid region.
\vskip 0.5in
\noindent{\bf Acknowledgments}
\vskip 0.04in
We acknowledge useful discussions with A. Chubukov, D. Freedman, M. Gingras, D. Johnston, T. McQueen and P. Orth. R.F. and J.S. were supported by NSF DMR-1555163. B. C. and D. K. were supported by SciDAC grant DE-FG02-12ER46875. This research is part of the Blue Waters sustained petascale computing project, which is supported by the National Science Foundation (award numbers OCI-0725070 and ACI-1238993) and the State of Illinois.  B.C. and R.F. also acknowledge the hospitality of the Aspen Center for Physics, supported by National Science Foundation Grant No. PHYS-1066293.

\bibliographystyle{apsrev4-1}
\bibliography{stuffed}

%\printbibliography
\end{document}